\RequirePackage{fix-cm} 
\documentclass[a4paper, twoside, reqno, 12pt]{amsart}
\usepackage{fixltx2e}   

\usepackage{etex}

\usepackage[latin1]{inputenc}
\usepackage[T1]{fontenc}

\usepackage{esint}
\usepackage{dsfont}
\usepackage{xspace}
\usepackage{amsgen}
\usepackage{amsthm}
\usepackage{amssymb}
\usepackage{amsmath}
\usepackage{wasysym}
\usepackage{upgreek}
\usepackage{amsfonts}
\usepackage{stmaryrd}
\usepackage{mathtools}

\usepackage[mathcal, mathscr]{euscript}

\usepackage{mathrsfs}
\DeclareMathAlphabet{\mathscrbf}{OMS}{mdugm}{b}{n}

\usepackage[lined, boxed, linesnumbered, english]{algorithm2e}

\usepackage{multicol}
\usepackage{multirow}

\usepackage{a4wide}

\usepackage[dvipsnames, table]{xcolor}
\definecolor{bckg}{RGB}{20.8, 20.8, 20.8}
\definecolor{oneblue}{rgb}{0.0, 0.0, 0.85}
\definecolor{Lightblue}{RGB}{214, 214, 214}
\definecolor{bluepigment}{rgb}{0.2, 0.2, 0.6}
\definecolor{charcoal}{rgb}{0.21, 0.27, 0.31}
\definecolor{denimblue}{rgb}{0.08, 0.38, 0.74}
\definecolor{Lightgray}{rgb}{0.89, 0.89, 0.89}
\definecolor{darkgrey}{rgb}{0.273, 0.281, 0.30}
\definecolor{darkelectricblue}{rgb}{0.33, 0.41, 0.47}

\usepackage[sort&compress, comma, square, numbers]{natbib}

\usepackage{psfrag}
\usepackage{graphicx}
\usepackage{subfigure}
\usepackage{morefloats}
\usepackage{indentfirst}

\usepackage{acronym}
\usepackage{microtype}
\usepackage[labelsep=period,%
            labelfont={bf,sf,color=bluepigment},%
            justification=raggedright]{caption}

\usepackage[perpage, symbol]{footmisc}

\usepackage[usenames, pdf]{pstricks}
\usepackage{epsfig}
\usepackage{pst-grad} 
\usepackage{pst-plot} 

\usepackage{url}
\usepackage[colorlinks,
           urlcolor=oneblue,
           linkcolor=denimblue,
           citecolor=NavyBlue,
           bookmarksopen=false,
           pdfpagemode=UseNone,
           pagebackref]{hyperref}

\usepackage[explicit]{titlesec}

\titleformat{\section}[block]
  {\color{NavyBlue}\Large\sffamily\bfseries}
  {}
  {0.0em}
  {\colorbox{bckg!5}{\strut\parbox{\dimexpr\linewidth-2\fboxsep\relax}{\thesection. #1}}}
  [\vspace*{0.33em}]

\titleformat{name=\section,numberless}[block]
  {\color{NavyBlue}\Large\sffamily\bfseries}
  {}
  {0.0em}
  {\colorbox{bckg!5}{\strut\parbox{\dimexpr\linewidth-2\fboxsep\relax}{#1}}}
  [\vspace*{0.33em}]

\titleformat{\subsection}
  {\color{NavyBlue}\large\sffamily\bfseries}
  {}
  {0.0em}
  {\colorbox{bckg!5}{\parbox{\dimexpr\linewidth-2\fboxsep\relax}{\thesubsection. #1}}}
  [\vspace*{0.33em}]

\titleformat{name=\subsection,numberless}
  {\color{NavyBlue}\Large\sffamily\bfseries}
  {}
  {0em}
  {\colorbox{bckg!5}{\parbox{\dimexpr\linewidth-2\fboxsep\relax}{#1}}}
  [\vspace*{0.33em}]

\titleformat{\subsubsection}
  {\color{bluepigment}\sffamily\normalsize\bfseries}
  {\thesubsubsection}
  {0.5em}
  {#1}
  [\vspace*{0.33em}]

\titleformat{\paragraph}[runin]
  {\color{bluepigment}\sffamily\small\bfseries}
  {}
  {0em}
  {#1}

\titlespacing{\section}{0.0em}{1.5em plus 2pt minus 2pt}%
{1.0em plus 2pt minus 2pt}[0em]
\titlespacing{\subsection}{0.5em}{1.5em plus 2pt minus 2pt}%
{1.0em}[0em]
\titlespacing{\subsubsection}{0.5em}{1.5em plus 2pt minus 2pt}%
{1.0em plus 2pt minus 2pt}[0em]

\usepackage{titletoc}

\contentsmargin{0.5em}
\newlength{\tocsep} 
\titlecontents{section}[\tocsep]
  {\addvspace{10pt}\bfseries\sffamily}
  {\contentslabel[\thecontentslabel]{\tocsep}}
  {}
  {\ \titlerule*[0.75pc]{.}\ \thecontentspage}
  []
\titlecontents{subsection}[\tocsep]
  {\addvspace{8pt}\sffamily}
  {\contentslabel[\thecontentslabel]{\tocsep}}
  {}
  {\ \titlerule*[0.5pc]{.}\ \thecontentspage}
  []
\titlecontents*{subsubsection}[\tocsep]
  {\addvspace{2pt}\footnotesize\sffamily}
  {}
  {}
  {\ \titlerule*[0.35pc]{.}\ \thecontentspage}
  [\\*]

\makeatletter
\def\@setauthors{%
  \begingroup
  \def\thanks{\protect\thanks@warning}%
  \trivlist
  \centering\footnotesize \@topsep30\p@\relax
  \advance\@topsep by -\baselineskip
  \item\relax
  \author@andify\authors
  \def\\{\protect\linebreak}%
  \textsc{\normalsize\textcolor{darkelectricblue}{\authors}}%
  \ifx\@empty\contribs
  \else
    ,\penalty-3 \space \@setcontribs
    \@closetoccontribs
  \fi
  \endtrivlist
  \endgroup
}
\def\@settitle{\begin{center}%
  \baselineskip14\p@\relax
    \bfseries
    \textsc{\Large\textcolor{charcoal}{\@title}}
  \end{center}%
}
\makeatother

\usepackage{enumitem}
\setlist[description]{%
  topsep=30pt,               
  itemsep=5pt,               
  font={\bfseries\sffamily\color{NavyBlue}}, 
}

\usepackage{fancyhdr}
\usepackage{lastpage}

\newcommand*\Title{\textcolor{bluepigment}{Estimation of sorption coefficients using the OED}}
\newcommand*\Authors{\textcolor{bluepigment}{J.~Berger, T.~Busser, D.~Dutykh \& N.~Mendes}}
\newcommand*{\plogo}{\textcolor{gray}{{\texttt{arXiv.org} / \textsc{hal}}}} 

\numberwithin{equation}{section}

\newcommand{\R}{\mathds{R}}

\newcommand*\unit[1]{ \, \mathsf{#1} \,}
\newcommand*\unitfrac[2]{\, \frac{\mathsf{#1}}{\mathsf{#2}} \,}

\newcommand{\scal}{\boldsymbol{\cdot}}
\newcommand*\pd[2]{\frac{\partial #1}{\partial #2}}
\renewcommand{\div}{\grad\scal}
\newcommand{\grad}{\boldsymbol{\nabla}}
\newcommand{\eqdef}{\mathop{\stackrel{\,\mathrm{def}}{:=}\,}}

\newcommand*\egal{\ = \ }
\newcommand*\plus{\ + \ }
\newcommand*\moins{\ - \ }
\newcommand{\f}{\mathrm{f}}
\newcommand*\e[1]{\cdot 10^{\,#1}}

\newcommand{\cref}{c_{\,0}}
\newcommand{\dref}{d_{\,0}}

\newcommand{\Ps}{P_{\,s}}
\newcommand{\Pv}{P_{\,v}}
\newcommand{\Pvi}{P_{\,v}^{\,i}}
\newcommand{\Pvref}{P_{\,v}^{\,\mathrm{ref}}}

\newcommand{\Rv}{R_{\,v}}
\newcommand{\TL}{T^{\,\infty}}
\newcommand{\tref}{t^{\,\mathrm{ref}}}

\newcommand{\as}{a^{\,\star}}
\newcommand{\ds}{d^{\,\star}}
\newcommand{\cs}{c^{\,\star}}
\newcommand{\xs}{x^{\,\star}}
\newcommand{\ts}{t^{\,\star}}
\newcommand{\Bi}{\mathrm{Bi}}
\newcommand{\Fo}{\mathrm{Fo}}
\newcommand{\Pe}{\mathrm{P\mbox{\'e}}}
\newcommand{\uL}{u_{\,\infty}}
\newcommand{\uc}{u_{\,\mathrm{c}}}
\newcommand{\ui}{u_{\,\mathrm{i}}}
\newcommand*\vP{\mathbf{P}}

\newcommand{\phiL}{\phi^{\,\infty}}

\newcommand{\RK}{\textsc{Runge}--\textsc{Kutta}}

\newcommand{\ie}{\emph{i.e.}\xspace}

\newcommand{\etc}{\emph{etc.}\xspace}

\begin{document}

\title[\Title]{An efficient method to estimate sorption isotherm curve coefficients}

\author[J.~Berger]{Julien Berger$^*$}
\address{\textbf{J.~Berger:} LOCIE, UMR 5271 CNRS, Universit\'e Savoie Mont Blanc, Campus Scientifique, F-73376 Le Bourget-du-Lac Cedex, France}
\email{Berger.Julien@univ-smb.fr}
\urladdr{https://www.researchgate.net/profile/Julien\_Berger3/}
\thanks{$^*$ Corresponding author}

\author[T.~Busser]{Thomas Busser}
\address{\textbf{T.~Busser:} LOCIE, UMR 5271 CNRS, Universit\'e Savoie Mont Blanc, Campus Scientifique, F-73376 Le Bourget-du-Lac Cedex, France}
\email{Thomas.Busser@univ-smb.fr}
\urladdr{https://www.researchgate.net/profile/Thomas\_Busser/}

\author[D.~Dutykh]{Denys Dutykh}
\address{\textbf{D.~Dutykh:} Univ. Grenoble Alpes, Univ. Savoie Mont Blanc, CNRS, LAMA, 73000 Chamb\'ery, France and LAMA, UMR 5127 CNRS, Universit\'e Savoie Mont Blanc, Campus Scientifique, F-73376 Le Bourget-du-Lac Cedex, France}
\email{Denys.Dutykh@univ-smb.fr}
\urladdr{http://www.denys-dutykh.com/}

\author[N.~Mendes]{Nathan Mendes}
\address{\textbf{N.~Mendes:} Thermal Systems Laboratory, Mechanical Engineering Graduate Program, Pontifical Catholic University of Paran\'a, Rua Imaculada Concei\c{c}\~{a}o, 1155, CEP: 80215-901, Curitiba -- Paran\'a, Brazil}
\email{Nathan.Mendes@pucpr.edu.br}
\urladdr{https://www.researchgate.net/profile/Nathan\_Mendes/}



\begin{titlepage}
\thispagestyle{empty} 
\noindent
{\Large Julien \textsc{Berger}}\\
{\it\textcolor{gray}{Polytech Annecy--Chamb\'ery, LOCIE, France}}
\\[0.02\textheight]
{\Large Thomas \textsc{Busser}}\\
{\it\textcolor{gray}{Polytech Annecy--Chamb\'ery, LOCIE, France}}
\\[0.02\textheight]
{\Large Denys \textsc{Dutykh}}\\
{\it\textcolor{gray}{CNRS--LAMA, Universit\'e Savoie Mont Blanc, France}}
\\[0.02\textheight]
{\Large Nathan \textsc{Mendes}}\\
{\it\textcolor{gray}{Pontifical Catholic University of Paran\'a, Brazil}}
\\[0.08\textheight]

\vspace*{0.7cm}

\colorbox{Lightblue}{
  \parbox[t]{1.0\textwidth}{
    \centering\huge\sc
    \vspace*{0.7cm}
    
    \textcolor{bluepigment}{An efficient method to estimate sorption isotherm curve coefficients}
    
    \vspace*{0.7cm}
  }
}

\vfill 

\raggedleft     
{\large \plogo} 
\end{titlepage}


\newpage
\thispagestyle{empty} 
\par\vspace*{\fill}   
\begin{flushright} 
{\textcolor{denimblue}{\textsc{Last modified:}} \today}
\end{flushright}


\newpage
\maketitle
\thispagestyle{empty}


\begin{abstract}

\vspace*{1.5cm}
This paper deals with an inverse problem applied to the field of building physics to experimentally estimate three sorption isotherm coefficients of a wood fiber material. First, the mathematical model, based on convective transport of moisture, the Optimal Experiment Design (OED) and the experimental set-up are presented. Then measurements of relative humidity within the material are carried out, after searching the OED, which is based on the computation of the sensitivity functions and \emph{a priori} values of the unknown parameters employed in the mathematical model. The OED enables to plan the experimental conditions in terms of sensor positioning and boundary  conditions out of $20$ possible designs, ensuring the best accuracy for the identification method and, thus, for the estimated parameter. Two experimental procedures were identified: i) single step of relative humidity from $10 \unit{\%}$ to $75 \unit{\%}$ and ii) multiple steps of relative humidity $10-75-33-75 \unit{\%}$ with an $8$-day duration period for each step. For both experiment designs, it has been shown that the sensor has to be placed near the impermeable boundary. After the measurements, the parameter estimation problem is solved using an interior point algorithm to minimize the cost function. Several tests are performed for the definition of the cost function, by using the $L^{\,2}$ or $L^{\,\infty}$ norm and considering the experiments separately or at the same time. It has been found out that the residual between the experimental data and the numerical model is minimized when considering the discrete Euclidean norm and both experiments separately. It means that two parameters are estimated using one experiment while the third parameter is determined with the other experiment. Two cost functions are defined and minimized for this approach. Moreover, the algorithm requires less than $100$ computations of the direct model to obtain the solution. In addition, the OED sensitivity functions enable to capture an approximation of the probability distribution function of the estimated parameters. The determined sorption isotherm coefficients calibrate the numerical model to fit better the experimental data. However, some discrepancies still appear since the model does not take into account the hysteresis effects on the sorption capacity. Therefore, the model is improved proposing a second differential equation for the sorption capacity to take into account the hysteresis between the main adsorption and desorption curves. The OED approach is also illustrated for the estimation of five of the coefficients involved in the hysteresis model. To conclude, the prediction of the model with hysteresis are compared with the experimental observations to illustrate the improvement of the prediction.

\bigskip
\noindent \textbf{\keywordsname:} Optimal Experiment Design (OED); parameter estimation problem; convective moisture transfer; sensitivity functions; sorption moisture coefficients; hysteresis \\

\pagestyle{empty}

\end{abstract}


\newpage
\tableofcontents
\thispagestyle{empty}


\newpage
\section{Introduction}

Moisture in buildings has been a subject of major concern since the eighties. It may affect energy consumption and demand so we can mention at least four International Energy Agency projects conducted in the last $30$ years to promote global research on this subject (Annexes $14\,$, $24\,$, $41$ and $55$) \cite{Administration2015}. Furthermore, moisture can also have a dramatic impact on occupants' health and on material deterioration. Several tools have been developed to simulate the moisture transport in constructions as described in \cite{Woloszyn2008}, which can be used to predict conduction loads associated to porous elements and mold growth risk in building enclosures. Nevertheless, those tools require input parameters containing temperature- and moisture-dependent hygrothermal properties.


\subsection{Moisture transport in constructions}

The following system of differential equations established by \textsc{Luikov} \cite{Luikov1966} represents the physical phenomenon of heat and mass transfer through capillary porous materials:
\begin{subequations}\label{eq:Luikov_eq}
\begin{align}\label{eq:Luikov_eq_b}
\pd{U}{t} & \egal \div \bigl(\, a_{\,m} \, \grad U \plus \delta \, a_{\,m} \grad T \,\bigr) \,, \\
c_{\,b} \, \rho_{\,0}\; \pd{T}{t} & \egal \div \bigl(\, \lambda \grad T \,\bigr) \plus r_{\,12}\, \div \biggl(\, a_{\,m1} \, \rho_{\,0}\, \bigl(\, \grad U \plus \delta_{\,1} \grad T \,\bigr) \,\biggr)\,,
\end{align}
\end{subequations}
where $U$ is the relative concentration of moisture in the porous body, $T$ the temperature, $a_{\,m}$ the mass transfer coefficient for vapor (denoted with the subscript $1$) and liquid inside the body, $\delta$ the thermal-gradient coefficient, $\rho{\,0}$ the specific mass of the dry body, $c_{\,b}$ the specific heat of the body and, $r_{\,12}$ the latent heat of vaporization.

In building physics, those equations represent the physics that occurs in the building porous envelope and indoor porous elements such as furniture, textiles, \etc. Regarding the envelope, the phenomenon is investigated to analyze the influence of moisture transfer on the total heat flux passing through the wall, with the objective of estimating the heat losses. They are also studied to analyze the durability of walls and to avoid disorders due to the presence of moisture as, for instance, mold growth, shrinking or interstitial condensation. This aspect is of major importance for wall configurations involving several materials with different properties, where moisture can be accumulated at the interface between two materials. Durability problems may also appear when considering important moisture sources as wind driven rain or rising damp problems. These analyses are performed by computing solutions to the partial differential equations. For this, numerical methods are used due to the nonlinearity of the material properties, depending on moisture content and temperature, and the non-stationary boundary conditions, defined as \textsc{Robin}-type and varying according to climatic data. Most of the numerical approaches consider standard discretization techniques. For the time discretisation, the \textsc{Euler} implicit \cite{Mendes1999, Mendes2005, IBP2005} or explicit \cite{Kalagasidis2007} schemes are adopted. Regarding the spatial discretisation, works in \cite{Mendes2017} are based on finite-differences methods, in \cite{Mendes2004, Mendes2005, BauklimatikDresden2011, IBP2005} on finite-volume methods and in \cite{Rouchier2013, Janssen2007} on finite-element methods. It is important to note that the solution of the equations requires the calculation of large systems of nonlinear equations (an order of $10^{\,6}$ for $3$-D problems). Furthermore, the problem deals with different time scales. The diffusive phenomena and the boundary conditions evolve on the time scale of seconds or minutes while the building performance usual analysis is done for a time interval of one year or even longer when dealing with durability or mold growth issues. Thus, the computation of heat and moisture transfer in porous material in building physics has a non-negligible computational cost. Recently, innovative and efficient methods of numerical simulation have been proposed. Some improved explicit schemes, enabling to overcome the stability restrictions of standard \textsc{Euler} explicit schemes, have been proposed in \cite{Gasparin2017, Gasparin2017b}. An accurate and fast numerical scheme based on the \textsc{Scharfetter}--\textsc{Gummel} idea has been proposed in \cite{Berger2017a} to solve the advection-diffusion moisture differential equation. Some attempts based on model reduction methods have been also proposed with an overview in \cite{Mendes2017}.


\subsection{Inverse problems in building physics}

While some research focuses on numerical methods to compute the solution of the so-called direct problem to analyze the physical phenomena, some studies aim at solving inverse problems of heat and mass transfer in porous materials. In this case, the focus is the estimation of material properties $\bigl(\, \gamma_{\,0}^{\,\circ} \,,\, c^{\,\circ} \,,\, a_{\,m}^{\,\circ} \,,\, \delta^{\,\circ} \,\bigr)$ using experimental data denoted as $\bigl(\, T_{\,\mathrm{exp}} \,,\, U_{\,\mathrm{exp}} \,\bigr)$ by minimizing a thoroughly chosen cost function $\mathrm{J}\,$:
\begin{align*}
\bigl(\, \gamma_{\,0}^{\,\circ} \,,\, c^{\,\circ} \,,\, a_{\,m}^{\,\circ} \,,\, \delta^{\,\circ} \,\bigr) 
& \egal \mathrm{arg} \min \; \mathrm{J} \,, \\
\text{with } \mathrm{J} 
 \egal \biggl| \biggl|\, T_{\,\mathrm{exp}} &\moins T \,,\, U_{\,\mathrm{exp}} \moins U \, \biggr| \biggr| \,,
\end{align*}
where $ \biggl| \biggl|\, \ldots \, \biggr| \biggr|$ is certain vector-norm in time.

Here, the inverse problem is an \emph{inverse medium problem}, as it aims at estimating the coefficient of the main equation \cite{Kabanikhin2008, Kabanikhin2011}.Two contexts can be distinguished. First, when dealing with existing buildings to be retrofitted, samples cannot be extracted from the walls to determine their material properties. Therefore, some \emph{in-situ} measurements are carried out according to a non-destructive design. The experimental data can be gathered by temperature, relative humidity, flux sensors and infrared thermography, among others. In most of the case, measurements are made at the boundary of the domain. From the obtained data, parameter estimation enables to determine the material thermo-physical properties. As mentioned before, the properties are moisture and temperature dependent. Therefore, the parameter identification problem needs to estimate functions that can be parameterized. Moreover, in such investigations, there is generally a few \emph{a priori} information on the material properties. In \cite{Berger2016b}, the thermal conductivity of an old historic building wall composed of different materials is presented. In \cite{Rouchier2016}, the thermo-physical properties of materials composing a wall are estimated. In \cite{Nassiopoulos2013}, the heat capacity and the thermal conductivity of a heterogeneous wall are determined. Once this parameter estimated, efficient simulations using the direct model can be performed to predict the wall conduction loads and at the end choose adequate retrofitting options.

\begin{figure}
\centering
\includegraphics[width=0.99\textwidth]{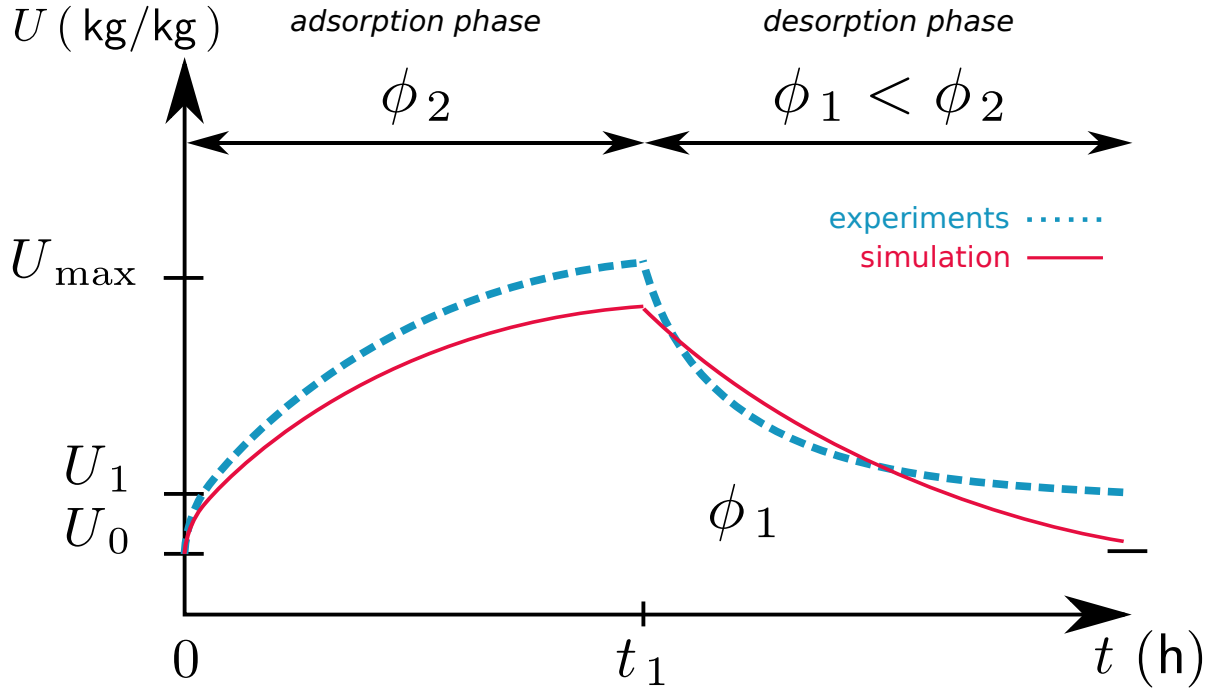}
\caption{\small\em Illustration of the discrepancies observed when comparing experimental data to results from numerical models of moisture transfer in porous material.}
\label{intro_fig:err_mod_exp}
\end{figure}

Another issue arises when comparing the numerical model results and experimental data. Some discrepancies were observed as reported in several studies \cite{James2010, Berger2017a} and illustrated in Figure~\ref{intro_fig:err_mod_exp}. A material, with an initial moisture content $U_{\,0}\,$, is submitted to an adsorption and desorption cycles. The moisture content raises up to $U_{\,\mathrm{max}}$ at the end of the adsorption phase. Then, during the desorption phase, the moisture content decreases until a value $U_{\,1} \ > \ U_{\,0}\,$ due to the hysteresis effects. When comparing the simulation results to the experimental data, it is observed that predictions of moisture content commonly underestimate the experimental observations of moisture adsorption processes. In other words, the numerical predictions are lower than the experimental values obtained during the adsorption phase. On the other hand, in the course of the desorption phase, the numerical predictions often overestimate the experimental observations, \ie the simulation values are greater than the experimental ones. At the end of the desorption process, the experimental values are greater again, when compared to both the prediced values and the initial moisture content since thehysteresis phenomenon significantly affects the material moisture sorption capacity. It means the experimental moisture front rushes faster than the numerical predictions. To answer this issue, models can be calibrated using \emph{in-situ} measurements for adapting the material properties to reduce the discrepancies between model predictions and real observations. In \cite{Kanevce2005}, moisture- and temperature-dependent diffusivity and thermophysical properties are estimated using only temperature measurements under a drying process. In \cite{Berger2017b}, the moisture permeability and advective coefficients are estimated using relative humidity measurements in a wood fiber material. In these cases, \emph{a priori} information on the material properties is known thanks to complementary measurements based on well-established standards.

In terms of methodology for the estimation of parameters, several approaches can be distinguished in literature. Descent algorithms, based on the \textsc{Levenberg}--\textsc{Marquardt} nonlinear least-square method, are used in \cite{Nassiopoulos2013}. Stochastic approaches, using \textsc{Bayesian} inferences and the \textsc{Markov} chain \textsc{Monte}-\textsc{Carlo} algorithms, are applied in \cite{Berger2016b, Rouchier2017, Biddulph2014, Dubois2014}. Approaches based on genetic algorithms based approaches are adopted to minimize the cost function in \cite{Xu2007, Czel2012}. Model reduction techniques, based on Proper Orthogonal Decomposition (POD), are employed in \cite{DelBarrio2003}.


\subsection{Problem statement}

This article presents an efficient ethod for the estimation of moisture sorption isotherm coefficients of a wood fiber material, represented by three parameters, using an experimental facility, aiming at reducing the discrepancies between model predictions and real performance. The estimation of the unknown parameters, based on observed data and identification methods, strongly depends on the experimental protocol. In particular, the boundary conditions imposed to the material and the location of sensors play a major role. In \cite{Berger2017}, the concept of searching the Optimal Experiment Design (OED) was used to determine the best experimental conditions in terms of imposed flux, quantity and location of sensors, aiming at estimating the thermophysical and hygrothermal properties of a material. The OED provides the best accuracy of the identification method and, thus, the estimated parameters.

Therefore, the main issue in this paper is to use the method to determine the OED considering the experimental set-up before starting the data acquisition process. First, the optimal boundary conditions and location of sensors are defined. Then, the experimental campaign is carried out, respecting the OED. From the experimental data, the parameters are estimated using an interior-point algorithm with constraints on the unknowns. To improve the fidelity of the physical phenomenon, the OED approach is extended to a model that takes into account the hysteresis effects on the sorption curves. Lastly, a comparison between the numerical predictions and the experimental observations reveals smaller discrepancies and a satisfactory representation of the physical phenomena.

This article is organized as follows. Section~\ref{sec:methodology} presents the physical problem with its mathematical formulation and the OED methodology. In Section~\ref{sec:exp_facility}, the existing experimental set-up is described. The OED search providing the different possibilities for the estimation of one or several parameters with the experimental set-up is presented in Section~\ref{sec:OED}. The parameters are estimated in Section~\ref{sec:parameter_estimation_problem} and then the main conclusions are finally outlined.


\section{Methodology}
\label{sec:methodology}

\subsection{Physical problem}
\label{sec:physical_problem}

In this section, the physical problem of moisture transport in porous material is described. The system of Eq.~\eqref{eq:Luikov_eq} proposed by \textsc{Luikov} has been established for non-isothermal conditions. In our study, the temperature variations in the material are assumed negligible. Therefore, in section~\ref{sec:moisture_transport}, the derivation of the unidimensional advective--diffusive moisture transport equation is detailed. Then, the assumption on the material properties are defined. In the last section~\ref{sec:boundary_conditions}, the boundary conditions of the problem are specified.


\subsubsection{Moisture transport}
\label{sec:moisture_transport}

The term moisture includes the water vapor, denoted by index $1$ and the liquid water, denoted by index $2\,$, both migrating through the porous matrix of the material. The differential equation describing the mass conservation for each species can be formulated as \cite{Luikov1966}: 
\begin{align}
\label{eq:conservation_mass}
\pd{w_{\,i}}{t} & \egal - \; \grad \scal \boldsymbol{j}_{\,i} \plus I_{\,i} \,, && i \ \in \ \bigl\{\, 1 \,,\, 2 \,\bigr\} \,,
\end{align}
where $w_{\,i}$ is the volumetric concentration of species $i\,$, $\boldsymbol{j}_{\,i}\,$, the total flux of the species $i$ and, $I_{\,i}\,$, the volumetric source. It is assumed that solid water is not present in the porous structure and $I_{\,1} \plus I_{\,2} \egal 0 \,$. Thus, by summing Eq.~\eqref{eq:conservation_mass} for $i \ \in \ \bigl\{\, 1 \,,\, 2 \,\bigr\}\,$, we obtain:
\begin{align}\label{eq:conservation_moisture}
  \pd{}{t} \, \bigl(\, w_{\,1} \plus w_{\,2} \,\bigr) & \egal - \; \grad \scal \bigl(\, \boldsymbol{j}_{\,1} \plus \boldsymbol{j}_{\,2} \,\bigr)
\end{align}
The total volumetric concentration of moisture is denoted by
\begin{align*}
  w \ \eqdef \ w_{\,1} \plus w_{\,2} \,.
\end{align*}
It can be noted that we have the following relation $w \egal \rho_{\,0} \, U$ between the potentials $w$ and $U$ from Eq.~\eqref{eq:Luikov_eq_b} derived by \textsc{Luikov}. Since the temperature variations in the material are assumed negligible, we can write:
\begin{align*}
  & \pd{w}{t} \egal \pd{w}{\phi} \, \cdot \, \pd{\phi}{\Pv} \, \cdot \, \pd{\Pv}{t}\,,
\end{align*}
where $\phi$ is the relative humidity. By its definition
\begin{align*}
  & \phi \egal \frac{\Pv}{\Ps} \,,
\end{align*}
thus, we have:
\begin{align*}
  & \dfrac{\partial \, \phi}{\partial \, \Pv} \egal \dfrac{1}{\Ps} \,,
\end{align*}
with $\Ps$ the saturation pressure. Considering the moisture sorption curve describing the material property between the moisture content $w$ and the relative humidity $\phi\,$, denoted as $w \egal \f \, (\,\phi\,)$, it can be written: 
\begin{align}\label{eq:transient_term}
  & \pd{w}{t} \egal \pd{\, \f \, (\phi)}{\phi} \; \frac{1}{\Ps} \; \pd{\Pv}{t} \,.
\end{align}
We define the moisture storage coefficient as 
\begin{align}\label{eq:coeff_c}
  c & \ \eqdef \ \pd{\, \f \, (\phi)}{\phi} \; \frac{1}{\Ps} \,.
\end{align}

Moreover, the moisture transfer occurs due to capillary migration, moisture diffusion and advection of the vapor phase. Here, advective transfer represents the movement of species under the presence of airflow through the porous matrix. The convection process designates both diffusion and advection transfer. Thus, the fluxes can be expressed as \cite{Berger2017a, Belleudy2016, Delleur2006}:
\begin{align}\label{eq:mass_flux}
  \boldsymbol{j}_{\,1} \plus \boldsymbol{j}_{\,2} \egal - \; d \, \grad \, \Pv \plus \boldsymbol{a} \, \Pv   \,,
\end{align}
where $\Pv$ is the vapor pressure, $d$ is the global moisture transport coefficient and $a$ the global advection coefficient defined as: 
\begin{align*}
  a & \ \eqdef \ \dfrac{\mathsf{\boldsymbol{\mathsf{v}}}}{\Rv \ T} \,,
\end{align*}
where $T$ is the fixed temperature, $\boldsymbol{\mathsf{v}}$ the constant mass average velocity and, $\Rv$ the water vapor gas constant.

Therefore, using Eq.~\eqref{eq:conservation_moisture} and the results from Eqs.~\eqref{eq:transient_term} and \eqref{eq:mass_flux}, the physical problem of unidimensional advective--diffusive moisture transport through a porous material can be mathematically described as:
\begin{align}\label{eq:HAM_equation2}
  & c \; \pd{\Pv}{t} \egal \pd{}{x} \Biggl[ \, d \, \pd{\Pv}{x} \, \Biggr] \moins a \, \pd{\Pv}{x} \,.
\end{align}


\subsubsection{Material properties}

The moisture capacity $c$ is assumed to be a second-degree polynomial of the relative humidity, while the moisture permeability $d$ is considered as a first-degree polynomial of the relative humidity:
\begin{subequations}
\begin{align}\label{eq:mat_hypothesis}
  & c \egal c_{\,0} \plus c_{\,1} \, \phi \plus c_{\,2} \, \phi^{\,2} \,, \\
  & d \egal d_{\,0} \plus d_{\,1} \, \phi \,.
\end{align}
\end{subequations}

These assumptions correspond to a third-order polynomial function of the relative humidity for the moisture sorption curve. It should be noted that other functions can be used to describe the moisture sorption curve $c$ and/or the moisture permeability $d\,$. However, for the material under investigation, these properties have been determined using the traditional cup method and the gravimetric approach, presented in preliminary studies \cite{Rafidiarison2015, Vololonirina2014} and expressed using these functions. These properties will be used as \emph{a priori} ones in the algorithm when searching the OED in Section~\ref{sec:OED}. It is important to stress out that the hysteresis of the sorption curve is not considered in the physical model.


\subsubsection{Boundary conditions}
\label{sec:boundary_conditions}

At $x \egal 0 \,$, the surface is in contact with the ambient air at temperature $\TL$ and relative humidity $\phiL \,$. Thus, the boundary condition is expressed as:
\begin{align}\label{eq:HAM_BC_L}
  d \; \pd{\Pv}{x} \moins a \, \Pv & \egal h \, \biggl(\, \Pv - \Ps\,\bigl(\, \TL \,\bigr) \cdot \phiL\,(\,t\,) \,\biggr) \,,
\end{align}
where $h$ is the convective vapor transfer coefficient, considered as constant. 
At $x \egal L\,$, the surface is impermeable. Thus, the total moisture flow vanishes at this boundary where the velocity $\mathsf{v}$ and the diffusion flow $d \, \pd{\Pv}{x}$ are null: 
\begin{align}\label{eq:HAM_BC_R}
  d \; \pd{\Pv}{x} \moins a \, \Pv & \egal 0 \,.
\end{align}
At $t \egal 0$, the vapor pressure is supposed to be uniform within the material
\begin{align}\label{eq:HAM_ic}
 \Pv &\egal \Pvi \,.
\end{align}
It is of great concern in the construction of the numerical model that the boundary conditions satisfies the initial condition.


\subsection{Dimensionless formulation}
\label{sec:dimensionless_formulation}

For building porous materials such as concrete, insulation and brick, the coefficients scales as $10^{\,2} \ \unit{kg/(m^{\,3}.Pa)} $ for the sorption curve $c$ and $10^{\,-11} \ \unit{s}$ for the moisture permeability $d$ and the advection coefficients $a \,$. Therefore, while performing a mathematical and numerical analysis of a given practical problem, it is of capital importance to obtain a unitless formulation of governing equations. For this, the vapor pressure is transformed to a dimensionless quantity: 
\begin{align*}
& u \egal \frac{\Pv}{\Pvref} \,,
&& \ui \egal \frac{\Pvi}{\Pvref} \,,
&& \uL \egal \frac{\Ps \bigl(\, \TL \,\bigr) \ \phiL}{\Pvref} \,.
\end{align*}
The time and space domains are also modified:
\begin{align*}
& \xs \egal \frac{x}{L} \,, 
&& \ts \egal \frac{t}{\tref} \,, 
\end{align*}
where $L$ is the thickness of the material and $\tref$ a characteristic time. The material properties are changed considering a reference value for each parameter: 
\begin{align*}
& \cs \egal \frac{c}{\cref} \,,
&& \ds \egal \frac{d}{\dref} \,.
\end{align*}
In this way, dimensionless numbers are highlighted: 
\begin{align*}
& \Pe \egal \frac{a \cdot L}{\dref} \,, 
&& \Bi \egal \frac{h \cdot L}{\dref}  \,,
&& \Fo \egal \frac{\tref \cdot \dref}{L^{\,2} \cdot \cref} \,.
\end{align*}

The dimensionless moisture \textsc{Biot} number $\Bi$ quantifies the importance of the moisture transfer at the bounding surface of the material. The transient transfer mechanism is characterised by the \textsc{Fourier} number $\Fo$ whereas the \textsc{P\'eclet} number $\Pe$ measures only the importance of moisture advection. The quantities $\cs\,(\,u\,)$ and $\ds\,(\,u\,)$ give the variation of storage and permeability coefficients from the reference state of the material. The dimensionless governing equations are finally written as:
\begin{subequations}\label{eq:moisture_dimensionlesspb_1D}
  \begin{align}
   \cs(\,u\,)  \ \pd{u}{\ts} &\egal \Fo \ \pd{}{\xs} \Biggl(\, \ds(\,u\,)  \ \pd{u}{\xs} \moins \Pe \ u \, \Biggr) \,,
  & \ts & \ > \ 0\,, \;&  \xs & \ \in \ \big[ \, 0, \, 1 \, \big] \,, \\[3pt]
   \ds(\,u\,) \, \pd{u}{\xs} & \moins \as \, u \egal \Bi \cdot \left( \, u \moins \uL \, \right) \,,
  & \ts & \ > \ 0\,, \,&  \xs & \egal 0 \,, \\[3pt]
   \ds(\,u\,) \, \pd{u}{\xs} & \moins \as \, u \egal 0 \,,
  & \ts & \ > \ 0\,, \,&   \xs & \egal 1 \,, \\[3pt]
   u & \egal \ui \,,
  & \ts & \egal 0\,, \,&  \xs & \ \in \ \big[ \, 0, \, 1 \, \big] \,.
  \end{align}
\end{subequations}
where functions $\cs(\,u\,)$ and $\ds(\,u\,)$ are given by:
\begin{align*}
  \cs(\,u\,) & \egal 1 \plus c^{\,\star}_{\,1} \, u \plus c^{\,\star}_{\,2} \, u^{\,2} \,, \\
  \ds(\,u\,) & \egal 1 \plus d^{\,\star}_{\,1} \, u \,.
\end{align*}

In this study, the reference parameters correspond to the conditions at $T \egal 23 \ \unit{^{\circ}C} $ and $\phi \egal 0.5\,$, which gives a vapor pressure of $\Pvref \egal 1404 \ \unit{Pa}\,$. Since the temperature is assumed as constant, a dimensionless value $u \egal 2$ corresponds to $\phi \egal 1\,$. The field varies within the interval $u \ \in \ \bigl[\,0\,,\,2\,\bigr]\,$.

The direct problem, defined by Eq.~\eqref{eq:moisture_dimensionlesspb_1D}, is solved using a finite-difference standard discretisation method. An embedded adaptive in time \textsc{Runge}--\textsc{Kutta} scheme  combined with a \textsc{Scharfetter}--\textsc{Gummel} spatial discretisation  approach, is used \cite{Berger2017a}. It is adaptive and embedded to estimate local error in time with low extra cost. The algorithm was implemented in the \texttt{Matlab\texttrademark} environment. For the sake of notation compactness, the upper-script $\star$ standing for dimensionless parameters, is no longer used.


\subsection{The Optimal Experiment Design}
\label{subsec:OED}

Efficient computational algorithms for recovering parameters $\vP$ given an observation $u_{\, \mathrm{exp}}$ of the field $u\, (x,\, t)$ have already been proposed. Readers may refer to \cite{Ozisik2000} for a primary overview of different methods. They are based on the minimization of the cost function $\mathrm{J} \, [ \, \vP \, ] \,$. For this purpose, it is required to equate to zero the derivatives of $\mathrm{J} \, [ \, \vP \, ] \,$, with respect to each of the unknown parameters $p_{\,m}$ to find critical points. Associated to this necessary condition for the minimization of $\mathrm{J} \, [ \, \vP \, ]$, the scaled dimensionless local sensitivity function \cite{Finsterle2015, Walter1990} is introduced:
\begin{align}\label{sec1_eq:sensitivity_matrix}
  & \Theta_{\,m} \,(x,t)\ =\ \frac{\sigma_{\,p}}{\sigma_{\,u}} \; \pd{u}{p_{\,m}} \,, && \forall\, m \ \in \   \bigl\{ 1, \ldots, M \, \bigr\} \,,
\end{align}
where $\sigma_{\,u}$ is the variance of the error measuring $u_{\, \mathrm{exp}}\,$. The parameter scaling factor $\sigma_{\,p}$ equals to $1$ as we consider that prior information on parameter $p_{\,m}$ has low accuracy. It is important to note that all algorithms have been developed considering the dimensionless problem in order to compare only the order of variation of parameters and observation, avoiding the effects of units and scales.

The function $\Theta_{\,m}$ measures the sensitivity of the estimated field $u$ with respect to changes in the parameter $p_{\,m}$ \cite{Nenarokomov2005, Ozisik2000,Artyukhin1985}. A small magnitude of $\Theta_{\,m}$ indicates that large changes in $p_{\,m}$ induce small changes in $u\,$. The estimation of parameter $p_{\,m}$ is therefore difficult, in this case. When the sensitivity coefficient $\Theta_{\,m}$ is small, the inverse problem is necessarily ill-conditioned. If the sensitivity coefficients are linearly dependent, the inverse problem is also ill-posed. Therefore, to get an optimal evaluation of parameters $\vP \,$, it is desirable to have linearly-independent sensitivity functions $\Theta_{\,m}$ with large magnitudes for all parameters $p_{\,m} \,$. These requirements ensure the best conditioning of the computational algorithm to solve the inverse problem and thus the better accuracy of the estimated parameter.

It is possible to define the experimental design in order to meet these requirements. The issue is to find the optimal sensor location $X^{\,\circ}$ and the optimal amplitude $\phi^{\,\infty, \,\circ}$ of the relative humidity of the ambient air at the material bounding surface, $x \egal 0 \,$. To search this optimal experiment design, we introduce the following measurement plan:
\begin{align}\label{sec1_eq:measurement_plan}
  & \pi\ \eqdef\ \bigl\{ \, X \ ,\ \phi^{\,\infty} \, \bigr\}\,.
\end{align}

In the analysis of optimal experiments for estimating the unknown parameter(s) $\vP \,$, a quality index describing the recovering accuracy is the $D-$optimum criterion \cite{Beck1977, VandeWouwer2000, Fadale1995, Emery1998, Anderson2005, Alifanov1995}:
\begin{align}\label{sec1_eq:D_optimum}
  \Psi \egal \det \bigl[ \, F\,(\,\pi\,) \, \bigr] \,,
\end{align}
where $F\,(\,\pi\,)$ is the normalized \textsc{Fisher} information matrix \cite{Karalashvili2015, Ucinski2004} defined as:
\begin{subequations}\label{sec1_eq:fisher_matrix}
  \begin{align}
    F(\,\pi\,) & \egal \bigl[\, \Phi_{\,i \, j} \,\bigr] \,, && \forall (i,j) \ \in \ \bigl\{ 1, \ldots, M \, \bigr\}^{\,2}\,, \\[3pt]
    \Phi_{\,i \, j} & \egal \sum_{\,n \,= \,1}^{\,N} \ \int_{\,0}^{\, \tau} \Theta_{\,i} \ (x_{\,n} \,, t) \ \Theta_{\,j} \ (x_{\,n} \,, t) \ \mathrm{dt} \,.
  \end{align}
\end{subequations}

The matrix $F\,(\,\pi\,)$ characterizes the total sensitivity of the system as a function of the measurement plan $\pi$ (Eq.~\eqref{sec1_eq:measurement_plan}). The OED search aims at finding a measurement plan $\pi^{\,\star}$ for which the objective function (Eq.~\eqref{sec1_eq:D_optimum}) reaches the maximum value:
\begin{align}\label{sec1_eq:optimal_experimental_design}
  & \pi^{\,\circ} \egal \bigl\{ \, X^{\,\circ} \ , \ \phi^{\,\infty, \,\circ} \, \bigr\} \egal \arg \max_{\,\pi} \ \Psi \,.
\end{align}

To solve Eq.~\eqref{sec1_eq:optimal_experimental_design}, a domain of variation $\Omega_{\,\pi}$ is considered for the sensor position $X$ and the amplitude $\phi^{\,\infty}$ of the boundary conditions. Then, the following steps are carried out for each value of the measurement plan $\pi \egal \bigl\{ \, X \ , \ \phi^{\,\infty} \, \bigr\} $ in the domain $\Omega_{\,\pi} \,$. First, the direct problem, defined by Eqs.~\eqref{eq:HAM_equation2} -- \eqref{eq:HAM_ic}, is solved. Then, given the solution $\Pv$ for a fixed value of the measurement plan, the next step consists of computing the sensitivity coefficients $\Theta_{\,m} \egal \dfrac{\partial u}{\partial p_{\,m}} \,$, using also an embedded adaptive time \textsc{Runge}--\textsc{Kutta} scheme combined with central spatial discretisation. Then, with the sensitivity coefficients, the \textsc{Fisher} matrix \eqref{sec1_eq:fisher_matrix}(a,b) and the $D-$optimum criterion \eqref{sec1_eq:D_optimum} are computed. The solution of the direct and sensitivity problems are obtained for a given \emph{a priori} parameter $\vP$ and, in this case, the validity of the OED depends on this knowledge. If there is no \emph{prior} information, the methodology of the OED can be done using an outer loop on the parameter $\vP$ sampled using, for instance, \textsc{Latin} hypercube or \textsc{Halton} or \textsc{Sobol} quasi-random samplings. Interested readers may refer to \cite{Berger2017} for further details on the computation of sensitivity coefficients.

An interesting remark with this approach is that the probability distribution of the unknown parameter $p_{\,m}$ can be estimated from the distribution of the measurements of the field $u$ and from the sensitivity $\Theta_{\,m} \,$. The probability $\mathcal{P}$ of $ u$ is given by:
\begin{align*}
  F\,(\, \bar{u}\,) \egal \mathcal{P} \, \biggl\{\, u(\,x \,, t \,, p_{\,m} \,) \ \leqslant \ \bar{u} \, \biggr\} \,.
\end{align*} 
Using the sensitivity function $\Theta_{\,m}$, the probability can be approximated by:
\begin{align*}
  F \, (\, \bar{u} \,) & \simeq \ \mathcal{P} \, \biggl\{\, u(\,x \,, t \,, p_{\,m}^{\,\circ} \,) \plus \Theta_{\,m} \cdot  \bigl(\, p_{\,m} \moins p_{\,m}^{\,\circ} \,\bigr) \ \leqslant \ \bar{u} \, \biggr\} \,,
\end{align*} 
Assuming $\Theta_{\,m} \ > \ 0\,$, we get: 
\begin{align*}
  F \, (\, \bar{u} \,) & \egal  \mathcal{P} \, \biggl\{\, p_{\,m} \ \leqslant \  p_{\,m}^{\,\circ} \plus \frac{\bar{u} \moins u(\,x \,, t \,, p_{\,m}^{\,\circ} \,) }{\Theta_{\,m}}\, \biggr\} \,.
\end{align*}
Therefore, using a change of variable, the cumulative derivative function of the probability of the unknown parameter $p_{\,m}$  is estimated by:
\begin{align*}
  F \, (\,\bar{p}_{\,m}\,) & \egal \mathcal{P} \, \biggl\{\, p_{\,m} \ \leqslant \ \bar{p}_{\,m} \, \biggr\}  \\
  & \egal F \, \biggl(\, u \plus \Theta_{\,m} \cdot \bigl(\, \bar{p}_{\,m} \moins p_{\,m}^{\,\circ} \,\bigr) \,\biggr) \,.
\end{align*}
When $\Theta_{\,m} \ < \ 0\,$, the cumulative derivative function of the probability is given by:
\begin{align*}
  F \, (\,\bar{p}_{\,m}\,) & \egal 1 \moins F \, \biggl(\, u \plus \Theta_{\,m} \cdot \bigl(\, \bar{p}_{\,m} \moins p_{\,m}^{\,\circ} \,\bigr) \,\biggr) \,.
\end{align*}
It gives a \emph{local} approximation of the probability distribution of the unknown parameter $p_{\,m} \,$, at a reduced computational cost. Moreover, the approximation is reversible. Thus, if one has the distribution of the unknown parameter, it is possible to get the one of field $u \,$.


\section{Experimental facility}
\label{sec:exp_facility}

\begin{figure}
  \begin{center}
    \includegraphics[width=0.99\textwidth]{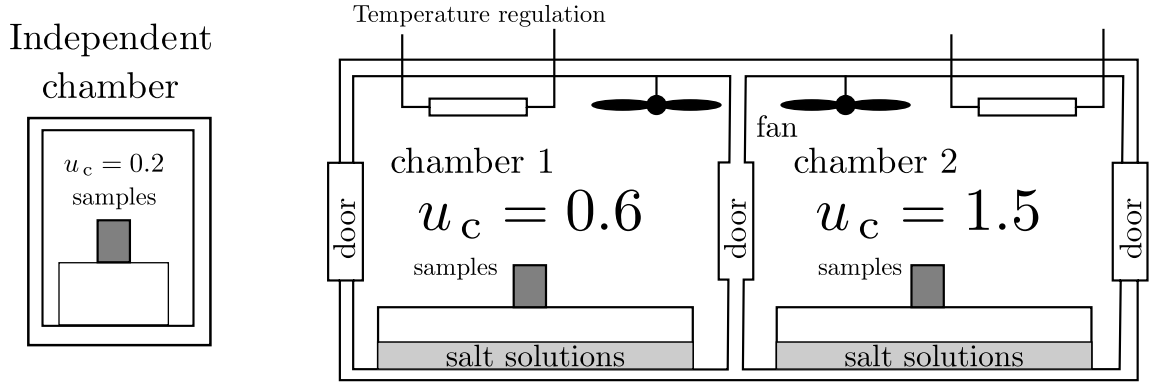}
    \caption{\small\em Illustration of the \emph{RH-box} experimental facility.}
    \label{fig:schema_RH_box}
  \end{center}
\end{figure}

The test facility used to carry out the experiment is illustrated in Figure~\ref{fig:schema_RH_box}. It is composed of two connected climatic chambers. The temperature of each chamber is controlled independently with a thermostatically-controlled water bath allowing water to recirculate in a heat exchanger. The relative humidity is kept fixed using saturated salt solutions of $\mathrm{MgCl}_{\,2}$ and $\mathrm{NaCl}$. Relative humidity values in chambers $1$ and $2$ are fixed to $\phi_{\,1} \egal 33 \ \unit{\%}$ and $\phi_{\,2} \egal 75 \ \unit{\%}$, respectively. Two door locks, at each side, allow the operator to insert or remove samples to minimize system disturbances. They enable easy and instantaneous change in humidity boundary conditions for the samples while passing from one chamber to another. Another climatic chamber is also available used to initially condition materials at $\phi_{\,0} \egal 10 \ \unit{\%}$.

\begin{figure}
  \centering
  \subfigure[\label{fig:sensors}]{\includegraphics[width=0.3\textwidth]{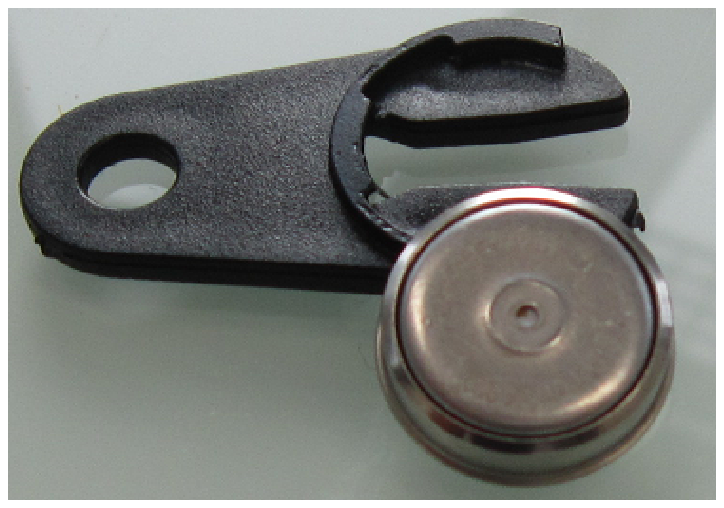}} 
  \subfigure[\label{fig:dessin_samples}]{\includegraphics[width=0.6\textwidth]{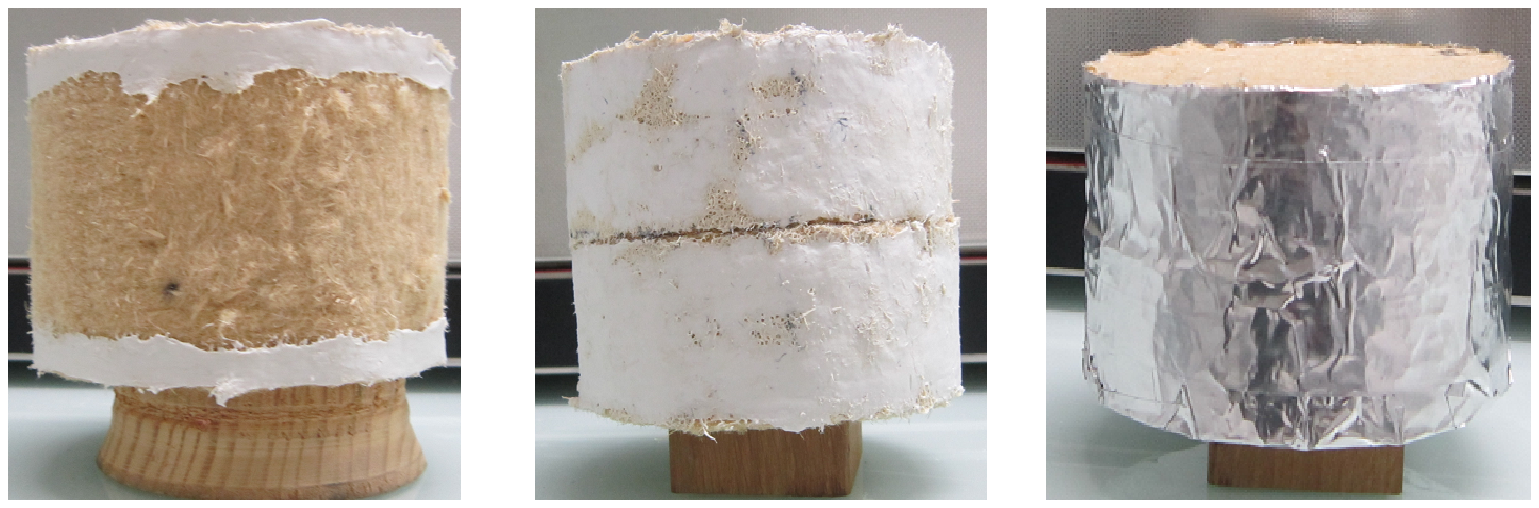}}
  \caption{\small\em Sensors of relative humidity and temperature (a) and wood fibre samples (b) with white acrylic seal and with aluminium tape.}
  \label{fig:samples_sensors}
\end{figure}

The temperature and relative humidity fields are measured within the samples with wireless sensors from the \texttt{HygroPuce} range (\texttt{Waranet} industry). The accuracy is $\pm \ 2 \ \unit{\%}$ for the relative humidity and $\pm \ 0.5 \ \unit{^\circ C}$ for the temperature and the dimensions are $0.6 \ \unit{cm}$ thickness and $1.6 \ \unit{cm}$ diameter, as illustrated in Figure~\ref{fig:sensors}. The sensors are placed within the material by cutting the samples. The total uncertainty on the measurement of relative humidity can be evaluated considering the propagation of the uncertainty due to sensor measurement and due to their location. In \cite{Berger2017b} the total uncertainty on the measurement has been evaluated to $\Delta \phi \egal 2 \ \unit{\%}$.

The material investigated is the wood fibre, which properties have been determined in \cite{Rafidiarison2015, Berger2017b} and are shown in its dimensionless form in Table~\ref{tab:mat_properties}. The reference parameter used to compute the unitless parameters are $\tref \egal 3600 \ \unit{s}$, $\dref \egal 5.17 \ \unit{s}$, $\cref \egal 2.85 \ \unit{kg/(m^{\,3}.Pa)}$ and $L \egal 0.08 \ \unit{m}\,$.
It constitutes \emph{a priori} information on the unknown parameters $\Fo \,$, $c_{\,1}$ and $c_{\,2}\,$. The samples are cylindrical, with a $10 \ \unit{cm}$ diameter and $8 \ \unit{cm}$ thickness in order to avoid border effects and to minimize perturbations by sensors placed within the sample. Moreover, to ensure unidimensional moisture transfer and a null flux condition at $x \egal 1$, the samples are covered with aluminium tape and glued on a white acrylic seal, as illustrated in Figure~\ref{fig:dessin_samples}. The convective moisture transport coefficient at $x \egal 0$ has been estimated experimentally in \cite{Busser2016, Berger2017b}. The corresponding \textsc{Biot} number is reported in Table~\ref{tab:mat_properties}.

Finally, the experimental facility is used to submit the samples to a single or multiple steps of relative humidity. For a single step, the boundary conditions are defined as: 
\begin{align*}
\uL \, (\,t\,) \egal \left. 
\begin{cases} 
\ui \,, & t \egal 0  \,, \\
\uc \,, & t \ > \ 0 \,. \\
\end{cases} \right.  
\end{align*}
For the case of multiple steps, we set:
\begin{align*}
\uL \, (\,t\,) \egal \left. 
\begin{cases} 
\ui \,, & t \egal 0 \\
\uc^{\,1} \,, & t \ \in \ \bigl(\, 0 \,, \, \tau \, \bigr] \,, \\
\uc^{\,2} \,, & t \ \in \ \bigl(\, \tau  \,, \, 2\, \tau \, \bigr] \,, \\
\uc^{\,3} \,, & t \ \in \ \bigl(\, 2\, \tau  \,, \, 3\, \tau \, \bigr] \,, \\
\end{cases} \right.
\end{align*}
where the initial condition belongs to $\ui \ \in \ \bigl\{\, 0.2 \, , \, 0.6 \, , \, 1.5 \,\bigr\} \,$, the climatic chamber boundary condition
$\bigl(\, \uc \,,\, \uc^{\,1} \,,\, \uc^{\,2} \,,\, \uc^{\,3}\,\bigr) \ \in \ \bigl\{\, 0.6 \, , \, 1.5 \,\bigr\}$ and  the duration of the step $\tau  \ \in \ \bigl\{\, 24 \, , \, 48 \,,\, 72 \,,\, 96 \,,\, 120 \,,\, 144 \,,\, 168 \,,\, 192\,\bigr\} $. A total of $20$ designs are possible for providing measurements to estimate the unknown parameters $\Fo \,$, $c_{\,1}$ and $c_{\,2} \,$. A synthesis of the possible designs is provided in Table~\ref{tab:possible_designs}. It should be noted that, according to the reference parameters, unitless values $u \egal 0.2\,$, $u \egal 0.66\,$, $u \egal 1.5$ correspond to $\phi \egal 0.1\,$, $\phi \egal 0.33$ and $\phi \egal 0.75\,$, respectively.

\begin{table}
\centering
\caption{\emph{A priori} dimensionless material properties of wood fiber from \cite{Berger2017b, Rafidiarison2015}.}
\bigskip
\setlength{\extrarowheight}{.3em}
\begin{tabular}[l]{@{} lc}
\hline\hline
Storage coefficient $c$ & 
$c(\,u\,) \egal c_{\,0} \plus c_{\,1} \, u \plus c_{\,2} \, u^{\,2} $ \\
& $c_{\,0} \egal 1 \,$, $c_{\,1} \egal -9.79 \e{-1} \,$, $c_{\,2} \egal 1.06$ \\
Diffusion coefficient  $d$ & 
$d(\,u\,) \egal d_{\,0} \plus d_{\,1} \, u$ \\
& $d_{\,0} \egal 1 \,$, $d_{\,1} \egal 0.29 \,$ \\
\textsc{P\'eclet} number & $\Pe \egal 1.1 \e{-2} $\\
\textsc{Fourier} number & $\Fo \egal 4 \e{-3} $\\
\textsc{Biot} number & $\Bi \egal 13.7 $\\
\hline\hline
\end{tabular}
\bigskip
\label{tab:mat_properties}
\end{table}

\begin{table}
\centering
\caption{\small\em Possible designs according to the experimental facility.}
\bigskip
\setlength{\extrarowheight}{.3em}
\begin{tabular}[l]{@{} ccc|cc|cc|cc|cc}
\hline\hline
\multicolumn{3}{c|}{ \multirow{2}{*}{\textit{Single step}}}
& \multicolumn{8}{c}{\textit{Multiple step}} \\
\multicolumn{3}{c|}{ }
& \multicolumn{4}{c|}{Case $1$} 
& \multicolumn{4}{c}{Case $2$} \\
\hline
Design & $\ui$
& $\uc$
&Design 
& $\tau$ 
& Design 
& $\tau$
& Design 
& $\tau$
& Design 
& $\tau$\\
1 & 0.2 & 0.66 &
5  & 1 & 
9  &  5 & 
13 &  1 & 
17 &  5 \\
2 &  0.2 & 1.5 &
6  &  2 & 
10 &  6 & 
14 &  2 & 
18 &  6 \\
3 & 0.66 & 1.5 &
7  &  3 & 
11 &  7 & 
15 &  3 & 
19 &  7 \\
4 & 1.5 & 0.66 &
8  &  4 & 
12 & 8 & 
16 & 4 & 
20 & 8 \\
\hline\hline
\\
\multicolumn{3}{c}{ } &
\multicolumn{8}{l}{Case 1: $\ui \egal 0.2\,$, $\uc^{\,1} \egal 0.66\,$, $\uc^{\,2} \egal 1.5\,$,  $\uc^{\,3} \egal 0.66$} \\
\multicolumn{3}{c}{ } &
\multicolumn{8}{l}{Case 2: $\ui \egal 0.2\,$, $\uc^{\,1} \egal 1.5\,$, $\uc^{\,2} \egal 0.66\,$,  $\uc^{\,3} \egal 1.5$} \\
\end{tabular}
\bigskip
\label{tab:possible_designs}
\end{table}


\section{Searching the OED}
\label{sec:OED}

\subsection{Estimation of one parameter}

This Section focuses on the estimation of one parameter within $\Fo \,$, $c_{\,1}$ or $c_{\,2}$ with experiments coming from single- or multiple-step designs. It should be noted that by estimating parameter $\Fo \,$, the complete sorption isotherm curve is defined, according to the dimensionless quantities defined in Section~\ref{sec:dimensionless_formulation}. The equations to compute the sensitivity functions are given in Appendix~\ref{sec:Eq_sensitivity_coeff}. In addition, the demonstration of structural identifiability of the three parameters is provided in Appendix~\ref{sec:Annex_identifiability}.


\subsubsection{Single step}

\begin{figure}
  \begin{center}
  \subfigure[\label{fig:1step_Psi_1param}]{\includegraphics[width=.48\textwidth]{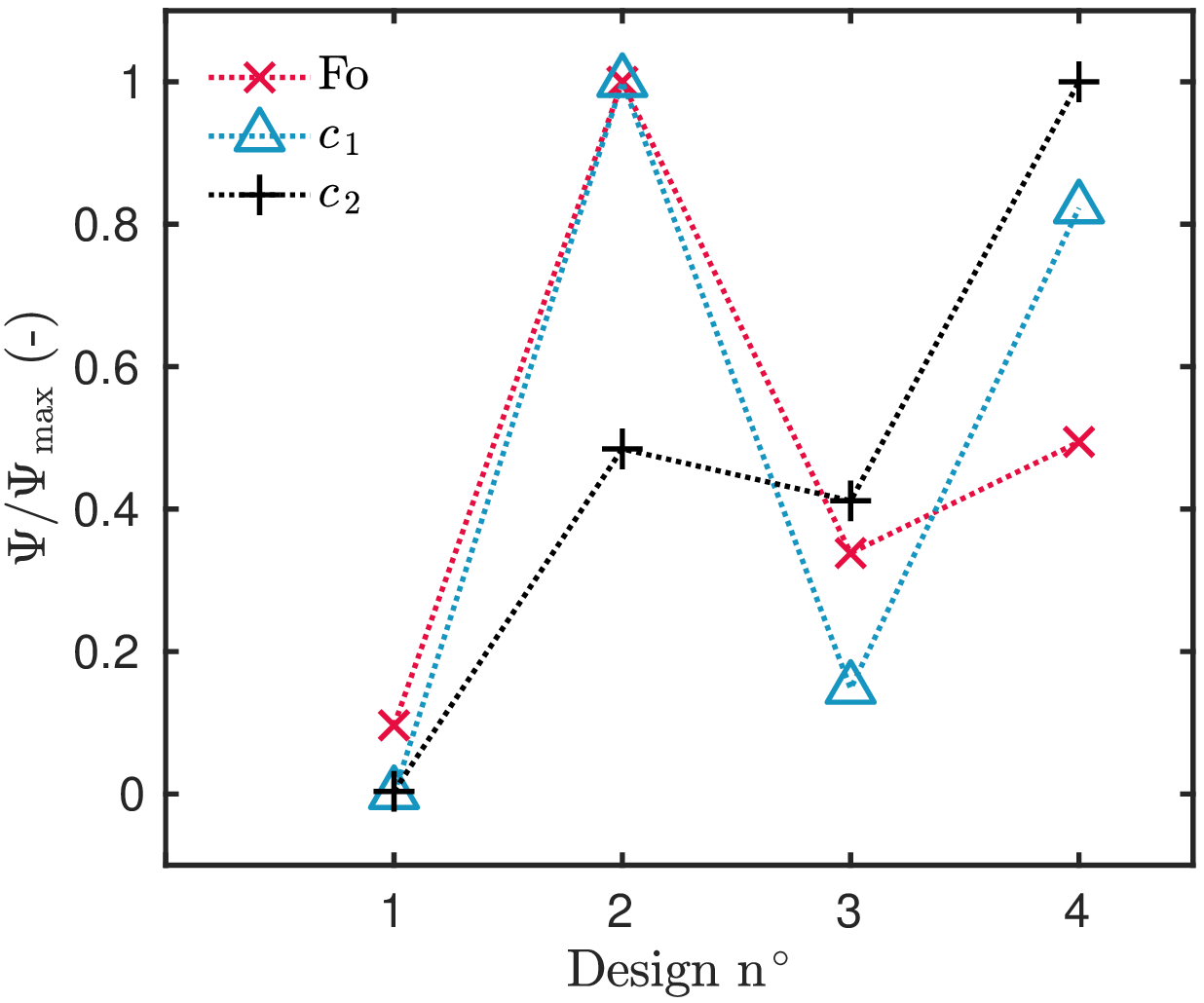}} \hspace{0.3cm}
  \subfigure[\label{fig:1step_Psi_fx_1param}]{\includegraphics[width=.48\textwidth]{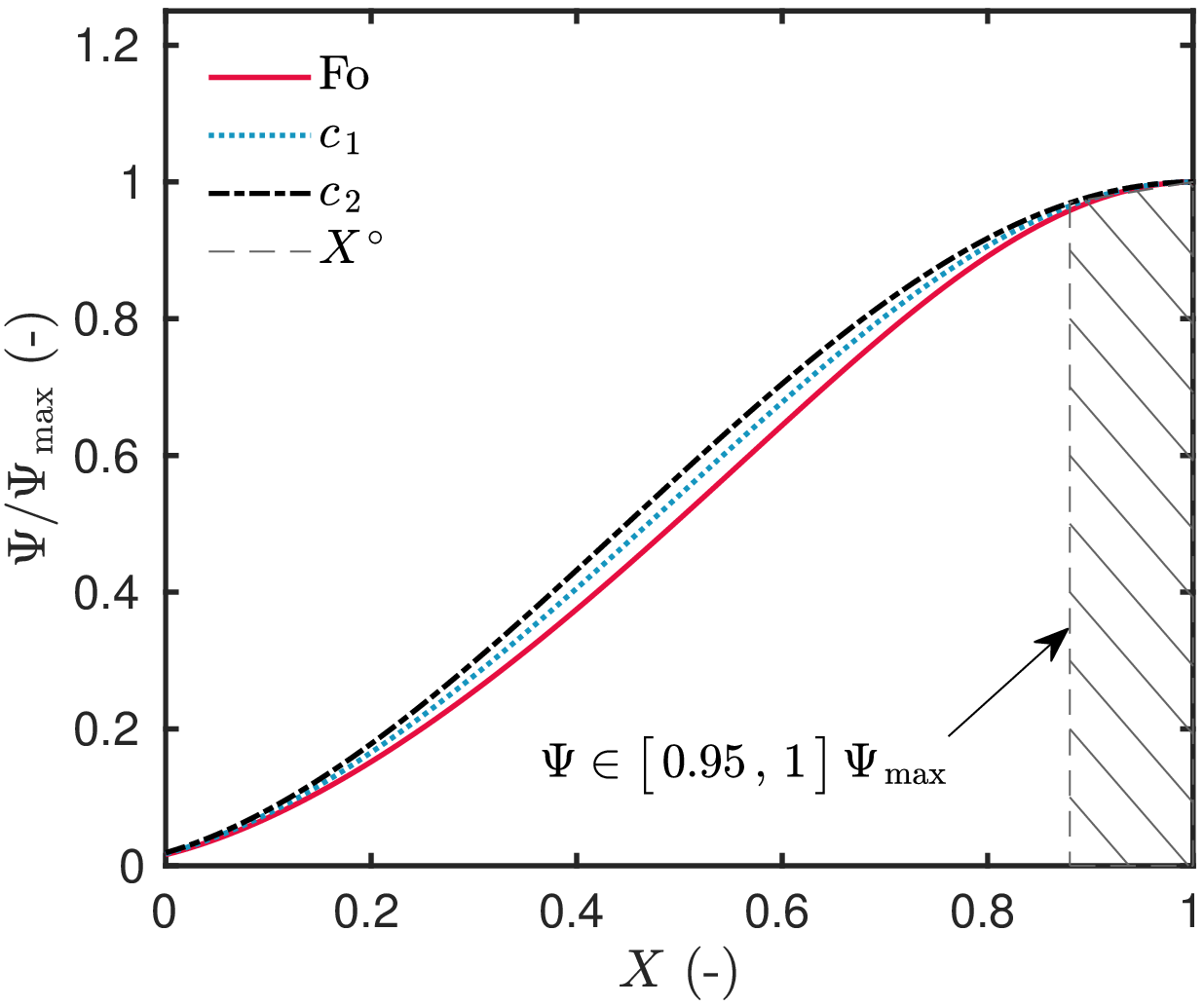}}
  \caption{\small\em Variation of the criterion $\Psi$ for the four possible single step designs (a) and as a function of the sensor position $X$ for the OED (b), in the case of estimating one parameter.}
  \label{fig:1step_Psi_design}
  \end{center}
\end{figure}

\begin{figure}
  \begin{center}
  \subfigure[\label{fig:1step_S_OED}]{\includegraphics[width=.48\textwidth]{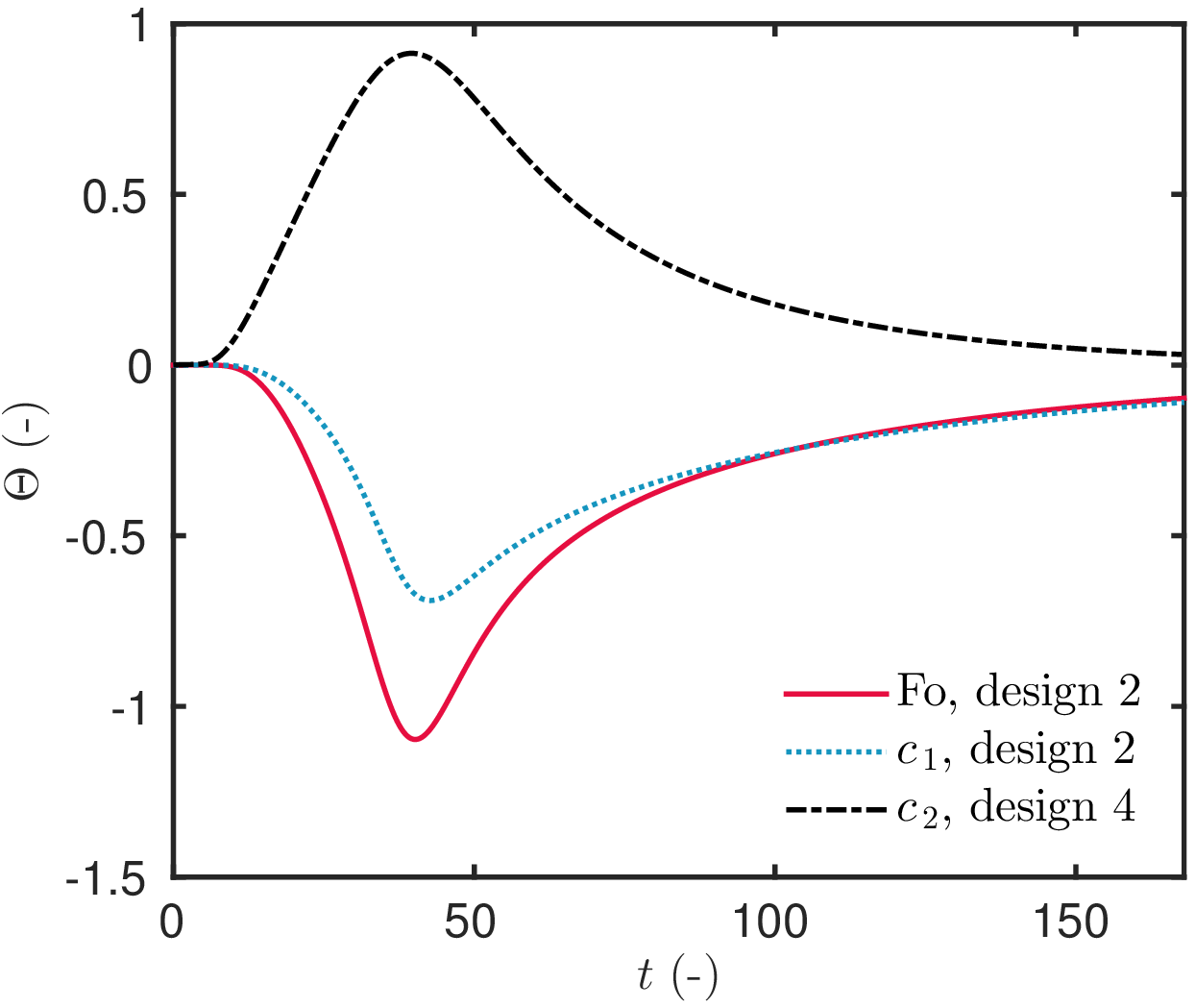}} \hspace{0.3cm}
  \subfigure[\label{fig:1step_S_anti_OED}]{\includegraphics[width=.48\textwidth]{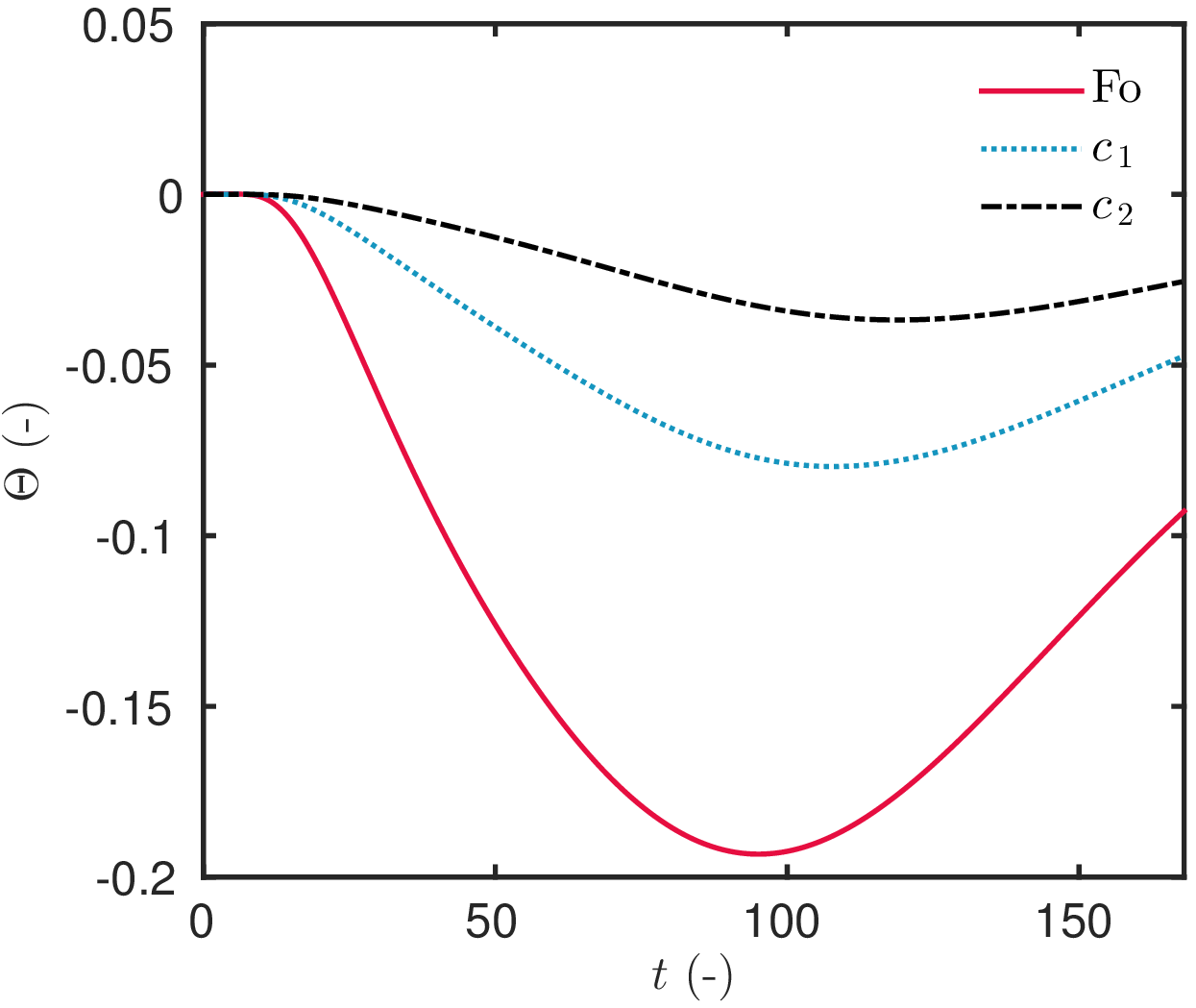}}
  \caption{\small\em Sensitivity coefficients $\Theta$ for parameters $\Fo \,$, $c_{\,1}$ and $c_{\,2}$ for the OED (a) and for design $1$ (b) ($X \egal X^{\,\circ}$).}
  \label{fig:1step_S_OED_anti_OED}
  \end{center}
\end{figure}

Figure~\ref{fig:1step_Psi_1param} gives the variation of the criterion $\Psi$ for the four single-step designs. For the estimation of parameters $\Fo$ or $c_{\,1} \,$, the criterion reaches its maximal value for design $2 \,$, corresponding to a step from $\ui \egal 0.1$ to $\uc \egal 1.5 \,$. For parameter $c_{\,2}$ , the design $4$ is the optimal one. It can be noted that, for parameter $c_{\,1}$, the relative criterion $\Psi$ attains $80\%$ for the design $4 \,$. It could be an interesting alternative to estimate this parameter. The variation of the criterion is related to the sensitivity function of each parameter. As noticed in Figures~\ref{fig:1step_S_OED} and \ref{fig:1step_S_anti_OED}, functions $\Theta$ have higher magnitudes of variation for the OED. The variation of $\Psi$ as a function of the sensor location $X$ is shown in Figure~\ref{fig:1step_Psi_fx_1param} for the OED. The optimal sensor location is at the boundary of the material opposite from the perturbations. If required for practical purpose, the sensor can be placed in the interval $X \ \in \ \bigl[\, 0.9 \,,\, 1 \,\bigr]$, ensuring to reach $95\%$ of the criterion $\Psi \,$. Results have similar tendencies for the three parameters.

\subsubsection{Multiple steps}

\begin{figure}
\begin{center}
  \subfigure[\label{fig:Mstep_Psi_1param}]{\includegraphics[width=.48\textwidth]{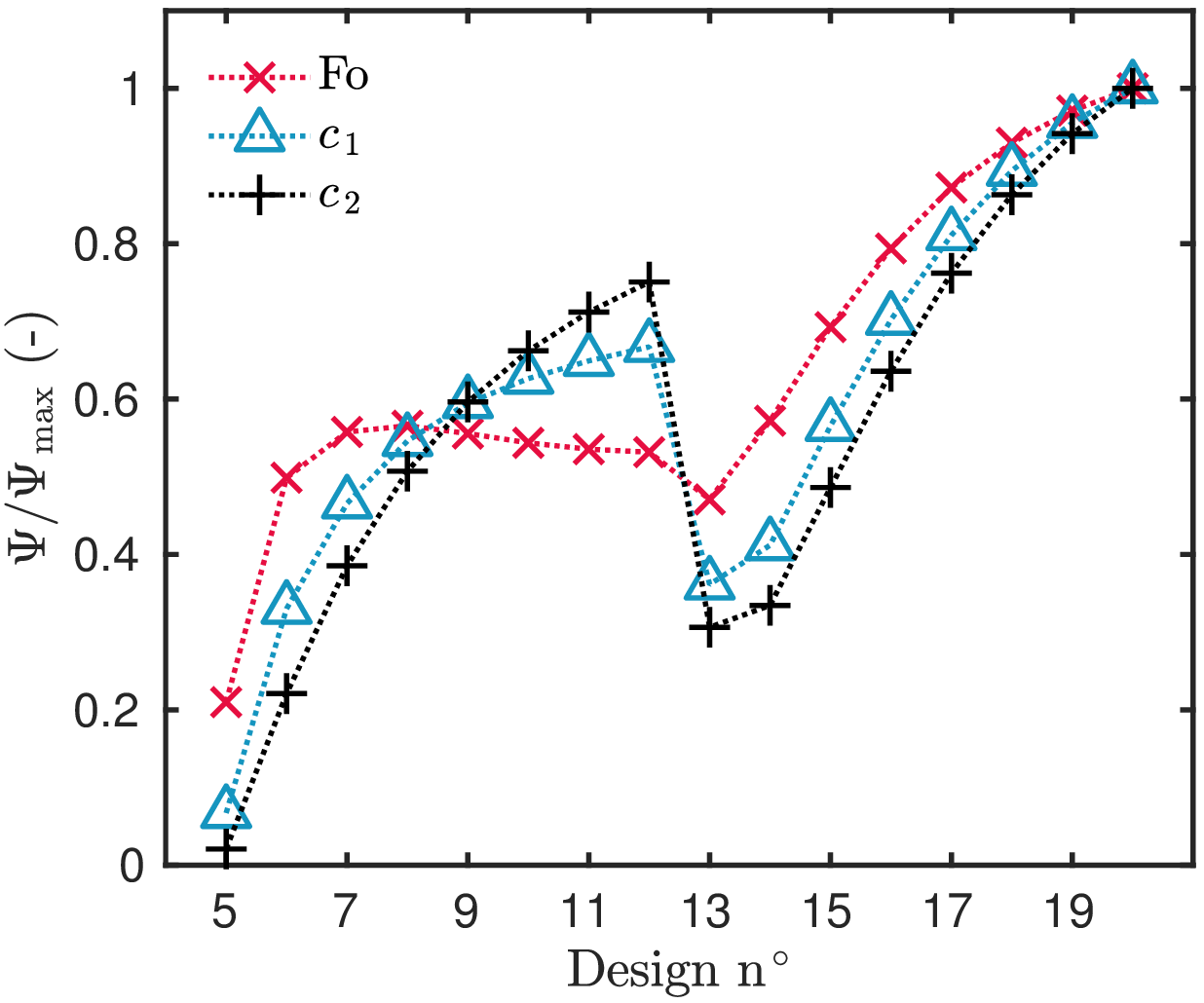}}
  \subfigure[\label{fig:Mstep_Psi_fx_1param}]{\includegraphics[width=.48\textwidth]{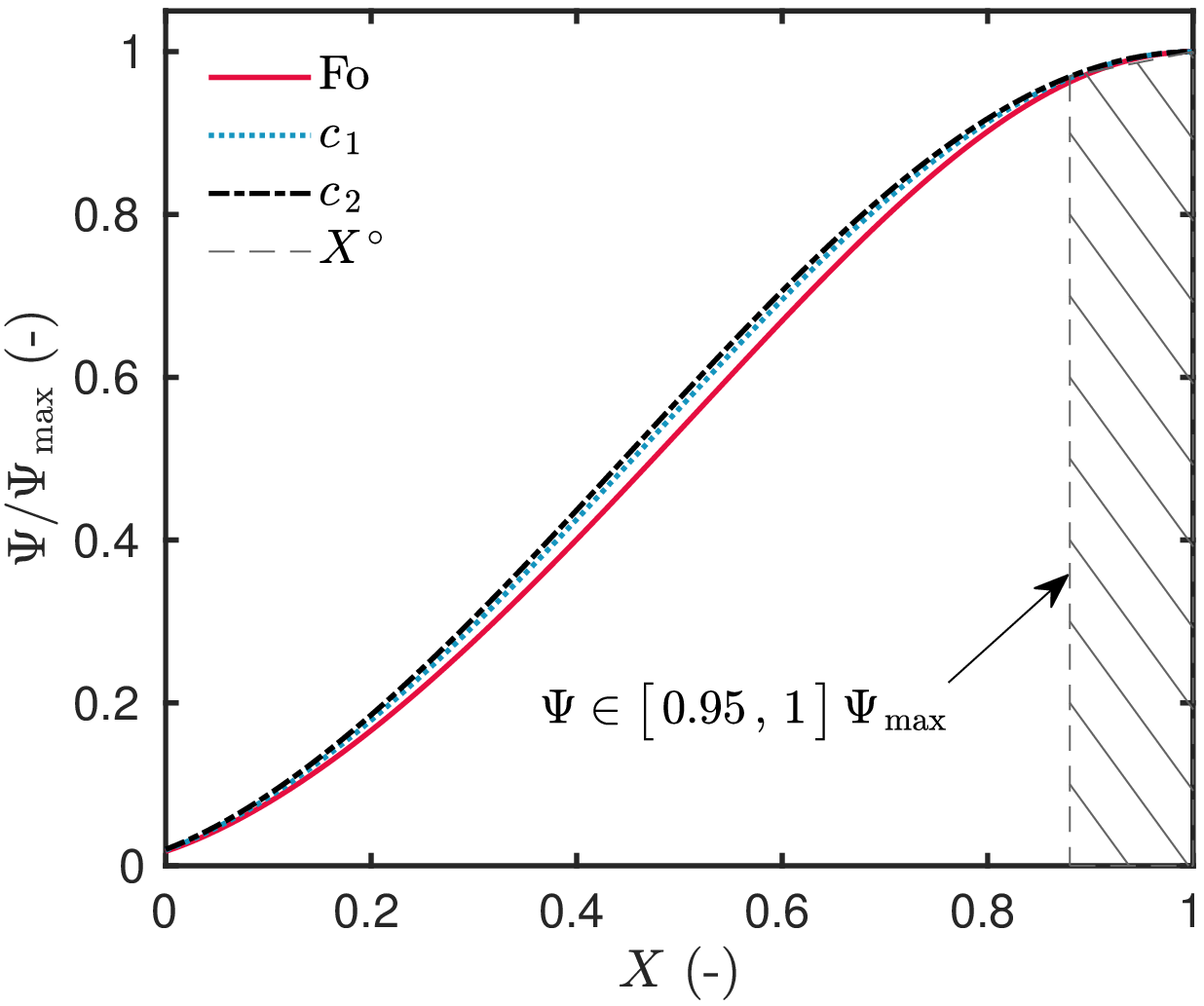}}
  \caption{\small\em Variation of the criterion $\Psi$ for the sixteen possible designs (a) and as a function of the sensor position $X$ for the OED (b), in the case of estimating one parameter.}
  \label{fig:Mstep_Psi_design}
\end{center}
\end{figure}

\begin{figure}
  \begin{center}
  \subfigure[\label{fig:Mstep_S_OED}]{\includegraphics[width=.48\textwidth]{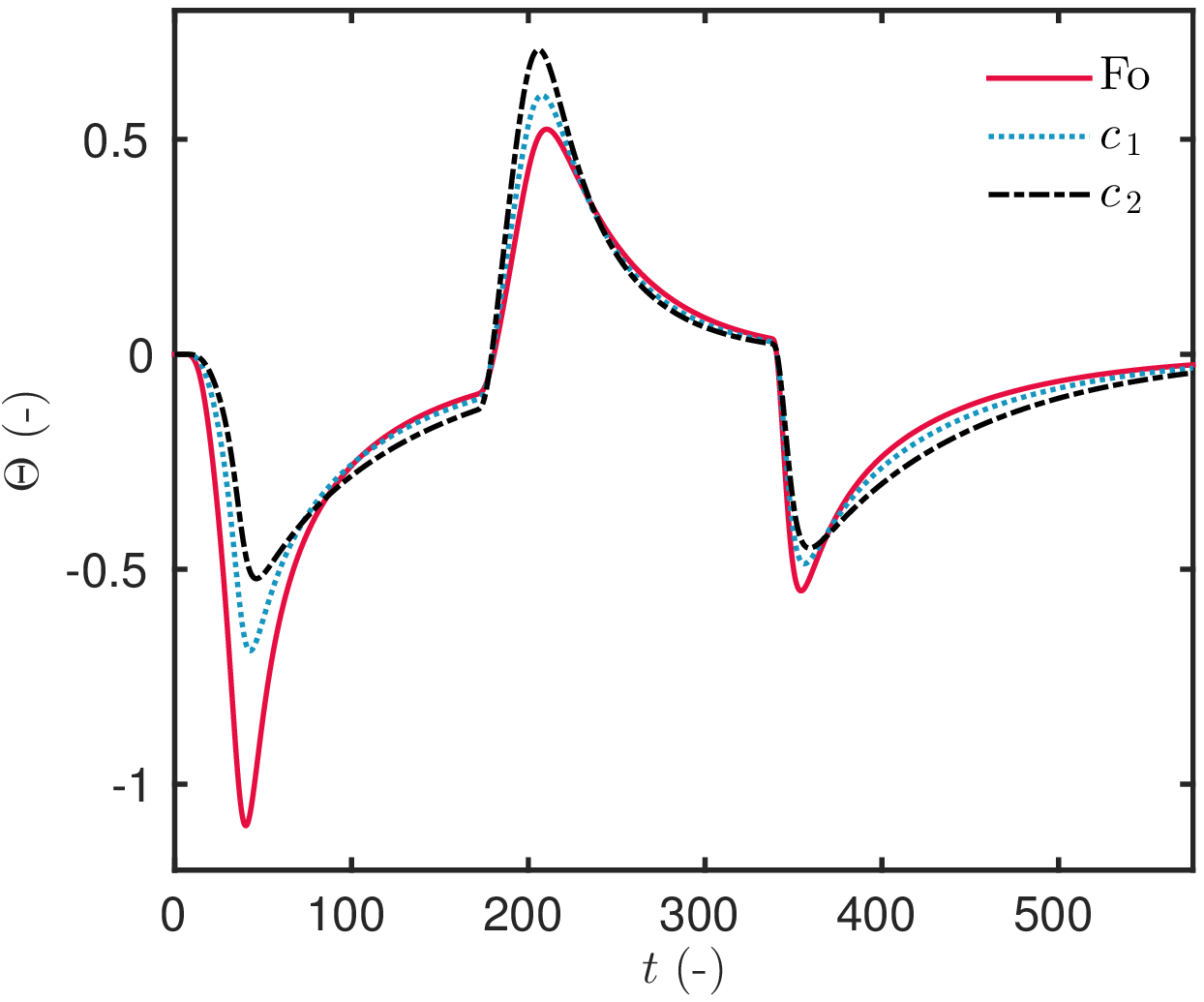}}
  \subfigure[\label{fig:Mstep_S_8}]{\includegraphics[width=.48\textwidth]{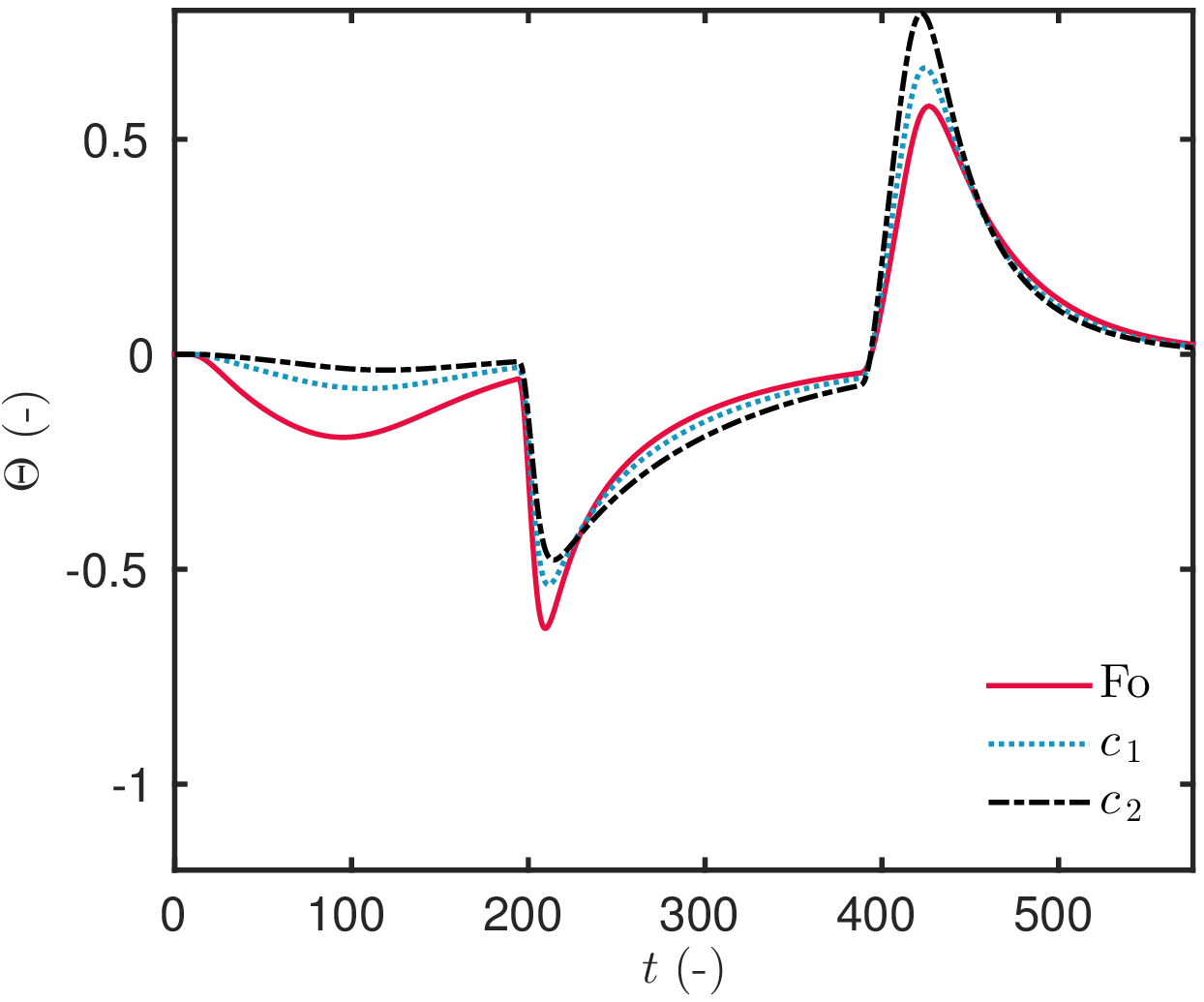}}
  \subfigure[\label{fig:Mstep_S_anti_OED}]{\includegraphics[width=.48\textwidth]{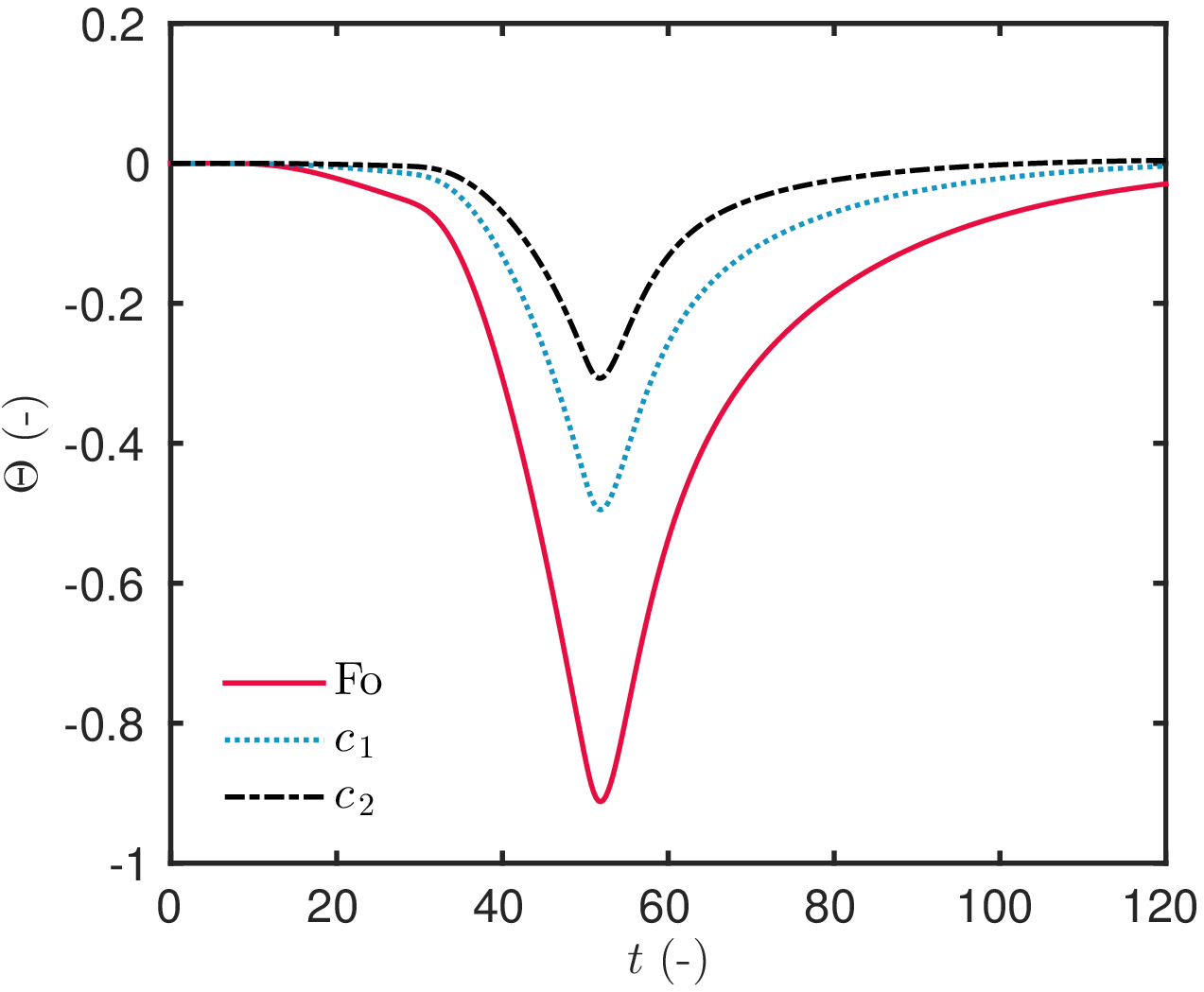}} 
  \subfigure[\label{fig:Mstep_S_anti2_OED}]{\includegraphics[width=.48\textwidth]{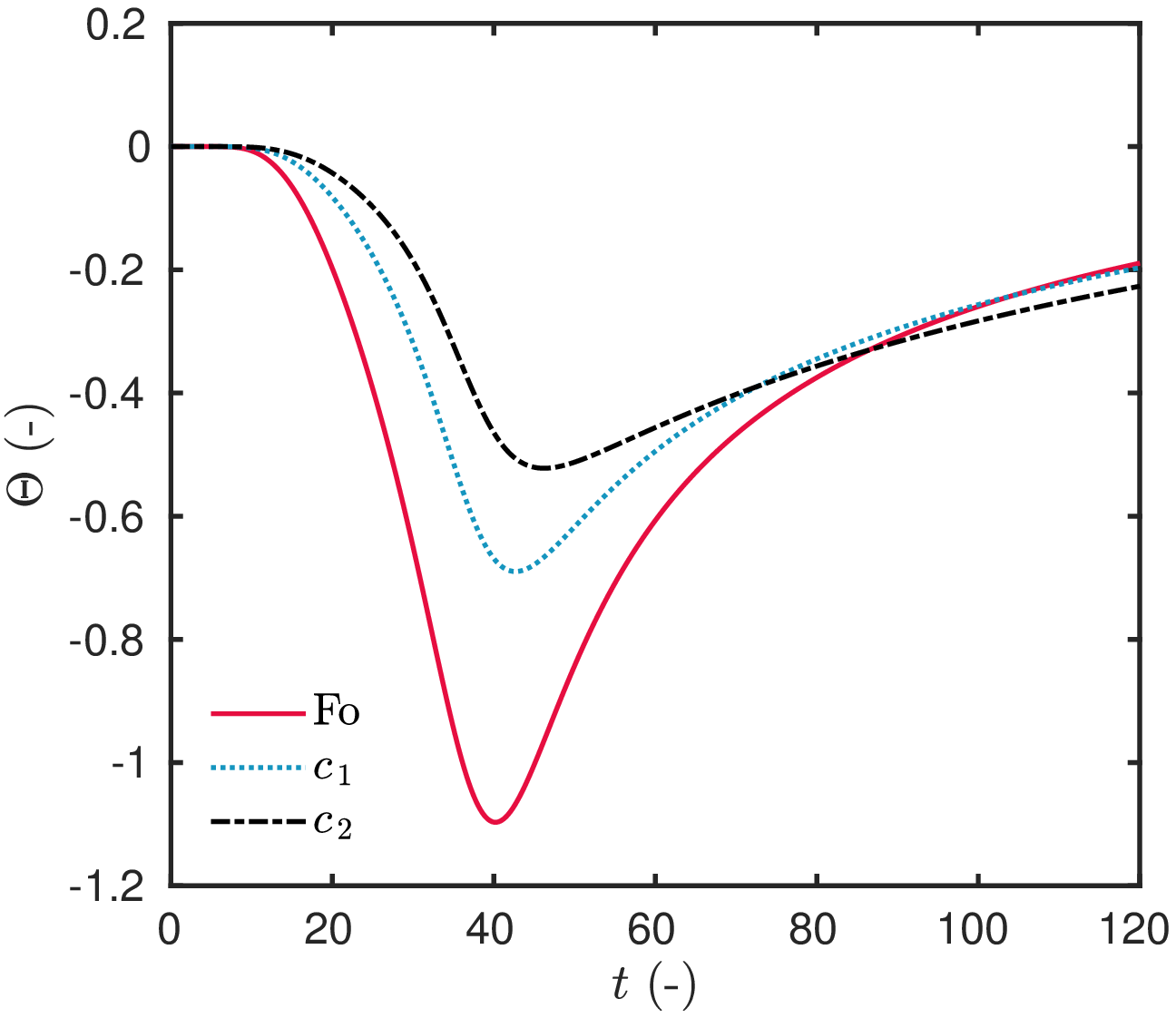}}
  \caption{\small\em Sensitivity coefficients $\Theta$ for parameters $\Fo \,$, $c_{\,1}$ and $c_{\,2}$ for the OED (design $20$) (a), design $12$ (b), design $5$ (c) and  design $13$ (d) ($X \egal X^{\,\circ}$).}
  \label{fig:Mstep_S_OED_anti_OED}
  \end{center}
\end{figure}

Figure~\ref{fig:Mstep_Psi_1param} shows the variation of the relative criterion $\Psi$ for the designs considering multiple steps of relative humidity. It increases with the duration $\tau$ of the steps. Thus, for the group of designs $5$ to $12$ and the group $13$ to $20\,$, the criteria reach their maximum for designs $12$ and $20\,$, respectively, corresponding to the step duration $\tau \egal 8 \,$. The group $5$ to $12$ corresponds to a multiple step $\uc^{\,1} \egal 0.66\,$, $\uc^{\,2} \egal 1.5\,$, $\uc^{\,3} \egal 0.66 \,$. For them, the criterion does not attain $80\%$ of the maximal criteria. Therefore, it is preferable to choose among designs $18$ to $20\,$, with a multiple step $\uc^{\,1} \egal 1.5\,$, $\uc^{\,2} \egal 0.66\,$, $\uc^{\,3} \egal 1.5\,$, and a duration $\tau \ \geqslant \ 6$ to estimate the parameters. Figures~\ref{fig:Mstep_S_OED}, \ref{fig:Mstep_S_8}, \ref{fig:Mstep_S_anti_OED} and \ref{fig:Mstep_S_anti2_OED} compare the sensitivity function of each parameter for three different designs. The quantity $\Theta$ has higher magnitude of variation for the OED than for the others. Moreover, for the design $5\,$, the duration of the step is so short that there is almost no variation on the sensitivity when occurring the first step for $t \ \in \ \bigl[\,0 \,,\,24 \,\bigr]\,$. As for the previous case, the optimal sensor position is $X \ \in \bigl[\, 0.9 \,,\, 1 \,\bigr]\,$.

\subsection{Estimation of several parameters}

The issue is now to estimate two or three parameters defining the moisture capacity $\Fo \,$, $c_{\,1}$ and $c_{\,2}$. First of all, it is important to notice in Figures~\ref{fig:1step_S_OED_anti_OED} and \ref{fig:Mstep_S_OED_anti_OED}, that the sensitivity function $\Theta$ of the parameters have a strong correlation. The interval of variation of the correlation coefficients for all the designs are:
\begin{align*}
  \mathrm{Cor} \, \bigl(\,\Fo \,,\, c_{\,1} \,\bigr) & \in \ \bigl[\, 0.94 \,,\, 0.99 \,\bigl] \,, \\
  \mathrm{Cor} \, \bigl(\,c_{\,1} \,,\, c_{\,2} \,\bigr) & \in \ \bigl[\, 0.92 \,,\, 0.99 \,\bigl] \,,\\
  \mathrm{Cor} \, \bigl(\,\Fo \,,\, c_{\,2} \,\bigr) & \in \ \bigl[\, 0.71 \,,\, 0.95 \,\bigl] \,.
\end{align*}
Therefore, the estimation of the three parameters at the same time using only one experiment might be a difficult task. In addition, over all the possible designs, the couple of parameters $\bigl(\,\Fo \,,\, c_{\,2}\, \bigr)$ is the one with the lower correlation. Therefore, the OED search will only consider their estimation.


\subsubsection{Single step}

\begin{figure}
\begin{center}
\subfigure[\label{fig:1step_Psi_2param}]{\includegraphics[width=.48\textwidth]{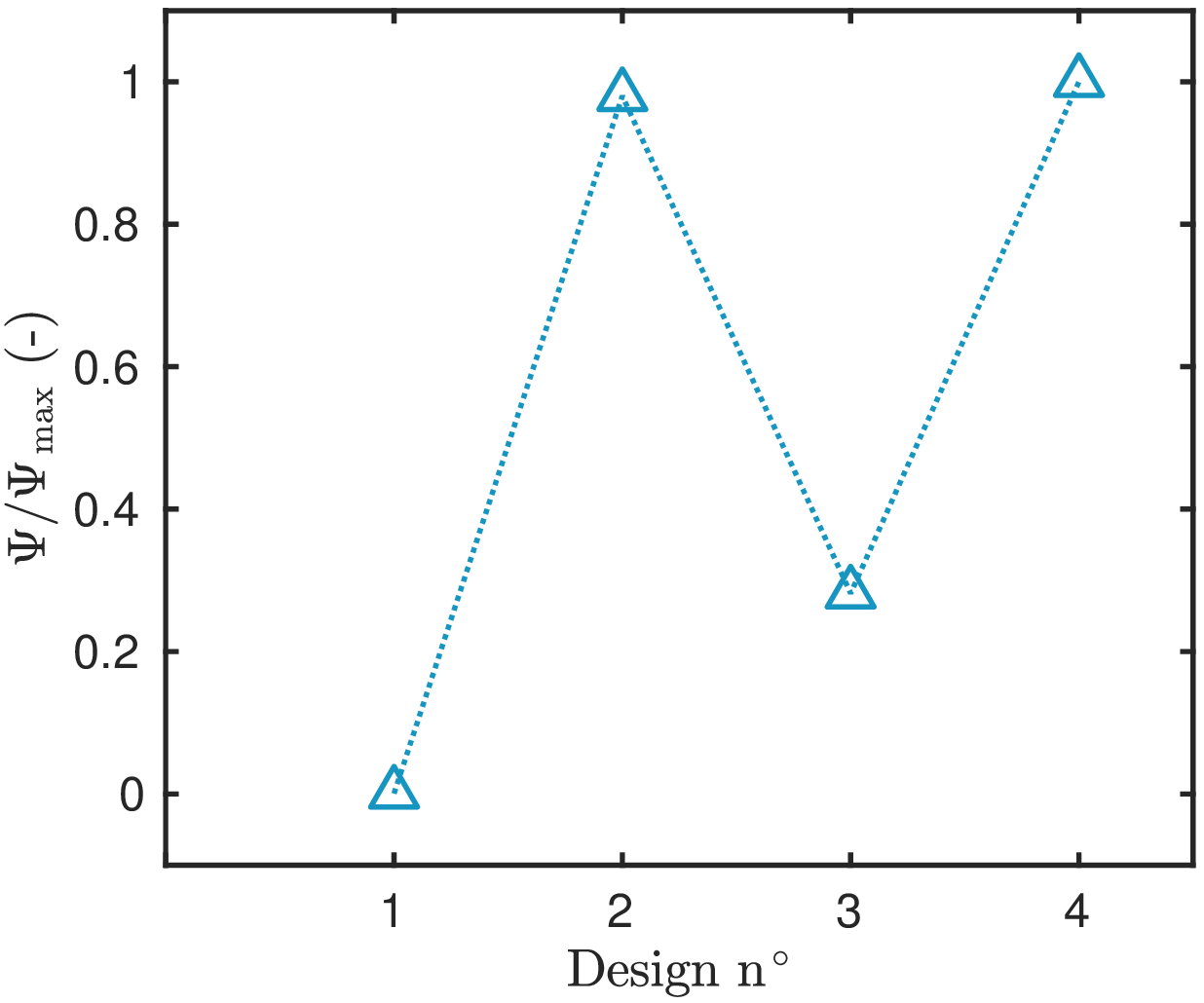}} \hspace{0.3cm}
\subfigure[\label{fig:1step_Psi_fx_2param}]{\includegraphics[width=.48\textwidth]{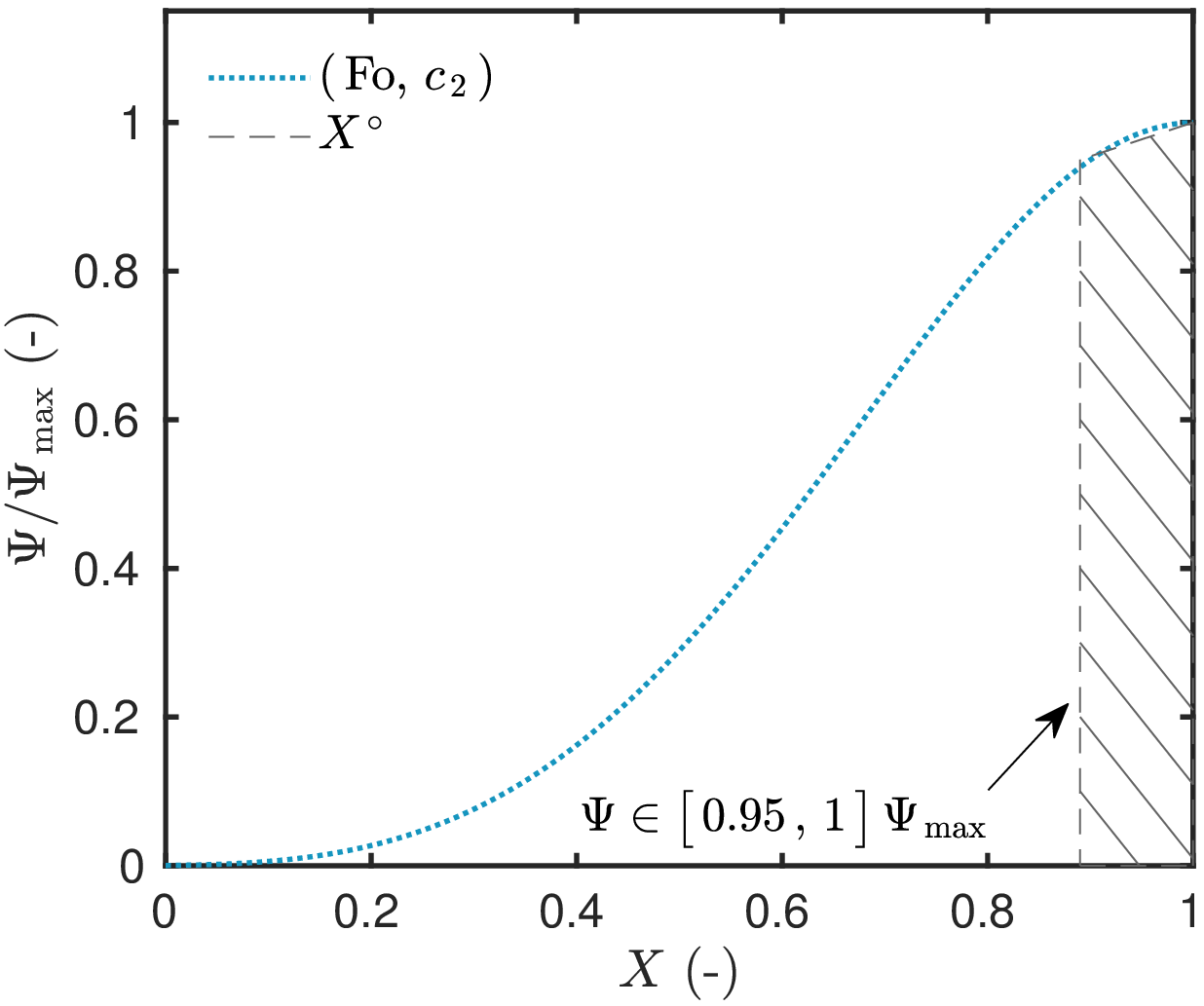}}
\caption{\small\em Variation of the criterion $\Psi$ for the four possible designs of single step of relative humidity (a) and as a function of the sensor position $X$ for the OED (b), in the case of estimating the couple of parameters $\bigl(\,\Fo \,,\, c_{\,2}\, \bigr)$.}
\label{fig_AN1:Psi_design_1}
\end{center}
\end{figure}

Figure~\ref{fig:1step_Psi_2param} gives the variation of the criterion $\Psi$ for the four possible designs, considering a single step of relative humidity. The OED is reached for design $4\,$. However, the design $2$ represents an interesting alternative as more than $95\%$ of the maximum criterion is reached. The sensor should be placed between  $X \ \in \bigl[\, 0.9 \,,\, 1 \,\bigr]\,$.

\subsubsection{Multiple steps}

The variation of the criterion $\Psi$ for the sixteen designs is shown in Figure~\ref{fig:Mstep_Psi_design}. It increases with the duration of the steps $\tau\,$. The OED is reached for the design considering a multiple step $\ui \egal 0.2\,$, $\uc^{\,1} \egal 1.5\,$, $\uc^{\,2} \egal 0.66\,$ and  $\uc^{\,3} \egal 1.5 \,$, with a duration $\tau \egal 8\,$. As for the previous case, the OED is defined for a sensor placed near the boundary of the material $x \egal 1 \,$.

\begin{figure}
  \begin{center}
  \subfigure[\label{fig:Mstep_Psi_2param}]{\includegraphics[width=.48\textwidth]{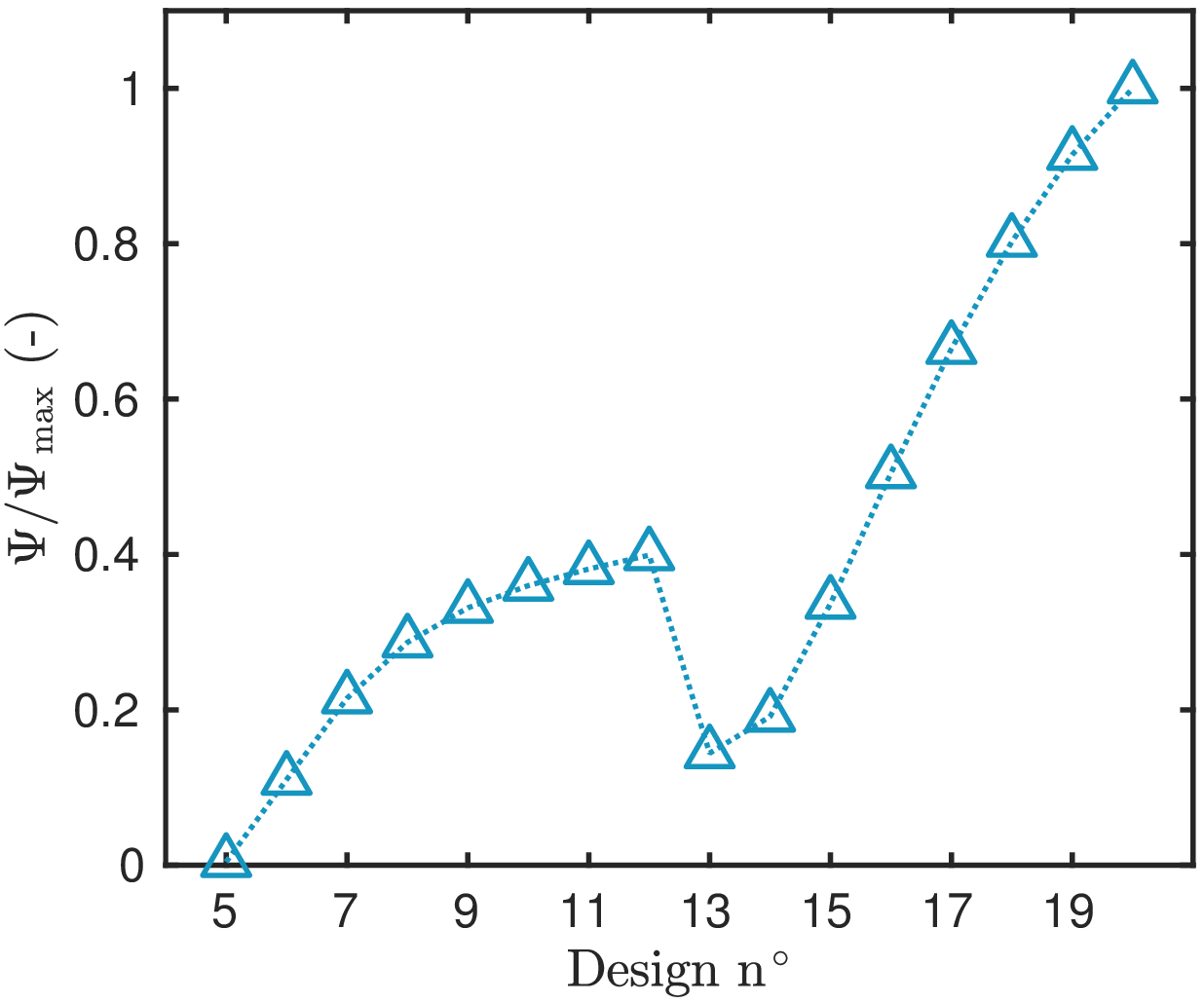}}
  \subfigure[\label{fig:Mstep_Psi_fx_2param}]{\includegraphics[width=.48\textwidth]{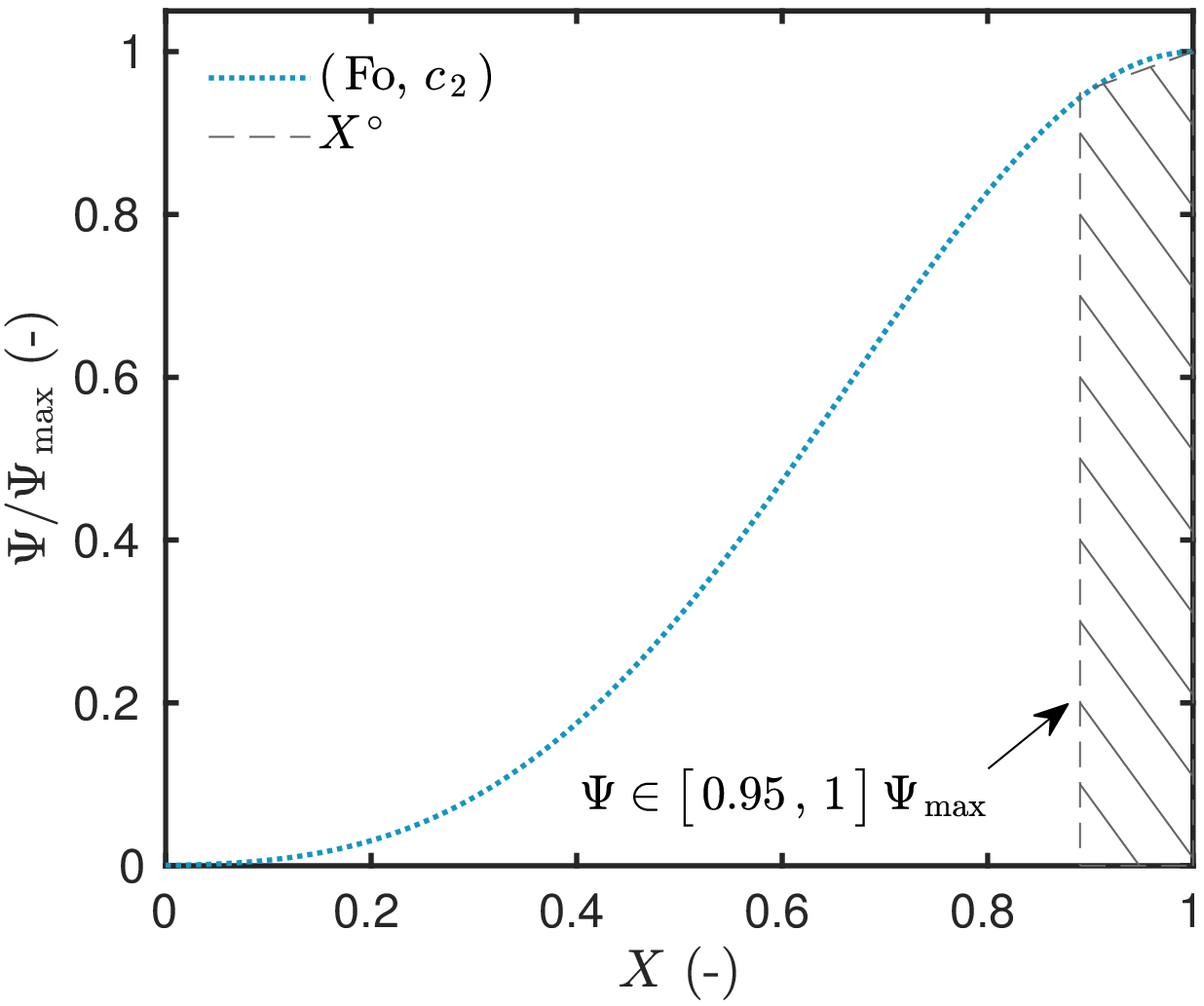}}
  \caption{\small\em Variation of the criterion $\Psi$ for the sixteen possible designs of multiple steps of relative humidity (a) and as a function of the sensor position $X$ for the OED (b), in the case of estimating the couple of parameters $\bigl(\,\Fo \,,\, c_{\,2}\, \bigr)$.}
  \label{fig:Mstep_Psi_design}
  \end{center}
\end{figure}


\section{Estimation of the unknown parameters}
\label{sec:parameter_estimation_problem}

\subsection{Methodology}

As mentioned before, the sensitivity functions of parameters $\Fo \,$, $c_{\,1}$ and $c_{\,2}$ are strongly correlated and the estimation of the three parameters using one single experiment might be a laborious task. To answer this issue, a single step, referenced as \emph{experiment $A \,$}, will be used for the estimation of the parameter $c_{\,1}$ and a multiple step, referenced as \emph{experiment $B\,$}, for the parameters $\bigr(\, \Fo \,,\, c_{\,2} \,\bigl)$, which sensitivity functions are less correlated. According to the results of the OED, the sensor is placed near the border $x \egal 1 $. For the boundary conditions, the single step will be operated from $\ui \egal 0.2$ to $\uc \egal 1.5 \,$ (design $2$ from Table~\ref{tab:possible_designs}). The OED multiple step is defined as $\ui \egal 0.2\,$, $\uc^{\,1} \egal 1.5\,$, $\uc^{\,2} \egal 0.66\,$,  $\uc^{\,3} \egal 1.5$ and a duration of each step $\tau \egal 8 \,$ (design $20$ from Table~\ref{tab:possible_designs}).

To estimate the unknown parameters, the following cost function is defined by minimizing the residual between the experimental data and the numerical results of the direct model: 
\begin{align*}
  \mathrm{J}^{\,n}_{\,i} & \egal \biggl|\biggl|\, u \moins u_{\,\mathrm{exp}\,,\, n} \, \,\biggr|\biggr|_{\,\mathcal{L}_{\,i}} \,,
  && n \ \in \ \bigl\{\,1 \,,\, 2 \,,\, 3 \, \bigr\} \,,
  && i \ \in \ \bigl\{\,2 \,,\, \infty \, \bigr\} \,.
\end{align*}

Several expressions of the cost function are tested. The subscript $i$ denotes the norm used for the cost function $\mathrm{J} \,$: $i \egal 2$ stands for the standard discrete $\mathcal{L}_{\,2}$ norm while $i  \egal \infty$ for the $\mathcal{L}_{\,\infty}$ (uniform) norm. The upper-script $n \egal 1$ implies that both experiments are considered separately. The single step is used for estimating parameter $c_{\,1}$ and the multiple step experiment for the parameters $\bigr(\, \Fo \,,\, c_{\,2} \,\bigl)$. In such case, there is a cost function according to each experiment:
\begin{align*}
  \mathrm{J}^{\,1 \,A}_{\,i} \, \bigr(\, \Fo \,,\, c_{\,2} \,\bigl) & \egal \biggl|\biggl|\, u \moins u_{\,\mathrm{exp}\,,\, A} \, \,\biggr|\biggr|_{\,\mathcal{L}_{\,i}} \,, && \text{for the single step experiment,}  \\
  \mathrm{J}^{\,1 \,B}_{\,i} \, \bigr(\, c_{\,1} \,\bigl) & \egal \biggl|\biggl|\, u \moins u_{\,\mathrm{exp}\,,\, B} \, \,\biggr|\biggr|_{\,\mathcal{L}_{\,i}} \,, && \text{for the multiple step experiment} \,.
\end{align*}

The estimation of the unknown parameters proceeds in an iterative approach as described in the Algorithm~\ref{alg:iteration_cost_function}. In this case, a tolerance $\eta \egal 10^{\,-6}$ has been chosen.

\IncMargin{1em}
\begin{algorithm}[H]
  \SetKwData{Left}{left}\SetKwData{This}{this}\SetKwData{Up}{up}
  \SetKwFunction{Union}{Union}\SetKwFunction{FindCompress}{FindCompress}
  \SetKwInOut{Input}{input}\SetKwInOut{Output}{output}
  $ \bigr(\, \Fo \,,\, c_{\,1} \,,\, c_{\,2} \,\bigl)^{\,k} \egal \bigr(\, \Fo \,,\, c_{\,1} \,,\, c_{\,2} \,\bigl)^{\,\mathrm{apr}}  $ \\
  \While{ $\big|\big| \, \bigr(\, \Fo \,,\, c_{\,1} \,,\, c_{\,2} \,\bigl)^{\,k} \moins \bigr(\, \Fo \,,\, c_{\,1} \,,\, c_{\,2} \,\bigl)^{\,k-1} \, \big|\big| \geqslant \eta$}%
  {
  $\bigr(\, \Fo \,,\, c_{\,2} \,\bigl)^{\,k} \egal \mathrm{arg} \min \mathrm{J}^{\,1 \,A}_{\,i}$ \\
  $\bigr(\, c_{\,1} \,\bigl)^{\,k} \egal \mathrm{arg} \min \mathrm{J}^{\,1 \,B}_{\,i}$ \\
  $ k \egal k \plus 1$ \\
  } 
  \caption{\small\em Estimation of the unknown parameters $\bigr(\, \Fo \,,\, c_{\,1} \,,\, c_{\,2} \,\bigl)$ considering both experiments separately with an iterative process.}
  \label{alg:iteration_cost_function}
\end{algorithm}
\DecMargin{1em}

When $n \ \in \ \bigl\{\, 2 \,,\, 3 \, \bigr\}$, both experiments of single and multiple steps are considered at the same time, without any distinction. For $n \egal 2 \,$, parameters $\bigr(\, \Fo \,,\, c_{\,1} \,,\, c_{\,2} \,\bigl)$ are estimated at once. An additional test, for $n \egal 3 $ is carried by considering both experiments to estimate all the material properties parameters $\bigr(\, \Fo \,,\, c_{\,1} \,,\, c_{\,2} \,,\, d_{\,1} \,,\, \Pe \,\bigl)$. Thus, in these cases, the cost functions are defined as:
\begin{align*}
  \mathrm{J}^{\,2}_{\,2} \ \bigr(\, \Fo \,,\, c_{\,1} \,,\, c_{\,2} \,\bigl) & \egal \biggl|\biggl|\, u \moins u_{\,\mathrm{exp}\,,\, A} \, \,\biggr|\biggr|_{\,\mathcal{L}_{\,2}} \plus \biggl|\biggl|\, u \moins u_{\,\mathrm{exp}\,,\, B} \, \,\biggr|\biggr|_{\,\mathcal{L}_{\,2}} \,, \\
  \mathrm{J}^{\,2}_{\,\infty \,,\, \mathrm{max}} \ \bigr(\, \Fo \,,\, c_{\,1} \,,\, c_{\,2} \,\bigl) & \egal \max \, \Biggl\{\, \biggl|\biggl|\, u \moins u_{\,\mathrm{exp}\,,\, A} \, \,\biggr|\biggr|_{\,\mathcal{L}_{\,\infty}}\,,\, \biggl|\biggl|\, u \moins u_{\,\mathrm{exp}\,,\, B} \, \,\biggr|\biggr|_{\,\mathcal{L}_{\,\infty}} \,\Biggr\} \,, \\
  \mathrm{J}^{\,2}_{\,\infty \,,\, +} \ \bigr(\, \Fo \,,\, c_{\,1} \,,\, c_{\,2} \,\bigl) & \egal \biggl|\biggl|\, u \moins u_{\,\mathrm{exp}\,,\, A} \, \,\biggr|\biggr|_{\,\mathcal{L}_{\,\infty}} \plus \biggl|\biggl|\, u \moins u_{\,\mathrm{exp}\,,\, B} \, \,\biggr|\biggr|_{\,\mathcal{L}_{\,\infty}}  \,, \\
\end{align*}
for $n \egal 2 \,$,  and for $n \egal 3 \,$:
\begin{align*}
  \mathrm{J}^{\,3}_{\,2} \ \bigr(\, \Fo \,,\, c_{\,1} \,,\, c_{\,2} \,,\, d_{\,1} \,,\, \Pe \,\bigl) & \egal \biggl|\biggl|\, u \moins u_{\,\mathrm{exp}\,,\, A} \, \,\biggr|\biggr|_{\,\mathcal{L}_{\,2}} \plus \biggl|\biggl|\, u \moins u_{\,\mathrm{exp}\,,\, B} \, \,\biggr|\biggr|_{\,\mathcal{L}_{\,2}} \,, \\
  \mathrm{J}^{\,3}_{\,\infty\,,\, \mathrm{max}} \ \bigr(\, \Fo \,,\, c_{\,1} \,,\, c_{\,2} \,,\, d_{\,1} \,,\, \Pe \,\bigl) & \egal \max \, \Biggl\{\, \biggl|\biggl|\, u \moins u_{\,\mathrm{exp}\,,\, A} \, \,\biggr|\biggr|_{\,\mathcal{L}_{\,\infty}} \,,\, \biggl|\biggl|\, u \moins u_{\,\mathrm{exp}\,,\, B} \, \,\biggr|\biggr|_{\,\mathcal{L}_{\,\infty}} \,\Biggr\} \,, \\
  \mathrm{J}^{\,3}_{\,\infty\,,\, \mathrm{+}} \ \bigr(\, \Fo \,,\, c_{\,1} \,,\, c_{\,2} \,,\, d_{\,1} \,,\, \Pe \,\bigl) & \egal \biggl|\biggl|\, u \moins u_{\,\mathrm{exp}\,,\, A} \, \,\biggr|\biggr|_{\,\mathcal{L}_{\,\infty}} \plus \biggl|\biggl|\, u \moins u_{\,\mathrm{exp}\,,\, B} \, \,\biggr|\biggr|_{\,\mathcal{L}_{\,\infty}}  \,. \\
\end{align*}

Table~\ref{tab:synthesis_tests} synthesizes all tests performed according to the definition of the cost function $\mathrm{J}$. The cost function $\mathrm{J}$ is minimized using function \texttt{fmincon} in the \texttt{Matlab\texttrademark} environment, providing an efficient interior-point algorithm with constraint on the unknown parameters \cite{Byrd2000}. Here, the box-type constraints are defined with upper and lower bound for the parameters:
\begin{align*}
  p^{\,\circ} \ \in \ \bigl[ \, 0.8 \,,\, 1.5 \, \bigl] \cdot p^{\,\mathrm{apr}} \,,
\end{align*}
where the upper-scripts $\circ$ and $\mathrm{apr}$ denote the estimated and \emph{a priori} values of the parameters, respectively. The bounds have been defined by performing previous tests and analyzing the parameter impact on the calibration.

In order to quantify the quality of measured data, we estimate the noise inherent to any real physical measurement. By assuming that the noise $\xi\,(\omega)$ is centered \textsc{Gaussian} (\ie $\xi\,(\omega)\ \sim\ \mathcal{N}\,(0,\,\sigma^{\,2})$), linear and additive, its variance $\sigma^{\,2}$ can be thus estimated. Moreover, we assume that the underlying signal is smooth. In order to extract the noise component, the signal is approximated locally by a low-order polynomial representing the trend. Then, the trend is removed by using a special filter, leaving us with the pure noise content, which can be further analyzed using the standard statistical techniques. For the considered data, the variance equals:
\begin{align*}
  & \sigma \ \simeq \ 0.01 \,, && \text{for the single-step experiments} \,, \\
  & \sigma \ \simeq \ 0.008 \,, && \text{for the multiple-step experiments} \,.
\end{align*}
The noise variance does not necessarily correspond to the measurement accuracy. This measure provides a lower bound of this error, \ie  the accuracy cannot be lower than the noise present in the measurement.

\begin{table}
\centering
\caption{\small\em Synthesis of the tests carried out with the expression of the cost function.}
\bigskip
\setlength{\extrarowheight}{.3em}
\begin{tabular}[l]{@{} l|ccccccccc}
\hline
\hline
\multirow{2}{*}{\textit{Definition of the cost function $\mathrm{J}$}} & \multicolumn{9}{c}{\textit{Tests}} \\
 & 1 & 2 & 3 & 4 & 5 & 6 & 7 & 8 & 9 \\
 \hline
Experiments considered separately
& x & x & x 
&   &   &  
&   &   &  \\
Experiments at the same time ($3$ parameters to estimate)
&   &   &  
& x & x & x 
&  &  &  \\
Experiments at the same time ($5$ parameters to estimate)
&   &   &  
&  &  & 
& x & x & x \\
Euclidean norm
& x &   &   
& x &   &  
& x &   &  \\
Infinite norm
&   & x &   
&   & x &  
&   & x &  \\
Sum of the infinite norms
&   &   & x
&   &   & x
&   &   & x\\
\hline
\hline
\end{tabular}
\bigskip
\label{tab:synthesis_tests}
\end{table}


\subsection{Results}

\begin{figure}
  \begin{center}
  \subfigure[\label{fig:EST_residual_ftests}]{\includegraphics[width=.48\textwidth]{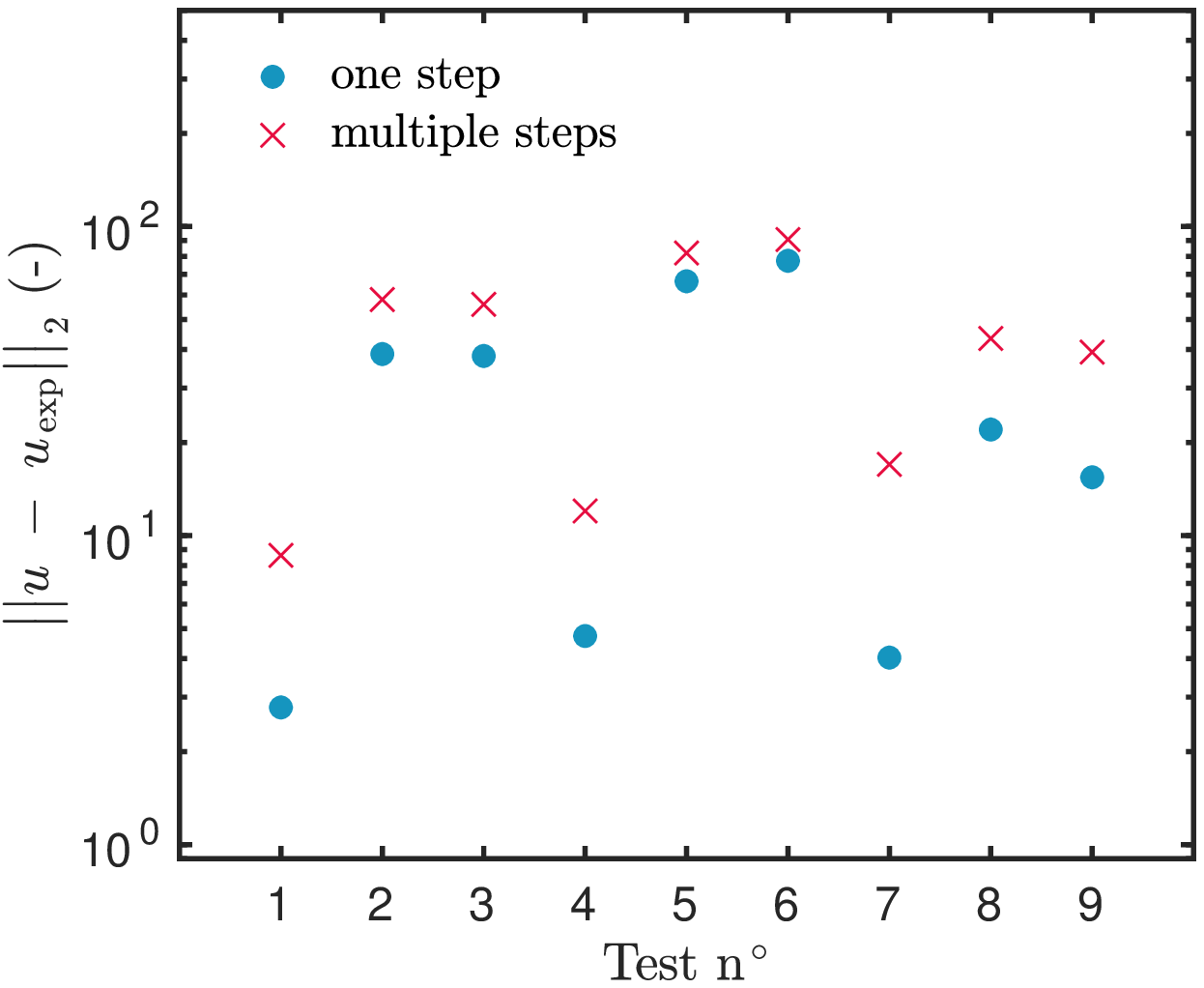}}
  \subfigure[\label{fig:EST_direct_computation_ftests}]{\includegraphics[width=.48\textwidth]{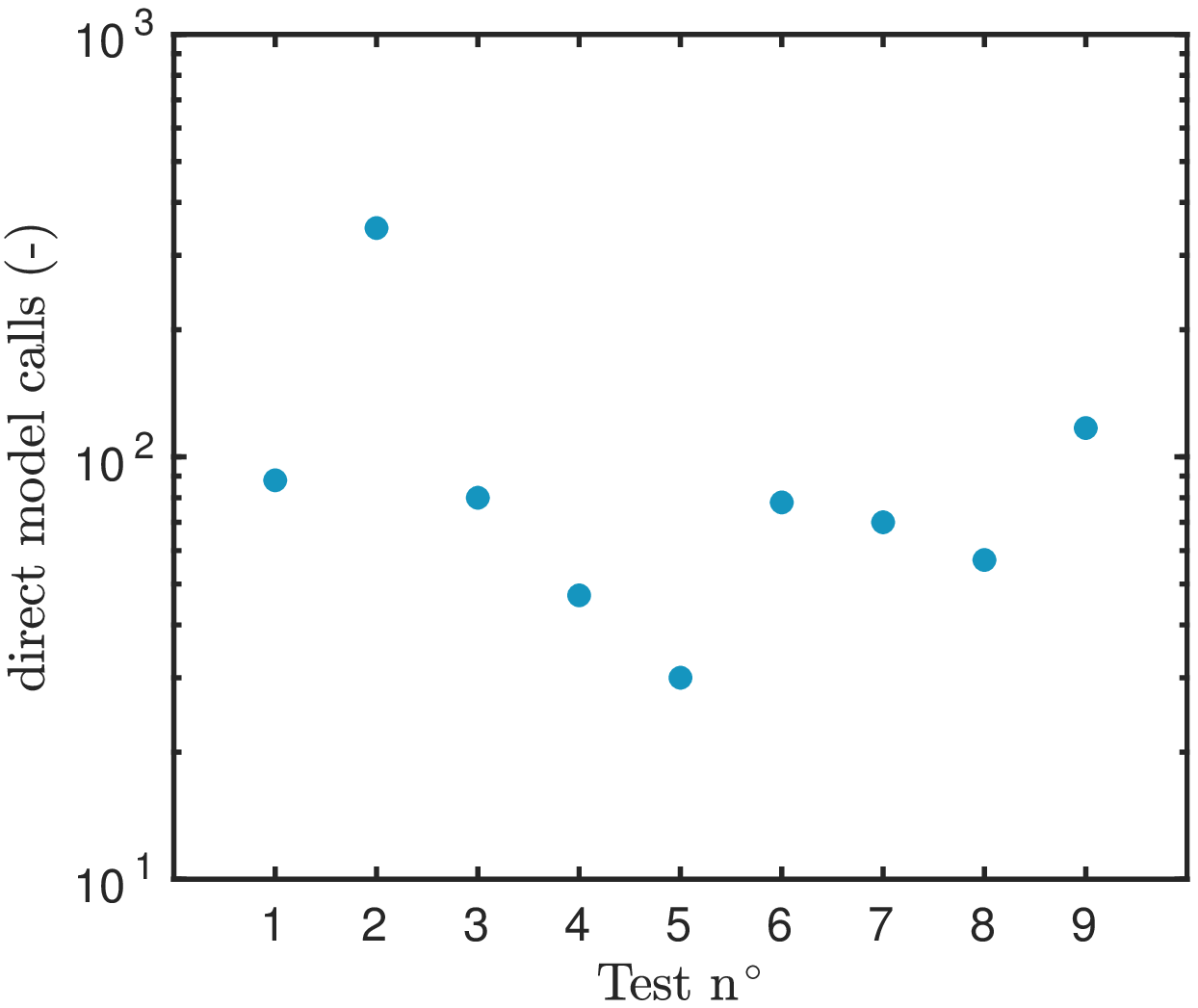}}
  \caption{\small\em Residual between the measured data and numerical results for both experiments (a) and number of computations of the direct problem (b) for the different definition of the cost function $\mathrm{J}\,$.}
  \label{fig:results_ftests}
  \end{center}
\end{figure}

Figure~\ref{fig:EST_residual_ftests} shows the variation of the residual between the measured data and the numerical results for different tests performed. The residual is minimized for tests $1$, $4$ and $7$, corresponding to the involving considering the \textsc{Euclidean} norm for the computation of the cost functions. The tendencies are similar for both experiments. It can be noted that the residual is lower when estimating only three parameters $\bigr(\, \Fo \,,\, c_{\,1} \,,\, c_{\,2} \,\bigl)$ and not all the parameters of the material properties. Figure~\ref{fig:EST_direct_computation_ftests} provides the number of computations of the direct problem. As expected, the tests $1$ to $3$, considering both experiments separately, require a few more computations of the direct problem, due to the iterative procedure. Globally, the algorithm requires less than $100$ computations, which is extremely low compared to stochastic approaches. For instance in \cite{Berger2016b}, $10^{\,4}$ computations are necessary to estimate the thermal conductivity of two materials by solving an inverse heat transfer problem.

\begin{figure}
  \begin{center}
  \subfigure[\label{fig:EST_comp_1step}]{\includegraphics[width=.48\textwidth]{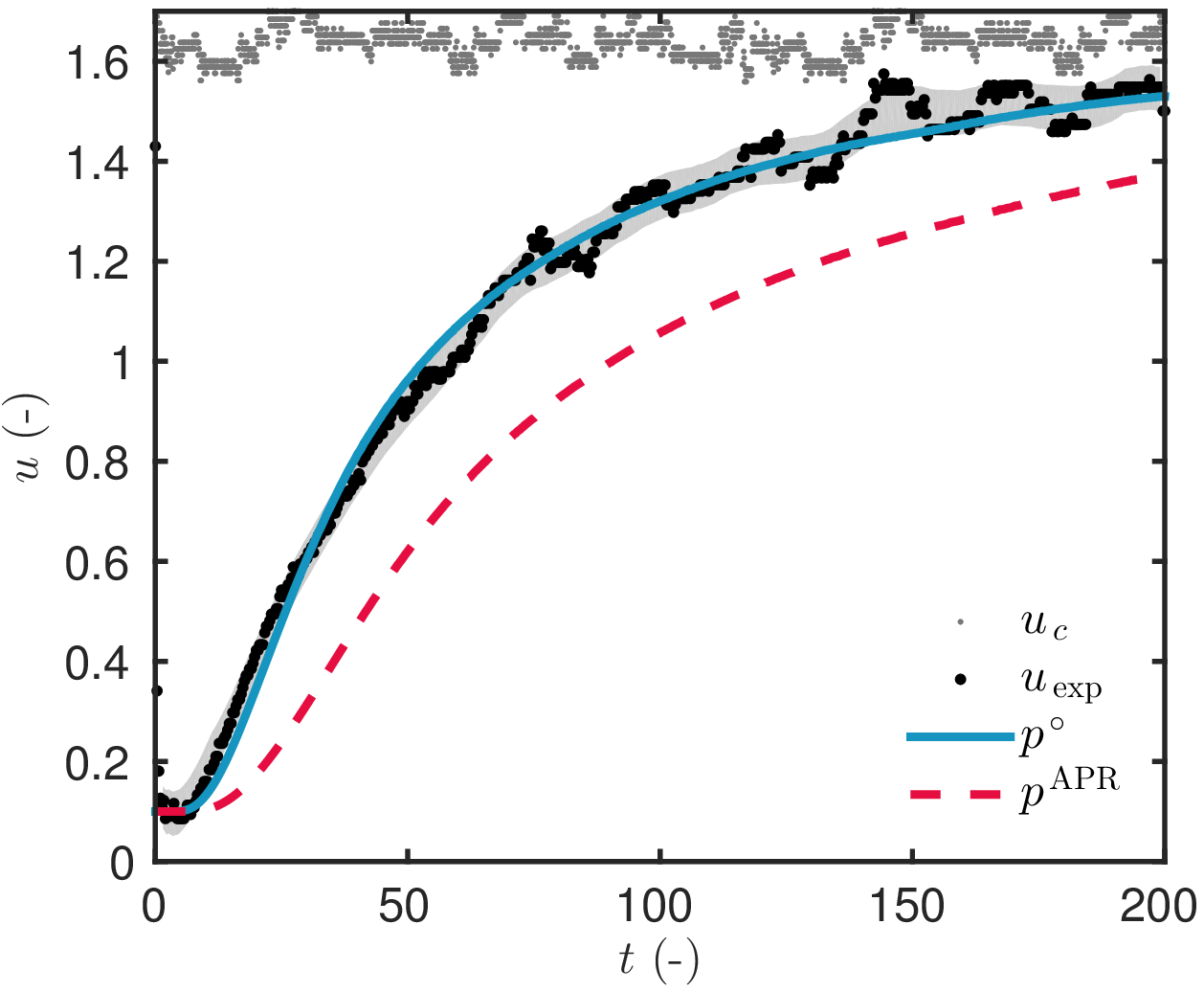}}
  \subfigure[\label{fig:EST_comp_Mstep}]{\includegraphics[width=.48\textwidth]{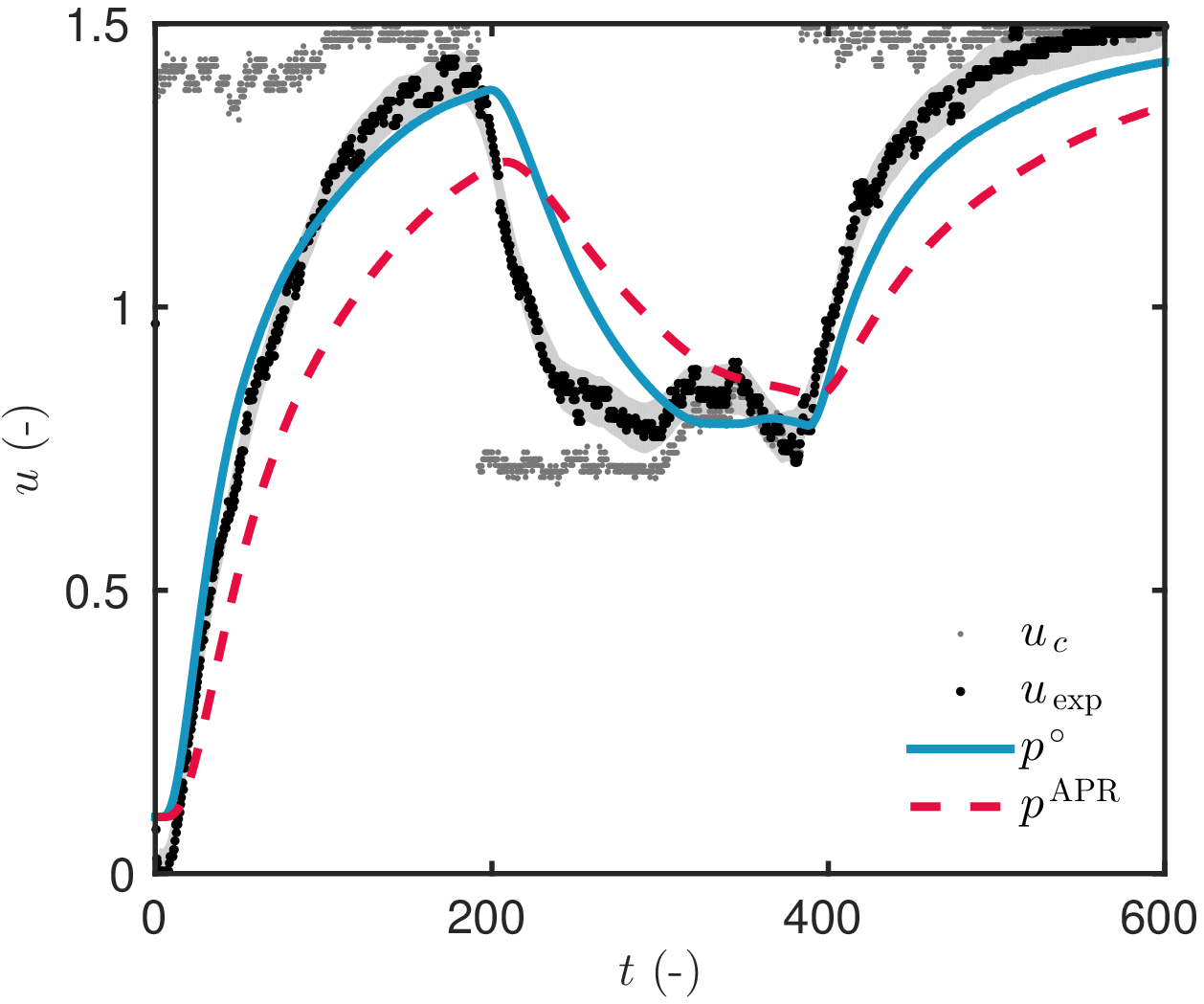}}
  \subfigure[\label{fig:EST_residual_1step}]{\includegraphics[width=.48\textwidth]{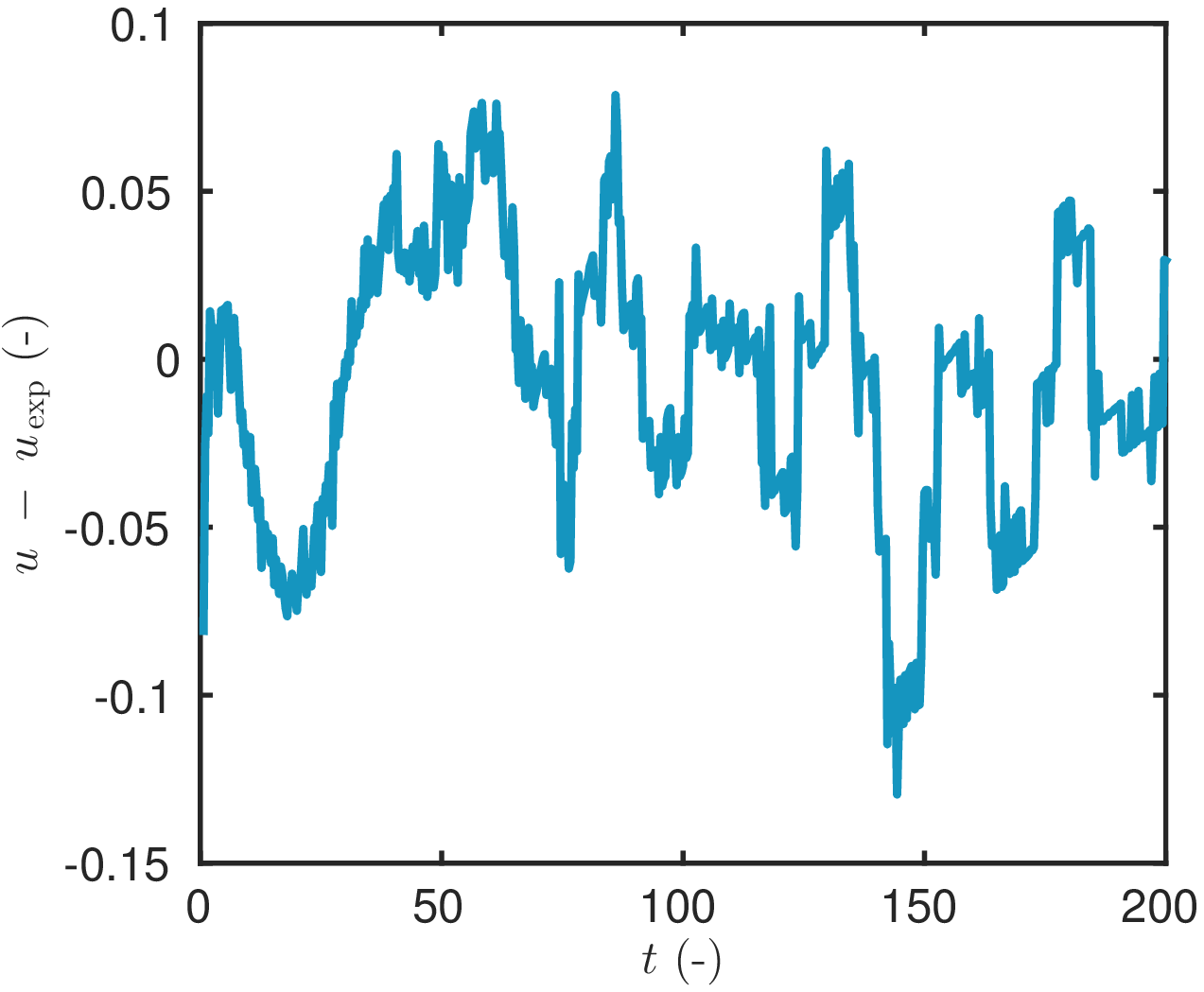}}
  \subfigure[\label{fig:EST_residual_Mstep}]{\includegraphics[width=.48\textwidth]{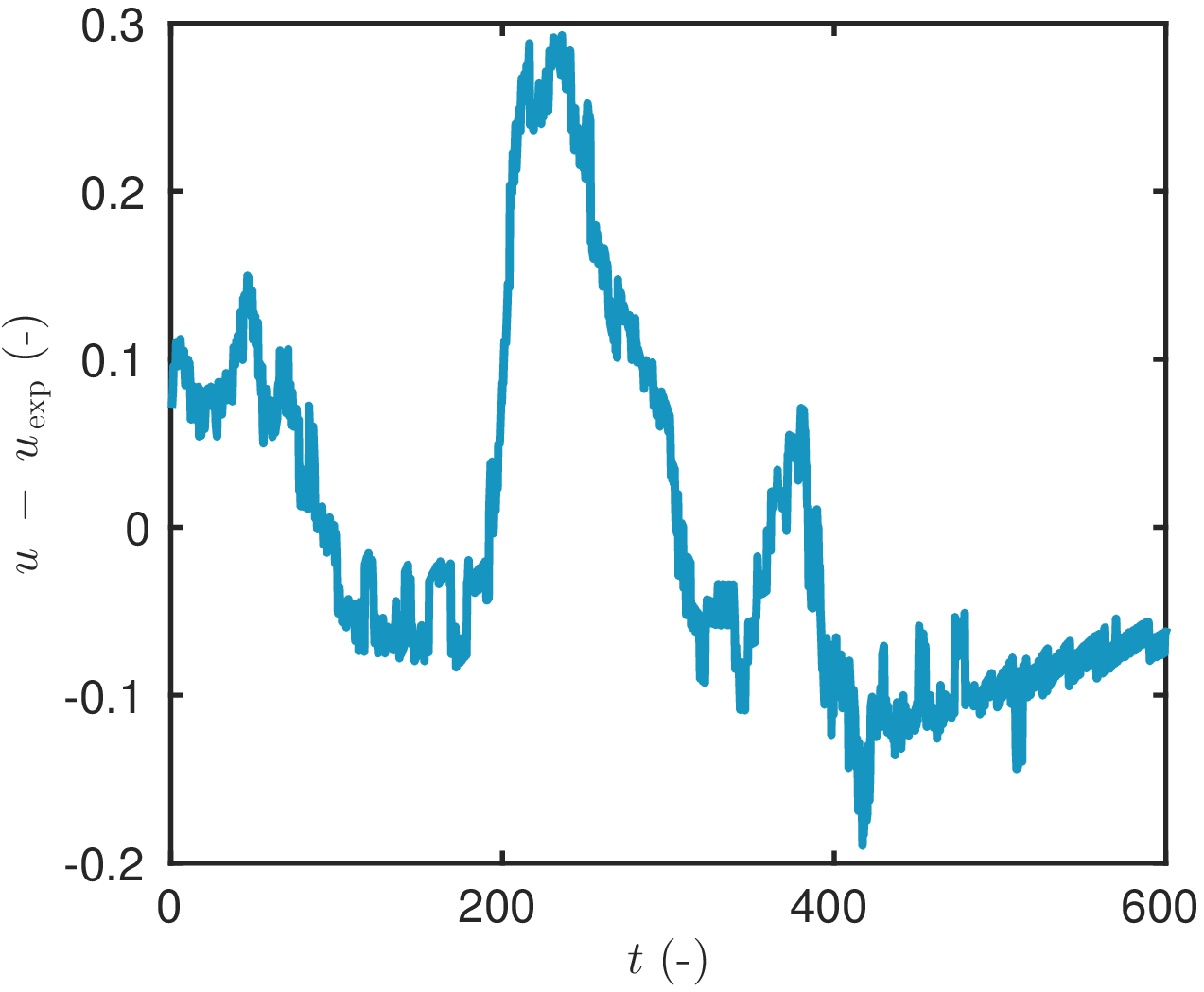}}
  \caption{\small\em Comparison of the numerical results with the experimental data (a-b) and their $98\%$ confidence interval for the single-step (a) and the multiple-step (b) experiments (test $1$). Comparison of the residual for the single-step (c) and multiple-step (d) experiments.}
  \label{fig:EST_results_test1}
  \end{center}
\end{figure}

\begin{figure}
  \begin{center}
  \subfigure[\label{fig:EST_convergence_test1}]{\includegraphics[width=.48\textwidth]{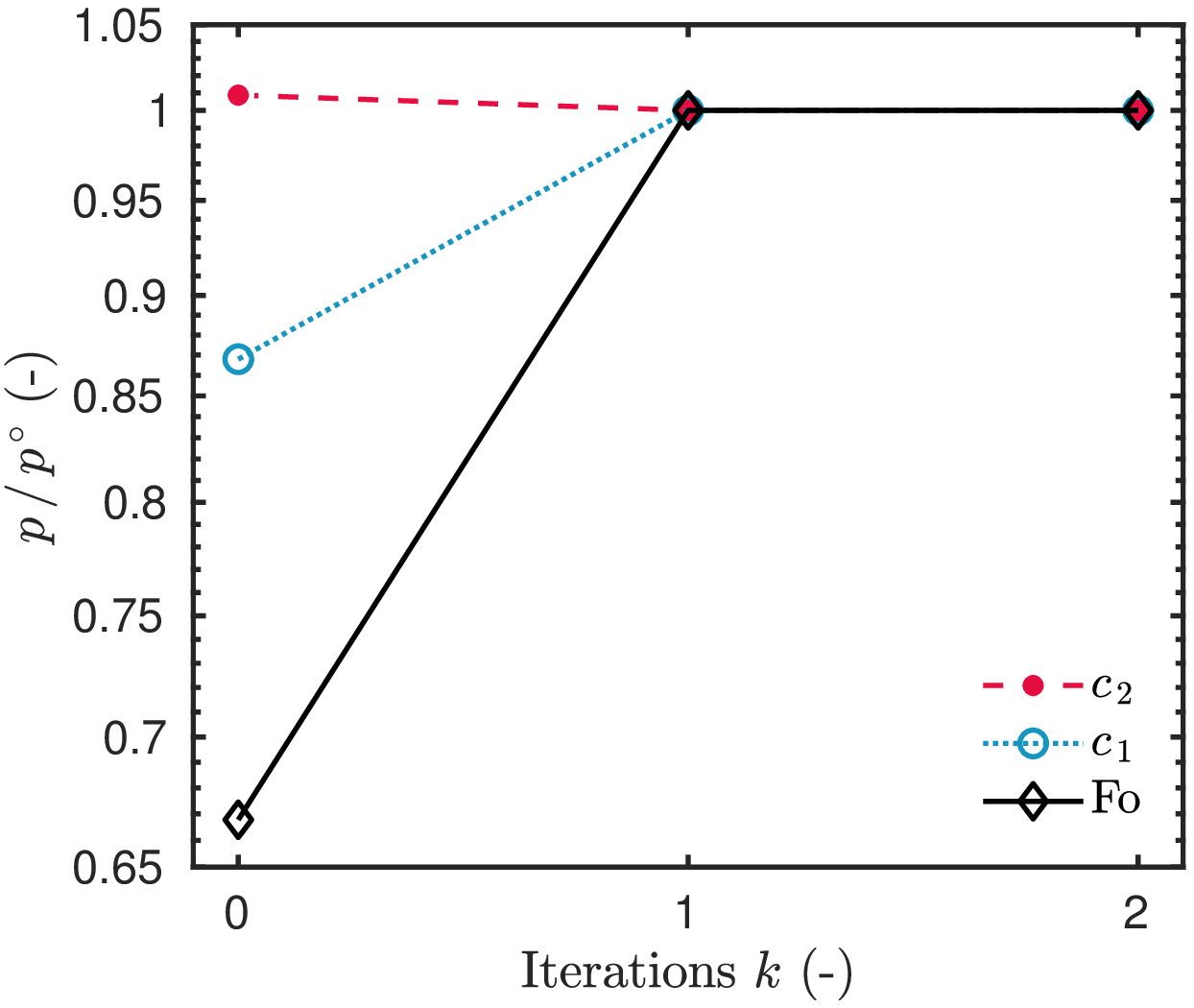}}
  \subfigure[\label{fig:EST_direct_computation_test1}]{\includegraphics[width=.48\textwidth]{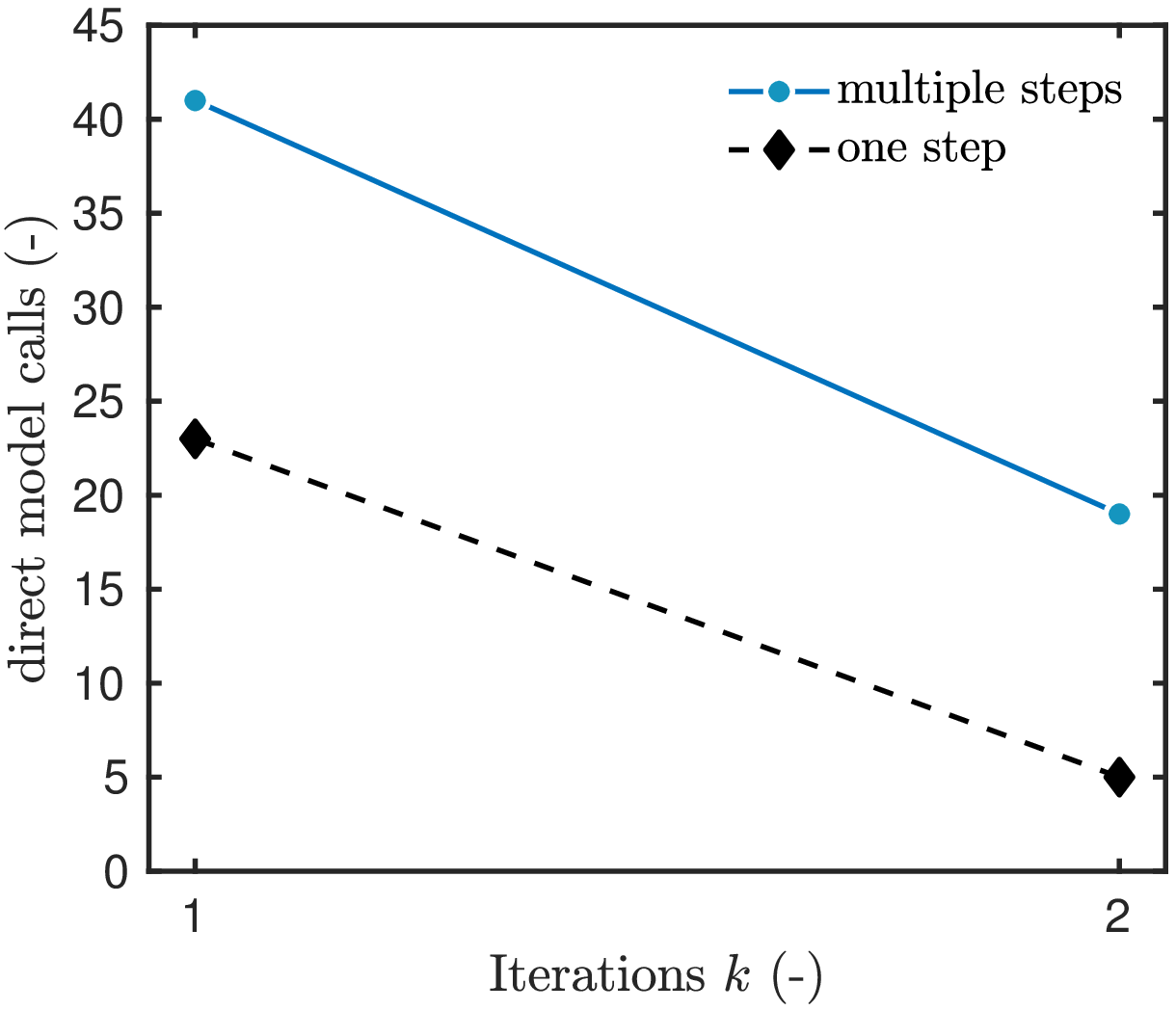}}
  \caption{\small\em Convergence of the parameter estimation, for the test $1$ as a function of the number of iterations (a) and number of computations of the direct model (b).}
  \end{center}
\end{figure}

The comparison of the measured data and numerical results is illustrated in Figure~\ref{fig:EST_comp_1step} for the one-step experiment. Figure~\ref{fig:EST_comp_Mstep} shows it for the multiple-step procedure, for the case $20\,$. Results of the numerical model are provided for the \emph{a priori} and estimated three parameters. As mentioned in the Introduction, the numerical model with \emph{a priori} parameters predicts values lower than those obtained from experiments during the moisture adsorption and greater than them along the desorption processes. With the calibrated model, \ie with the estimated parameters, there is a better agreement between the numerical results and the experimental data. Figures~\ref{fig:EST_residual_1step} and \ref{fig:EST_residual_Mstep} show the residual. It is uncorrelated, highlighting a satisfactory estimation of the parameters. Nevertheless, it can be noted that some discrepancies remain between the experimental data and the numerical results. This can be specifically observed at $t \egal 200 \,$, in Figures~\ref{fig:EST_comp_Mstep} and \ref{fig:EST_residual_Mstep}, for which some explanations are possible. First, the mathematical model may fail in representing the physical phenomenon. Some assumptions such as isothermal conditions, unidimensional transport and constant velocity might contribute to the differences observed between experimental and numerical results. Although the experiment has been conceived to be isothermal, slight variations in the temperature field occurs due to mechanisms of phase change that may affect the profile of vapor pressure, which is highly temperature dependent. An important assumption, very often considered in literature, is the disregard of the hysteresis phenomenon, which may considerably affect the results. This issue can be noted in Figure~\ref{fig:EST_comp_Mstep} for $t \ \in \ \bigl[\,0\,,\,300 \,\bigr]$. A good agreement is noted during the adsorption process $t \ \in \ \bigl[\,0\,,\,200 \,\bigr]\,$, since the coefficients have been estimated for a similar experiment. However, during the desorption process, the field $u$ is underestimated by the numerical predictions. On the other hand, despite the low relative humidity uncertainty, other uncertainties appear such as the interference of sensors on the moisture transfer through the sample, contact resistances and no perfect impermeabilization. Another possible explanation is associated to the parametrization of the material properties that can be improved. An interesting alternative could be to search for time varying parameters by adding a regularization term in the cost function $\mathrm{J}$. The convergence of the parameter estimation is shown in Figure~\ref{fig:EST_convergence_test1}. On the contrary to parameters $\Fo$ and $c_{\,1}\,$, the \emph{a priori} values of $c_{\,2}$ is not far from the estimated one. After one iteration, the algorithm almost estimates the parameters. The number of computations of the direct model for Test $1$ is given in Figure~\ref{fig:EST_direct_computation_test1}. Only two global iterations are required to compute the solution of the inverse problem. More computations are required at the first iteration as the unknown parameters are more distant from the estimated optimal ones.

\begin{figure}
  \begin{center}
  \subfigure[\label{fig:EST_pdfF0_ft}]{\includegraphics[width=.48\textwidth]{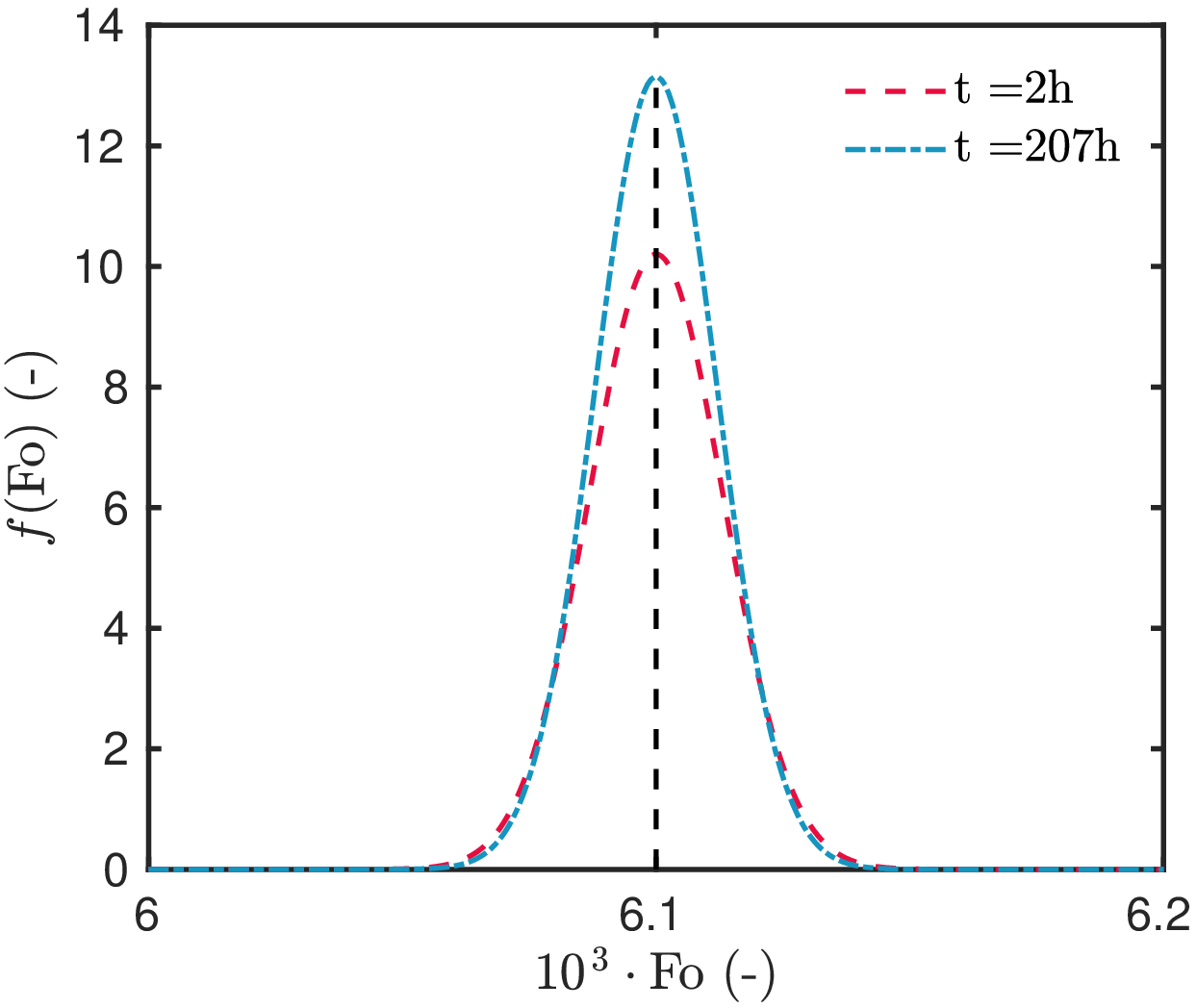}}
  \subfigure[\label{fig:EST_pdfc2_ft}]{\includegraphics[width=.48\textwidth]{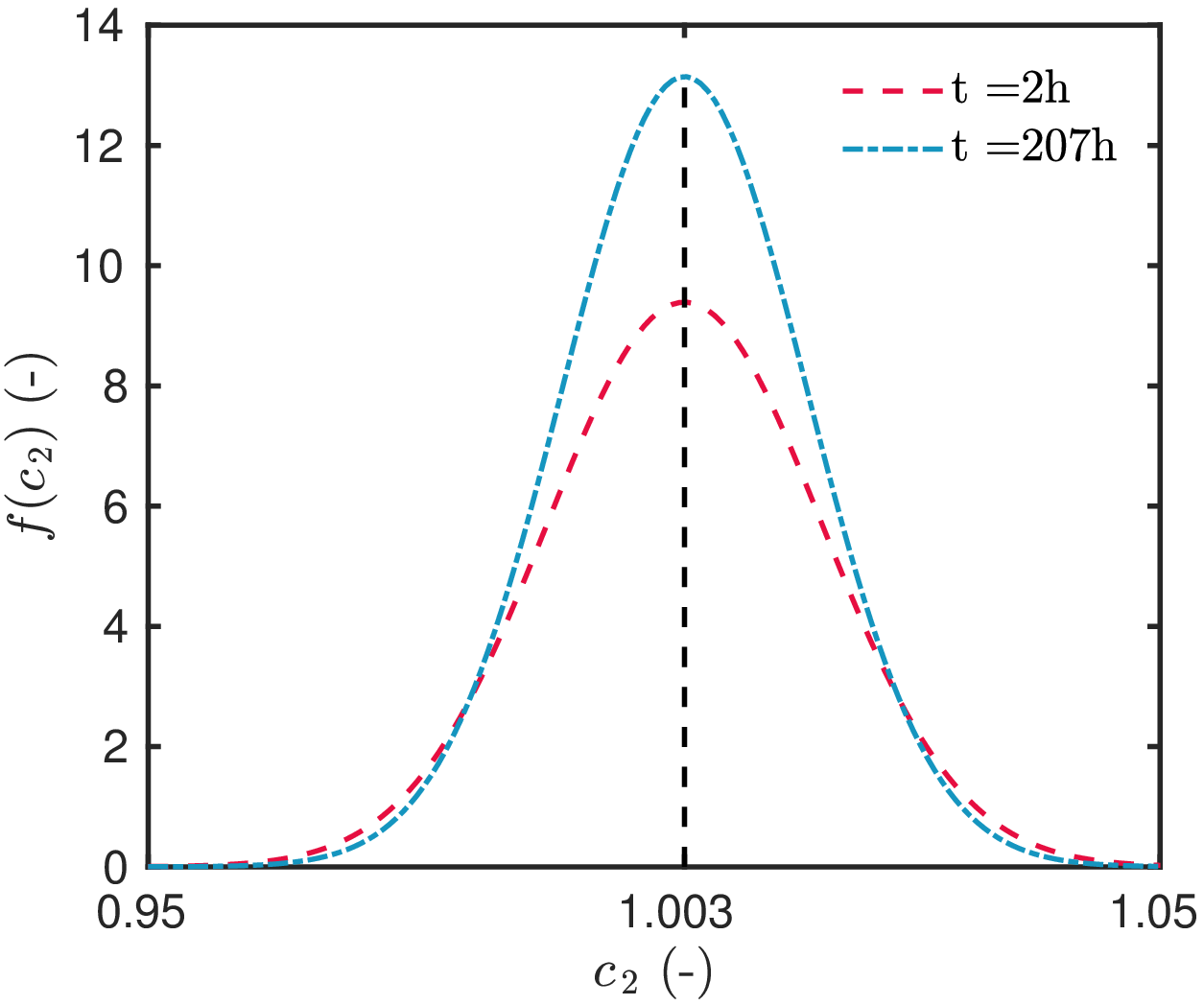}}
  \subfigure[\label{fig:EST_pdfc1_ft}]{\includegraphics[width=.48\textwidth]{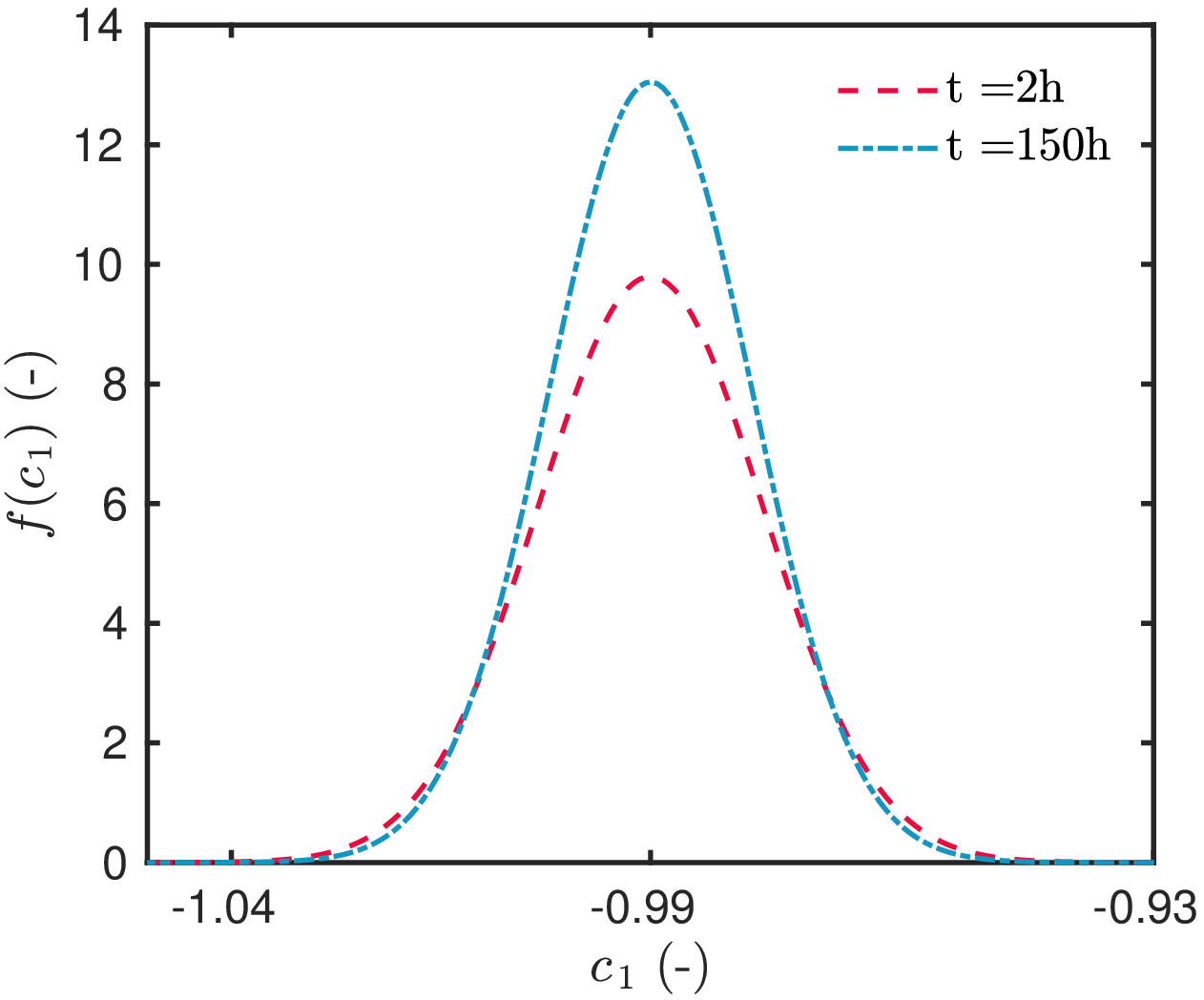}}
  \caption{\small\em Probability density function approximated for the estimated parameters, computed using the sensitivity of single-step (c) and a multiple-step (a-b) experiments.}
  \label{fig:EST_pdf_ft}
  \end{center}
\end{figure}

The computation of the sensitivity functions of the parameters to be estimated enables to approximate their probability density functions. The error is assumed as a normal distribution $\mathcal{N} (\, 0 \,,\, \sigma_{\,\mathrm{eff}}^{\,2} \,)$ with its standard deviation $\sigma_{\,\mathrm{eff}}$ computed by adding the ones due to uncertainty and to the noise:
\begin{align*}
  \sigma_{\,\mathrm{eff}} \egal \sigma_{\,\mathrm{noise}} \plus \sigma_{\,\mathrm{error}} \,.
\end{align*}
Thus, the probability distribution is computed for different times as illustrated in Figure~\ref{fig:EST_pdf_ft}. As reported in Figure~\ref{fig:Mstep_S_OED}, the sensitivity function of parameters  $\Fo$ and $c_{\,1}$ is maximum and minimum at $t \egal 207$ and $t \egal 4\,$, respectively. It explains why the probability function is maximum at $t \egal 207$ for these parameters. Similar results can be observed when comparing the sensitivity function of parameter $c_{\,2}$ in Figure~\ref{fig:1step_S_OED} and the probability function in Figure~\ref{fig:EST_pdfc1_ft}.

\section{Accounting for hysteresis}

As mentioned in the description of the physical model, the hysteresis effects are not considered in the sorption capacity of the material. This assumption impacts the comparison between numerical and experimental observations, particularly for the cases of multiple steps of relative humidity (Figure~\ref{fig:EST_comp_Mstep}). In this section, the physical model is changed to revise this assumption, which is commonly considered.


\subsection{Model for hysteresis}

The hysteresis impacts the sorption curve $f(\,\phi\,)$ and consequently on the moisture capacity coefficient $c$ defined in Eq.~\eqref{eq:coeff_c}. To account for hysteresis, several model are available in literature as proposed in \cite{Mualem2009} or in \cite{Derluyn2012}. Recently, in \cite{Berger2018a}, a new model of hysteresis has been proposed and validated using experimental data. This model is referred as a smoothed \emph{bang}--\emph{bang} model in control literature. It enables intermediary sorption states between the main adsorption $c^{\,\star}_{\,\mathrm{ads}}$ and desorption $c^{\,\star}_{\,\mathrm{des}}$ curves illustrated in Figure~\ref{fig:EST_wH_c_fu}. The coefficient $c$ is computed using the following dimensionless differential equation:
\begin{align}\label{eq:model_hysteresis}
  \pd{\cs}{\ts} \egal \beta \ \mathrm{sign} \ \biggl(\, \pd{u}{\ts} \,\biggr) \, \biggl(\, \cs \moins c^{\,\star}_{\,\mathrm{ads}} \,\biggr) \ \biggl(\, \cs \moins c^{\,\star}_{\,\mathrm{des}} \,\biggr) \,.
\end{align}
As for the previous case, these two properties are assumed to be second-degree polynomials of the relative humidity:
\begin{align*}
  c^{\,\star}_{\,\mathrm{ads}} & \egal 1 \plus c^{\,\star}_{\,a\,,1} \, u \plus c^{\,\star}_{\,a\,,2} \, u^{\,2} \,, \\
  c^{\,\star}_{\,\mathrm{des}} & \egal c^{\,\star}_{\,d\,,0} \plus c^{\,\star}_{\,d\,,1} \, u \plus c^{\,\star}_{\,d\,,2} \, u^{\,2} \,.
\end{align*}
It should be noted that here, the dimensionless coefficient $c^{\,\star}$ is defined as $\cs \egal \displaystyle \frac{c}{c^{\,\star}_{\,a\,,0}}\,$. The coefficient $\beta$ is a numerical parameter, controlling the transition velocity between the two curves. The function $\mathrm{sign} \,: \R \ \longrightarrow \ \bigl\{\,-1 \,,\, 0 \,,\, 1 \,\bigr\} $ is defined as:
\begin{align*}
\mathrm{sign} \, \bigl(\,x\,\bigr) \egal 
\begin{cases} 
\, 1 \,, & \quad  x \ > \ 0 \,, \\
\, 0 \,, & \quad  x \egal 0 \,, \\
\, -1 \,, & \quad  x \ < \ 0 \,.
\end{cases}
\end{align*}
The mathematical model of moisture transfer is now defined by a system of one partial differential equation coupled with an ordinary differential equation:
\begin{subequations}
\begin{align}
   \cs\ \pd{u}{\ts} &\egal \Fo \ \pd{}{\xs} \Biggl(\, \ds(\,u\,)  \ \pd{u}{\xs} \moins \Pe \ u \, \Biggr) \,, \label{eq:moisture_model_wH} \\
   \pd{\cs}{\ts} & \egal \beta \ \mathrm{sign} \ \biggl(\, \pd{u}{\ts} \,\biggr) \cdot \biggl(\, \cs \moins c^{\,\star}_{\,\mathrm{ads}} \,\biggr) \cdot \biggl(\, \cs \moins c^{\,\star}_{\,\mathrm{des}} \,\biggr) \,, \label{eq:hysteresis_model}
\end{align}
\end{subequations}
associated with initial and boundary conditions.

The issue is now to estimate the five coefficients of the adsorption and desorption curves $\bigl(\,c^{\,\star}_{\,a,\,1}\,,\,c^{\,\star}_{\,a,\,2}\,,\,c^{\,\star}_{\,d,\,0}\,,\,c^{\,\star}_{\,d,\,1}\,,\,c^{\,\star}_{\,d,\,2}\,\bigr)\,$. It should be noted that the demonstration of structural identifiability of the five parameters is provided in Appendix~\ref{sec:Annex_identifiability}. When searching the OED, one need to compute the partial derivative of $u$ according to each coefficient $c^{\,\star}_{\,a,\,d,\,m}$. Since the model is now composed of two differential equations, for each coefficients, two sensitivity functions have to be computed. For instance, for the coefficient $c^{\,\star}_{\,d,\,0}$, it is required to compute:
\begin{align*}
  & \Theta_{\,u} \egal \pd{u}{c^{\,\star}_{\,d,\,0}} && \text{and} && \Theta_{\,c} \egal \pd{c}{c^{\,\star}_{\,d,\,0}} \,. 
\end{align*}
From a mathematical point of view, the computation of $\Theta_{\,c}$ is not possible since the function $\mathrm{sign} (\,x\,)$ is not differentiable in the classical sense for $x \egal 0\,$. It is differentiable in the sense of distributions. Therefore, a regularized version of the hysteresis model~\eqref{eq:model_hysteresis}:
\begin{align*}
  \pd{\cs}{\ts} \egal \beta \ \mathrm{Rsign} \ \biggl(\, \pd{u}{\ts} \,\biggr) \cdot \biggl(\, \cs \moins c^{\,\star}_{\,\mathrm{ads}} \,\biggr) \cdot \biggl(\, \cs \moins c^{\,\star}_{\,\mathrm{des}} \,\biggr) \,,
\end{align*}
where the function $\mathrm{Rsign} \,: \mathbb{R} \ \rightarrow \ \mathbb{R} $ is defined as:
\begin{align*}
  \mathrm{Rsign}\,(\,x\,)\ =\ \tanh\,\bigl(\,\delta\, x\,\bigr)\,,
\end{align*}
with $\delta$ a sufficiently large real parameter. In this study, the following value is taken $\delta \egal 10^{\,8}\,$. A numerical validation of the regularized hysteresis model is provided in Appendix~\ref{sec:Annex1}.


\subsection{Searching the OED}

The OED is now searched among the experimental designs described in Section~\ref{sec:exp_facility} for both single and multiple steps. The issue is to estimate one or several parameters among the five coefficients $\bigl(\,c^{\,\star}_{\,a,\,1}\,,\,c^{\,\star}_{\,a,\,2}\,,\,c^{\,\star}_{\,d,\,0}\,,\,c^{\,\star}_{\,d,\,1}\,,\,c^{\,\star}_{\,d,\,2}\,\bigr)\,$. For this, the sensitivity functions $\displaystyle \Theta_{\,u,\,a\,,\,d\,,\,m} \egal \pd{u}{c^{\,\star}_{\,a,\,d,\,m}}$ and $\displaystyle \Theta_{\,c,\,a,\,d,\,m} \egal \pd{c}{c^{\,\star}_{\,a,\,d,\,m}} \,, m \ \in \ \bigl\{\,0 \,,1 \,, 2\,\bigr\}$ are computed using central finite-difference and a \RK ~approach provided by \texttt{ode45} solver in \texttt{Matlab\texttrademark} \cite{Shampine1997}. The equations to compute the sensitivity functions are given in Appendix~\ref{sec:Eq_sensitivity_coeff}.

\begin{figure}
  \begin{center}
  \subfigure[\label{fig:Mstep_wH_Psica_fdesign}]{\includegraphics[width=.48\textwidth]{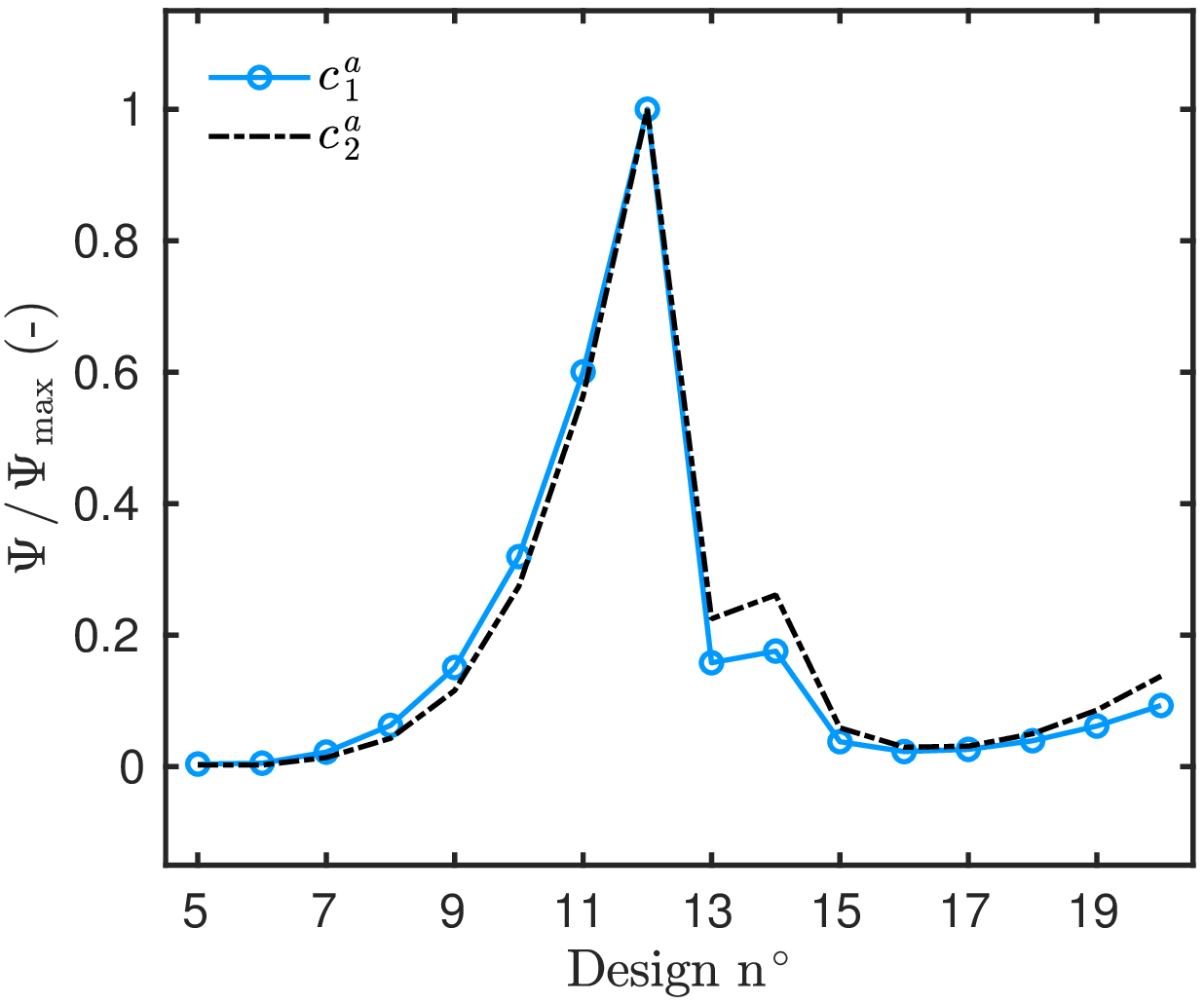}}
  \subfigure[\label{fig:Mstep_wH_Psicd_fdesign}]{\includegraphics[width=.48\textwidth]{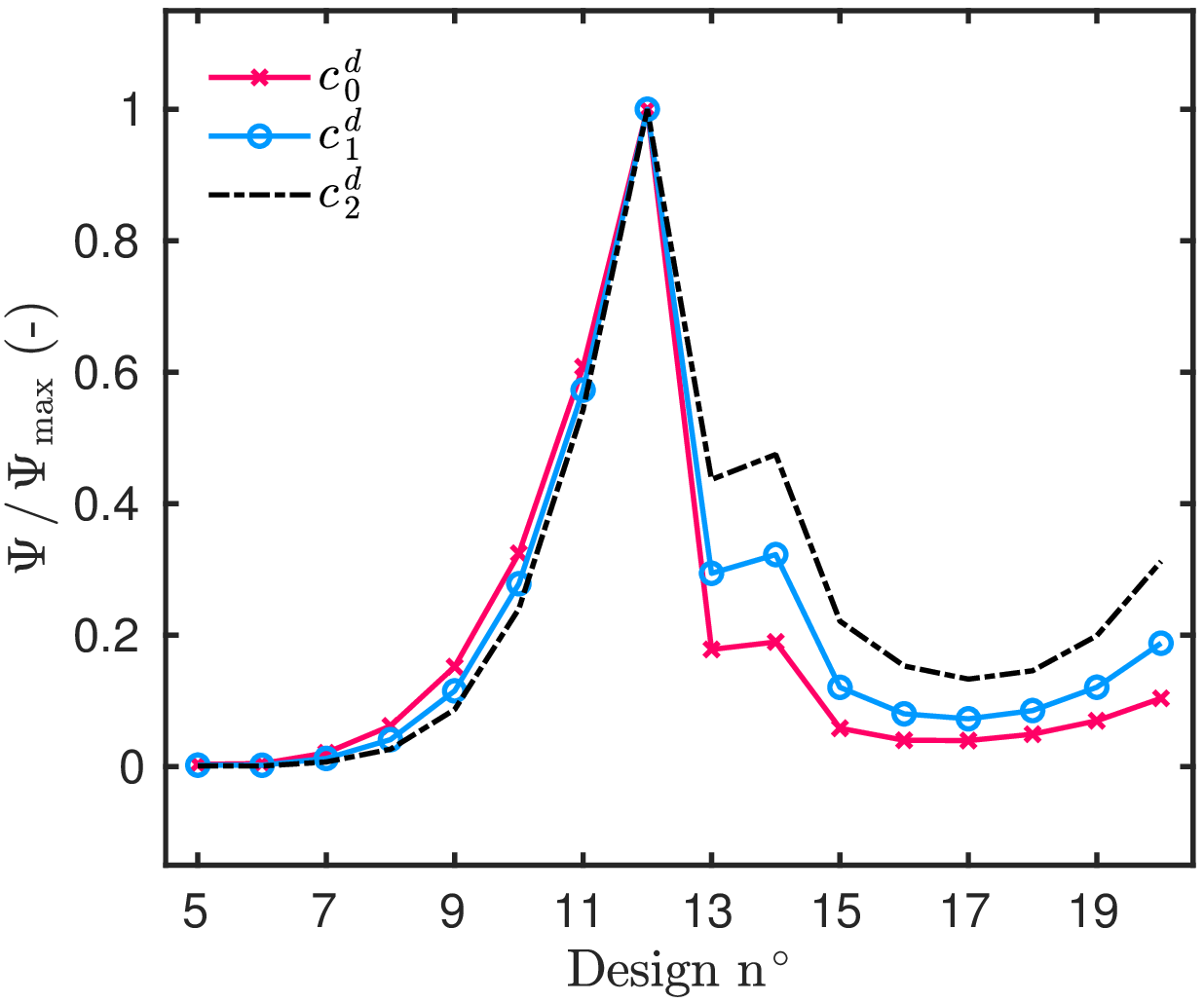}}
  \caption{\small\em Variation of the criterion $\Psi$ for the sixteen possible designs for the adsorption coefficients (a) and the desorption coefficients (b).}
  \label{fig:Mstep_wH_Psi_fdesign}
  \end{center}
\end{figure}

\begin{figure}
\begin{center}
\subfigure[\label{fig:Mstep_wH_Xa_OED}]{\includegraphics[width=.48\textwidth]{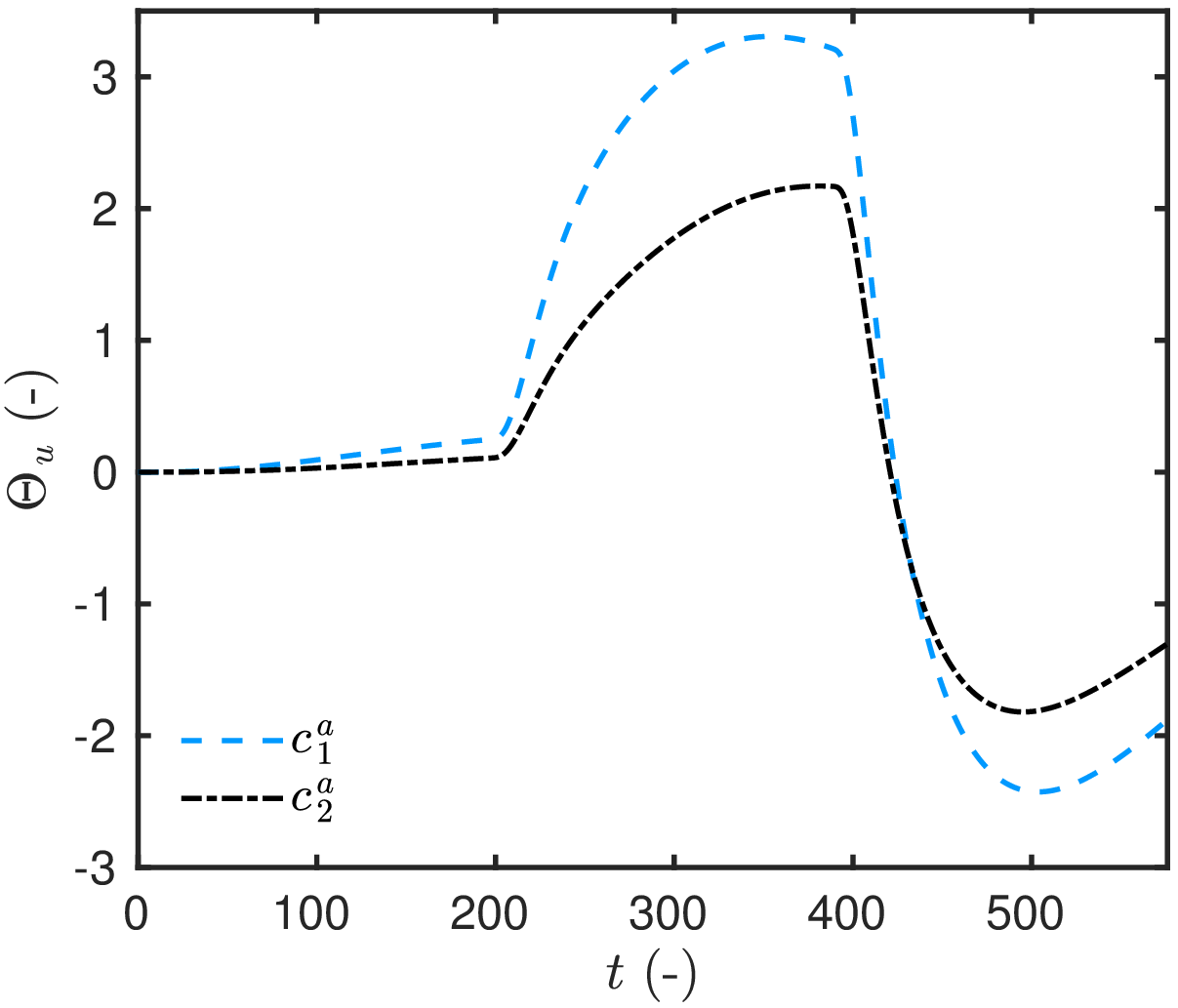}}
\subfigure[\label{fig:Mstep_wH_Xd_OED}]{\includegraphics[width=.48\textwidth]{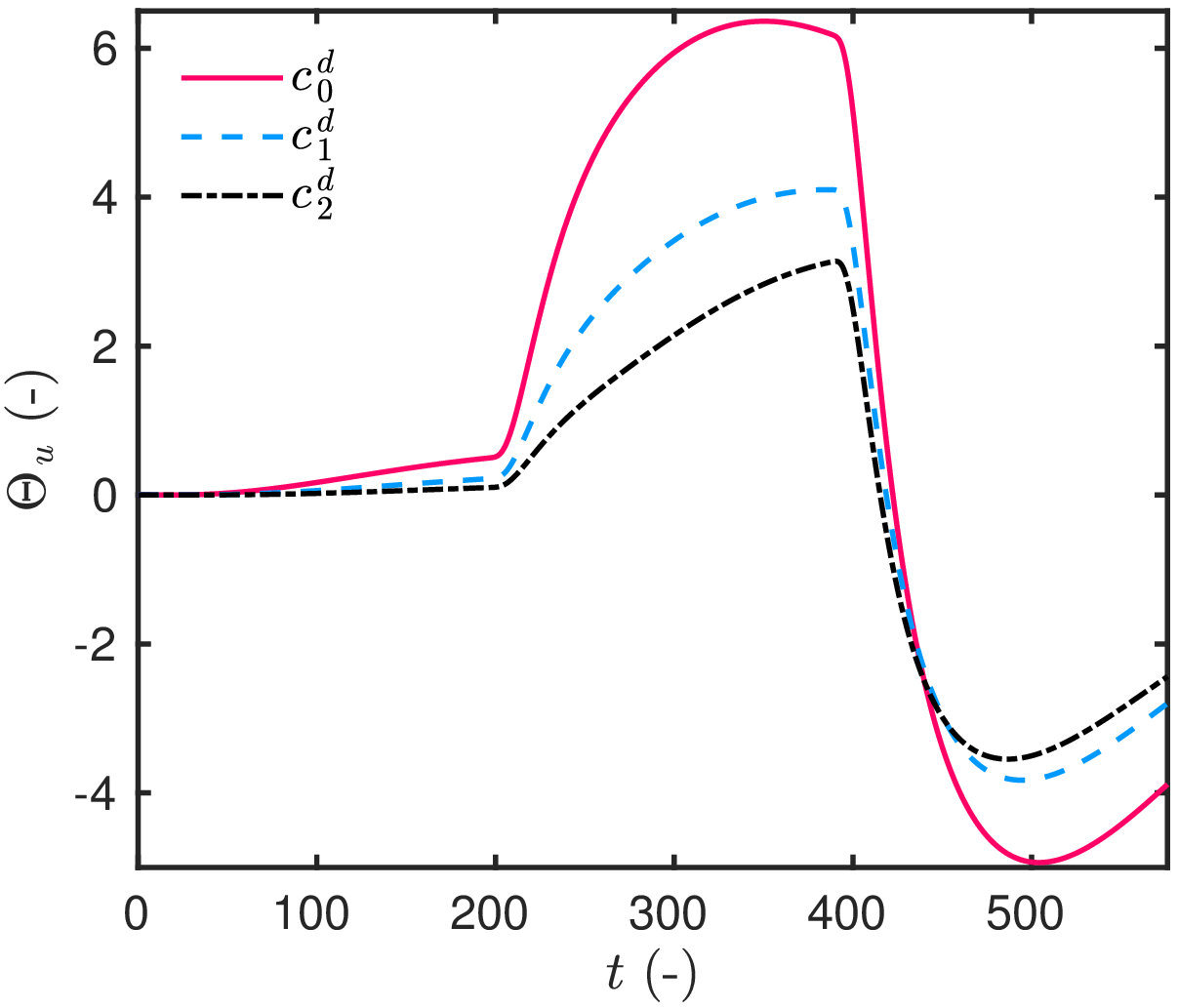}}
\subfigure[\label{fig:Mstep_wH_Ya_OED}]{\includegraphics[width=.48\textwidth]{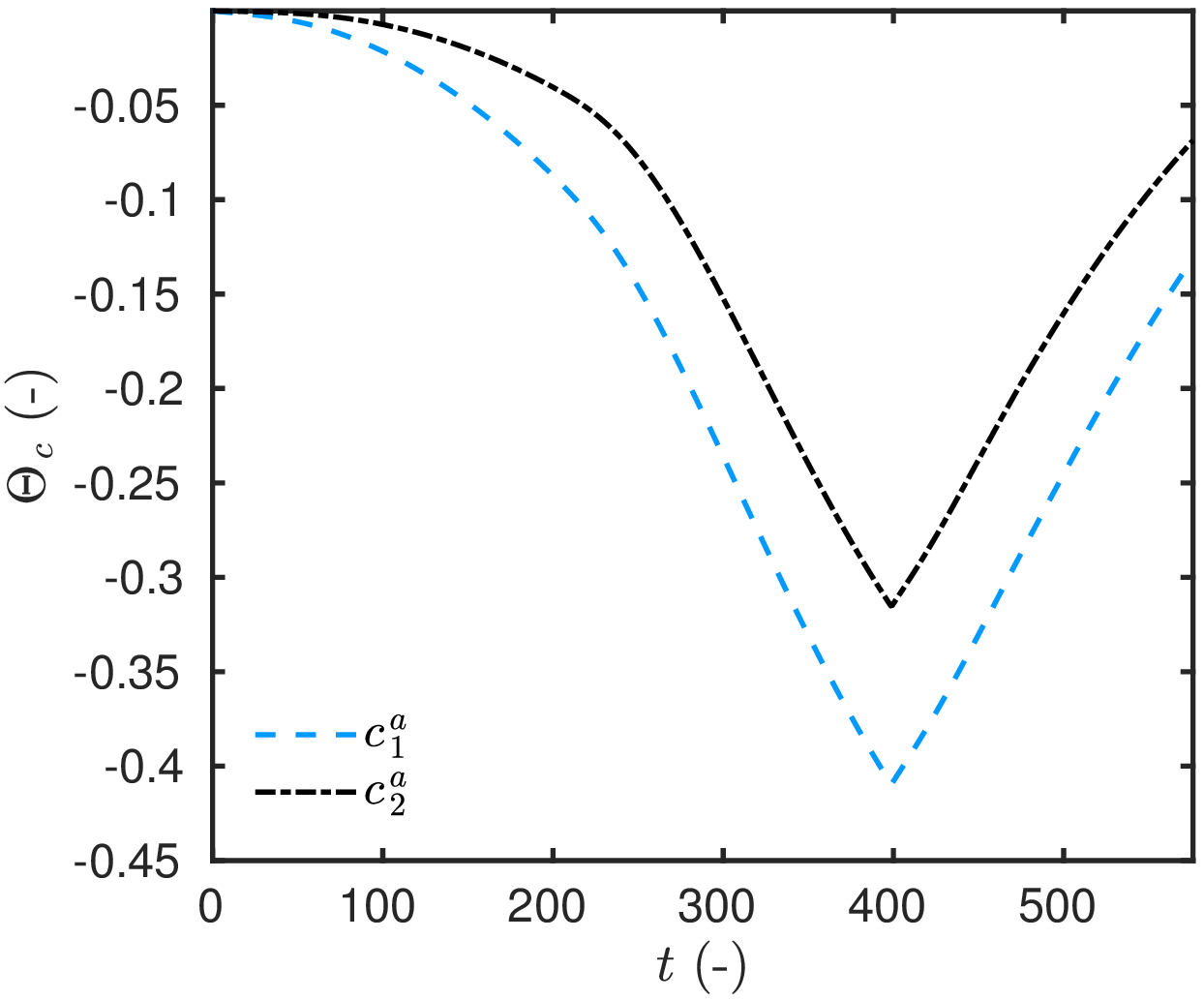}}
\subfigure[\label{fig:Mstep_wH_Yd_OED}]{\includegraphics[width=.48\textwidth]{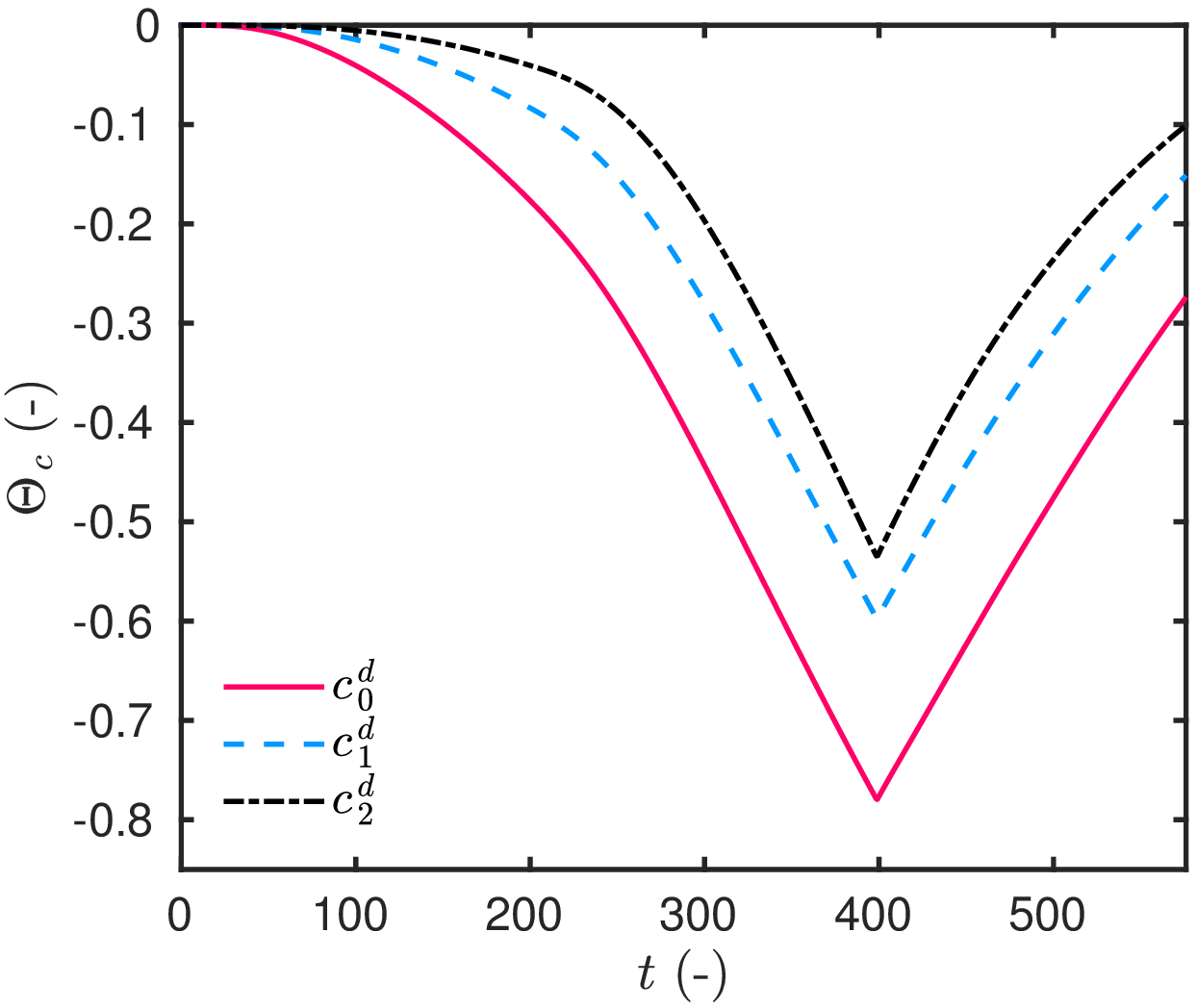}} 
\caption{\small\em Sensitivity coefficients $\Theta_{\,u}$ (a,b)  and $\Theta_{\,c}$ (c,d) for the OED (design $12\,$, $X \egal X^{\,\circ} \egal 1$).}
\label{fig:Mstep_wH_OED}
\end{center}
\end{figure}

\begin{figure}
\begin{center}
\subfigure[\label{fig:Mstep_wH_Xa_aOED}]{\includegraphics[width=.48\textwidth]{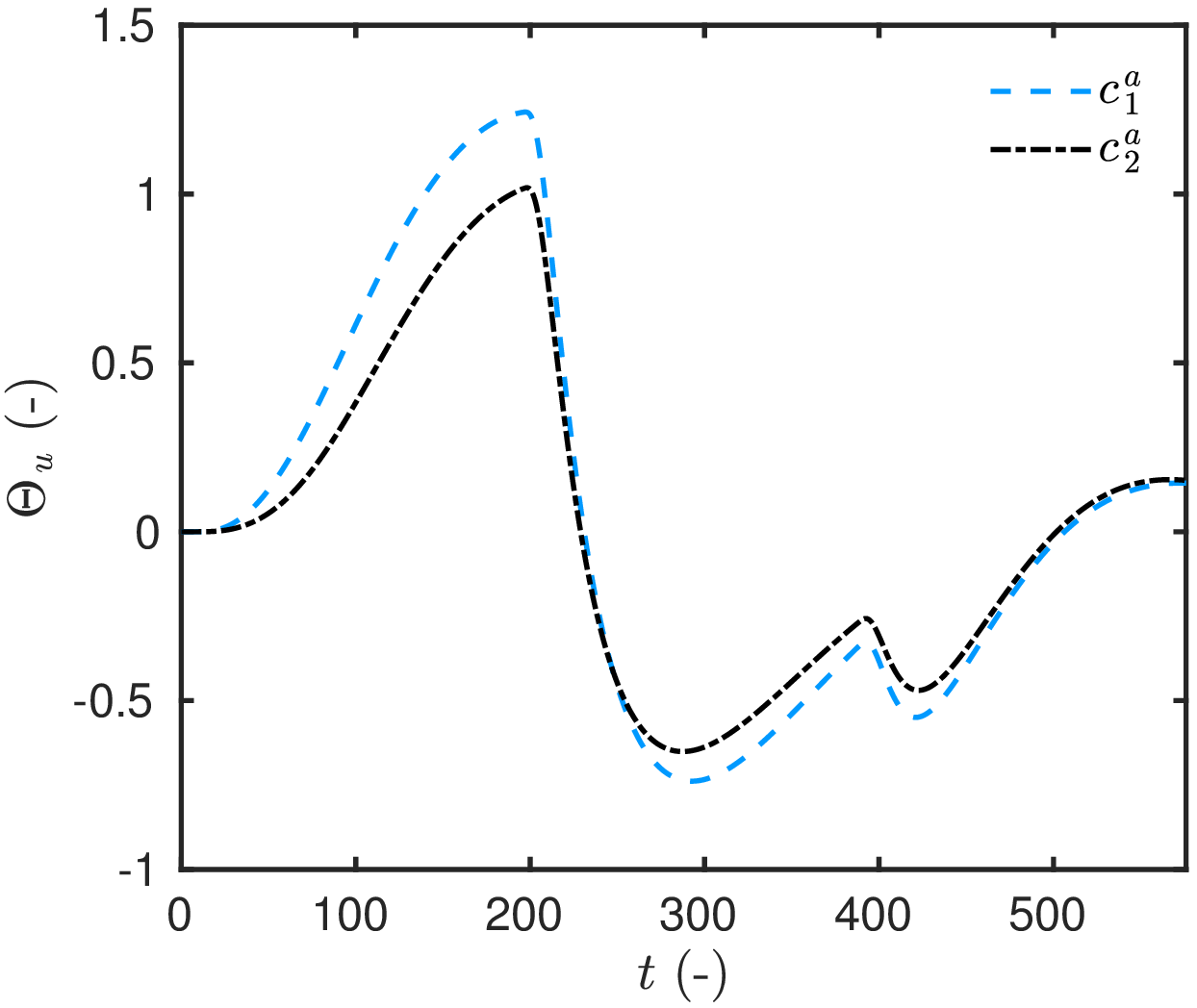}}
\subfigure[\label{fig:Mstep_wH_Xd_aOED}]{\includegraphics[width=.48\textwidth]{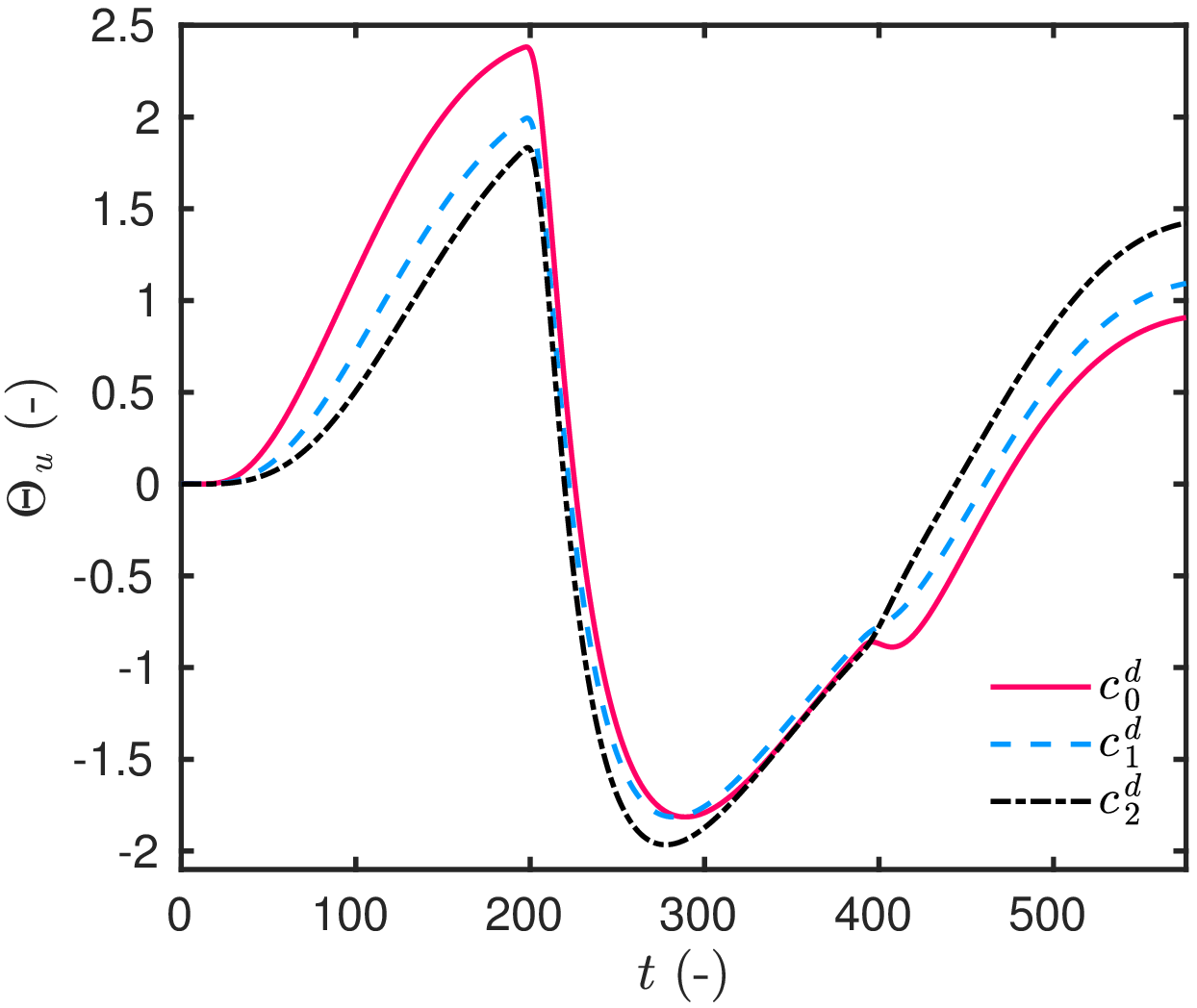}}
\subfigure[\label{fig:Mstep_wH_Ya_aOED}]{\includegraphics[width=.48\textwidth]{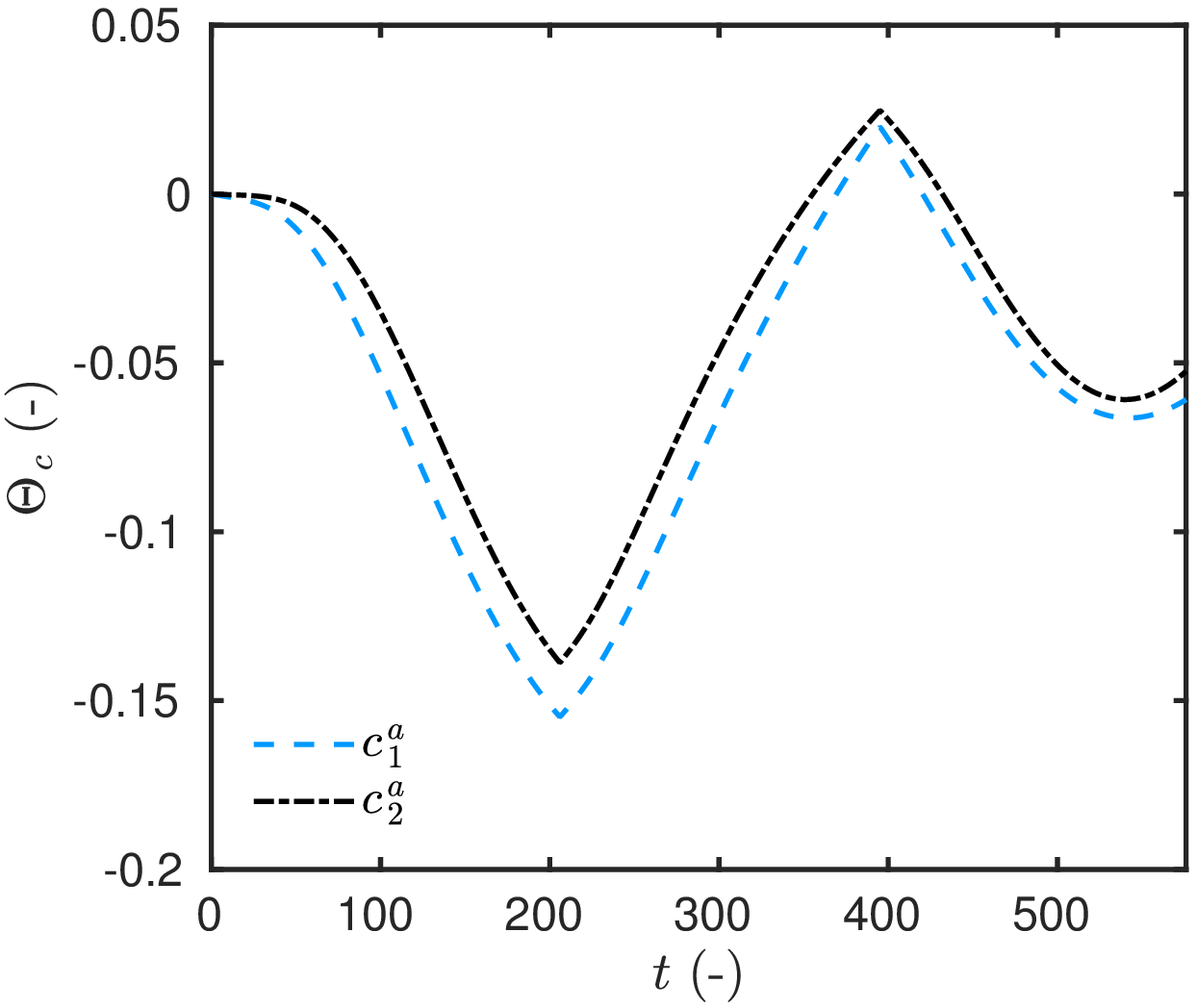}}
\subfigure[\label{fig:Mstep_wH_Yd_aOED}]{\includegraphics[width=.48\textwidth]{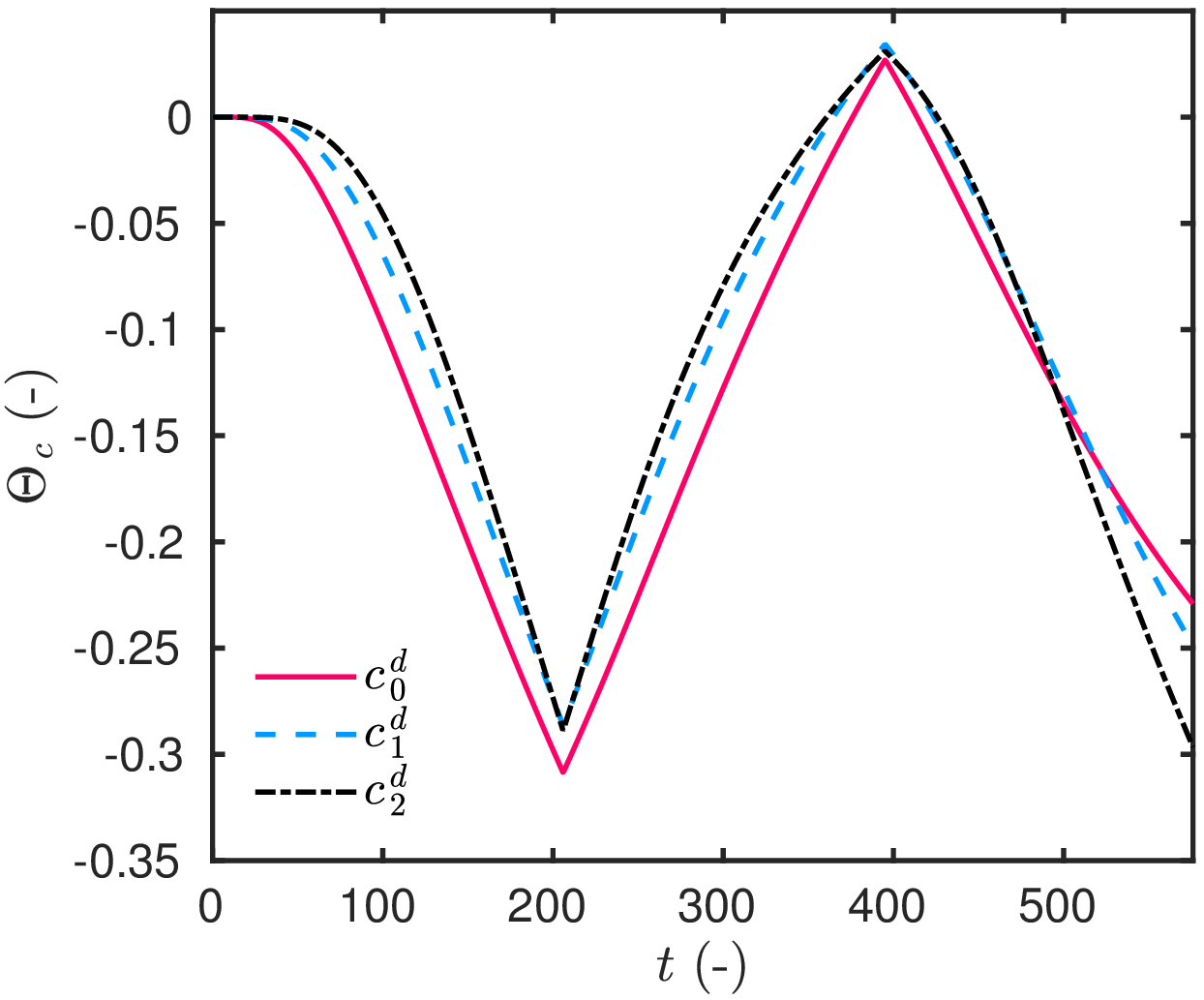}} 
\caption{\small\em Sensitivity coefficients $\Theta_{\,u}$ (a,b) and $\Theta_{\,c}$ (c,d) for the design $20\,$, ($X \egal X^{\,\circ} \egal 1$).}
\label{fig:Mstep_wH_anti_OED}
\end{center}
\end{figure}

First, the OED is searched using the adsorption coefficients estimated in Section~\ref{sec:parameter_estimation_problem} and the desorption  coefficients obtained from literature \cite{Rafidiarison2015}:
\begin{align*}
& c^{\,\star}_{\,a,\,1} \egal -0.99 \,,
&& c^{\,\star}_{\,a,\,2} \egal 1.003 \,, \
&& c^{\,\star}_{\,d,\,0} \egal 0.75 \,,
&& c^{\,\star}_{\,d,\,1} \egal -1.24 \,,
&& c^{\,\star}_{\,d,\,2} \egal 0.89 \,.
\end{align*}
The \textsc{Fourier} number corresponds to the estimated one in Section~\ref{sec:parameter_estimation_problem} and equals $\Fo \egal 6.1 \times 10^{\,-3} \,$.
The investigations are performed for both the single and multiple steps of relative humidity, identified in Table~\ref{tab:possible_designs}. For the sake of compactness, the results are illustrated only for the multiple steps experiments. By improving the model with hysteresis effects, one has to compute the sensitivity coefficients for both equations~\eqref{eq:moisture_model_wH} and \eqref{eq:hysteresis_model}. The sensitivity coefficients $\Theta_{\,c}$ and $\Theta_{\,u}$ are  shown in Figures~\ref{fig:Mstep_wH_Xa_OED} and \ref{fig:Mstep_wH_Ya_OED} for the adsorption coefficients $\,c^{\,\star}_{\,a,\,1}\,$ and $c^{\,\star}_{\,a,\,2}\,$. The variation of the sensitivity coefficients follows the changes in the boundary conditions $u_{\,\infty}\,$. The design~$12$ corresponds to the OED for each parameter to be estimated, as shown in Figure~\ref{fig:Mstep_wH_Psi_fdesign}. On the contrary to the results obtained in Section~\ref{sec:OED}, the design $20$ does not allow to estimate the unknown parameters with accuracy. The influence of including the hysteresis effects in the mathematical model can be remarked in the determination of the OED. Indeed, by comparing Figures~\ref{fig:Mstep_wH_OED} and \ref{fig:Mstep_wH_anti_OED}, it can be noticed that the sensitivity coefficients $\Theta_{\,c}$ and $\Theta_{\,u}$ have larger magnitudes for the OED than for the design $20\,$. For multiple-step experiments, a strong correlation is observed between the five coefficients $\bigl(\,c^{\,\star}_{\,a,\,1}\,,\,c^{\,\star}_{\,a,\,2}\,,\,c^{\,\star}_{\,d,\,0}\,,\,c^{\,\star}_{\,d,\,1}\,,\,c^{\,\star}_{\,d,\,2}\,\bigr)\,$:
\begin{align*}
\begin{array}{c|ccccc}
& c^{\,\star}_{\,a,\,1}
& c^{\,\star}_{\,a,\,2}
& c^{\,\star}_{\,d,\,0} 
& c^{\,\star}_{\,d,\,1}
& c^{\,\star}_{\,d,\,2} \\[3pt] \hline 
c^{\,\star}_{\,a,\,1}
& 1
& 0.99669
& 0.99974
& 0.99473
& 0.97735
\\
c^{\,\star}_{\,a,\,2}
& 
& 1
& 0.99663
& 0.999
& 0.99
\\
c^{\,\star}_{\,d,\,0} 
& 
& 
& 1
& 0.99567
& 0.97941
\\
c^{\,\star}_{\,d,\,1}
& 
& 
& 
& 1
& 0.99388
\\
c^{\,\star}_{\,d,\,2}
&
&
&
&
& 1
\\
\end{array}
\end{align*}
indicating that it is not possible to estimate more than two parameters of the model for such experiments. For single case experiments, the correlation is lower:
\begin{align*}
\begin{array}{c|ccccc}
& c^{\,\star}_{\,a,\,1}
& c^{\,\star}_{\,a,\,2}
& c^{\,\star}_{\,d,\,0} 
& c^{\,\star}_{\,d,\,1}
& c^{\,\star}_{\,d,\,2} \\[3pt] \hline 
c^{\,\star}_{\,a,\,1}
& 1
& 0.98981
& 0.97193
& 0.89302
& 0.78741
\\
c^{\,\star}_{\,a,\,2}
& 
& 1
& 0.98349
& 0.93466
& 0.85096
\\
c^{\,\star}_{\,d,\,0} 
& 
& 
& 1
& 0.9713
& 0.90551
\\
c^{\,\star}_{\,d,\,1}
& 
& 
& 
& 1
& 0.98031
\\
c^{\,\star}_{\,d,\,2}
&
&
&
&
& 1
\\
\end{array}
\end{align*}
Thus, for this case, it is possible to estimate a couple of two parameters among $\bigl(\,c^{\,\star}_{\,a,\,1}\,,\,c^{\,\star}_{\,d,\,2}\,\bigr)\,$ or $\bigl(\,c^{\,\star}_{\,a,\,2}\,,\,c^{\,\star}_{\,d,\,2}\,\bigr)\,$. The OED results are synthesized in Table~\ref{tab:synthesis_OED_wH}. For all experiments, the sensor should be located at the bottom of the material, near the impermeable face. The designs~$2$ ($\ui \egal 0.2$ to $\uc \egal 1.5 \,$) and $12$ ($\ui \egal 0.2\,$, $\uc^{\,1} \egal 0.66\,$, $\uc^{\,2} \egal 1.5\,$,  $\uc^{\,3} \egal 0.66$ and a duration of each step $\tau \egal 8 \,$) provide the highest accuracy to estimate the parameters. Due to high correlation between the coefficients, it is important to note that it is not possible to estimate all five coefficients considering the possible designs described in Table~\ref{tab:possible_designs}. If one aims at estimating the five coefficients, additional designs have to be planned with the experimental facility, including for instance other levels of relative humidity.

\begin{table}
\centering
\setlength{\extrarowheight}{.3em}
\caption{\small\em Synthesis of OED for the estimation of one or several parameters when the model account for hysteresis.}
\bigskip
\begin{tabular}[l]{@{} cc|ccc}
\hline
\hline
 & \multirow{2}{*}{\textit{Parameter(s) to be estimated}} 
 & \multicolumn{3}{c}{\textit{Optimal Experimental Design}} \\
 & & Single-step & Multiple-step & Sensor position \\
 \hline
\textit{One parameter}
& $c^{\,\star}_{\,a,\,1}\,$, $c^{\,\star}_{\,a,\,2}\,$, $c^{\,\star}_{\,d,\,0}\,$, $c^{\,\star}_{\,d,\,1}\,$ or $c^{\,\star}_{\,d,\,2}\,$
& design $2$
& design $12$
& $X^{\,\circ} \, \in \, \bigl[\,0.85 \,,\, 1 \,\bigr]$ \\
\hline
\textit{Multiple parameter}
& $\bigl(\,c^{\,\star}_{\,a,\,1}\,,\,c^{\,\star}_{\,d,\,2}\,\bigr)\,$ or 
  $\bigl(\,c^{\,\star}_{\,a,\,2}\,,\,c^{\,\star}_{\,d,\,2}\,\bigr)\,$
& design $2$
& -
& $X^{\,\circ} \, \in \, \bigl[\,0.85 \,,\, 1 \,\bigr]$ \\
\hline
\hline
\end{tabular}
\bigskip
\label{tab:synthesis_OED_wH}
\end{table}

\subsection{Comparing the numerical results with the experimental observations}

Previous section aimed at illustrating the possibility of searching the OED with an improved model that includes the hysteresis effects. Since it is rather difficult to estimate the five coefficients due to strong correlations of the sensitivity functions, the purpose is now to show the influence of taking into account the hysteresis effects in the model. It should be noted that the regularized model is not needed so that the normal hysteresis model given by Equation~\eqref{eq:hysteresis_model} is used for the present case study. To avoid stability restrictions, an implicit--explicit numerical scheme, detailed in Appendix~\ref{sec:Annex2}, is employed to compute numerically the solution.

\begin{figure}
  \begin{center}
  \subfigure[\label{fig:EST_wH_v_ft}]{\includegraphics[width=.48\textwidth]{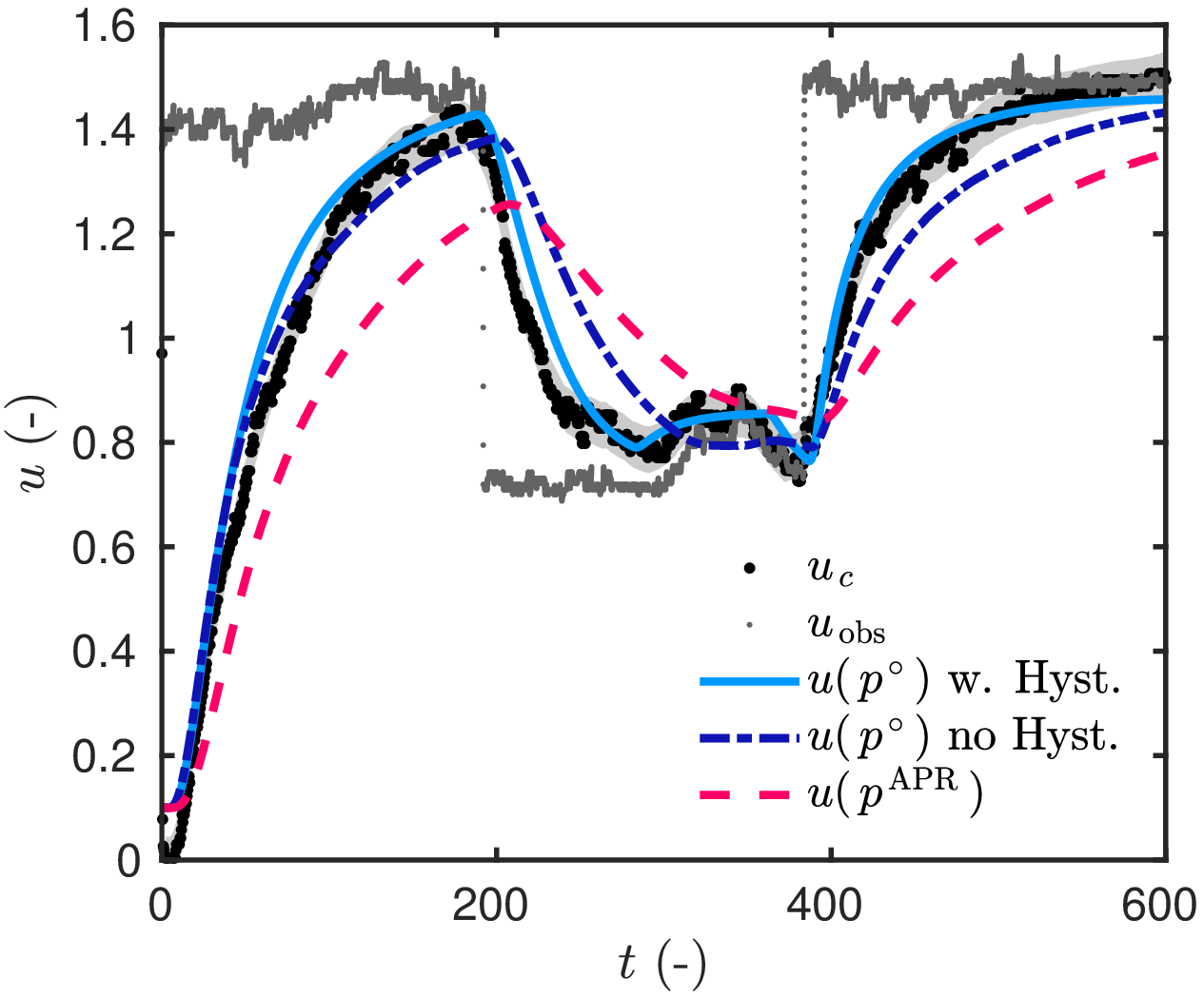}}
  \subfigure[\label{fig:EST_wH_res_ft}]{\includegraphics[width=.48\textwidth]{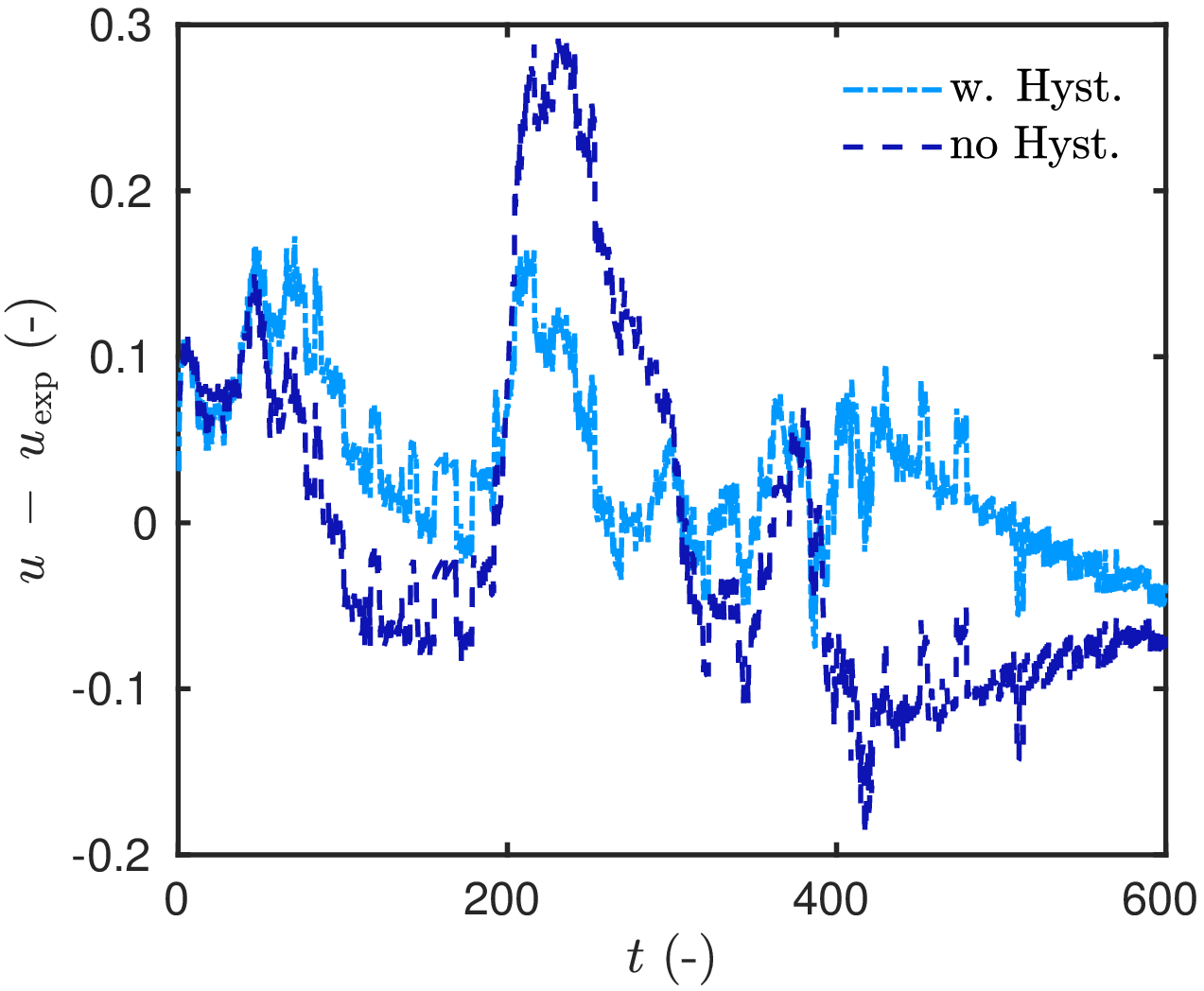}}
  \caption{\small\em Comparison of the numerical results with the experimental data (a) and their residual (b).}
  \label{fig:EST_wH_v}
  \end{center}
\end{figure}

\begin{figure}
  \begin{center}
  \subfigure[\label{fig:EST_wH_c_ft}]{\includegraphics[width=.48\textwidth]{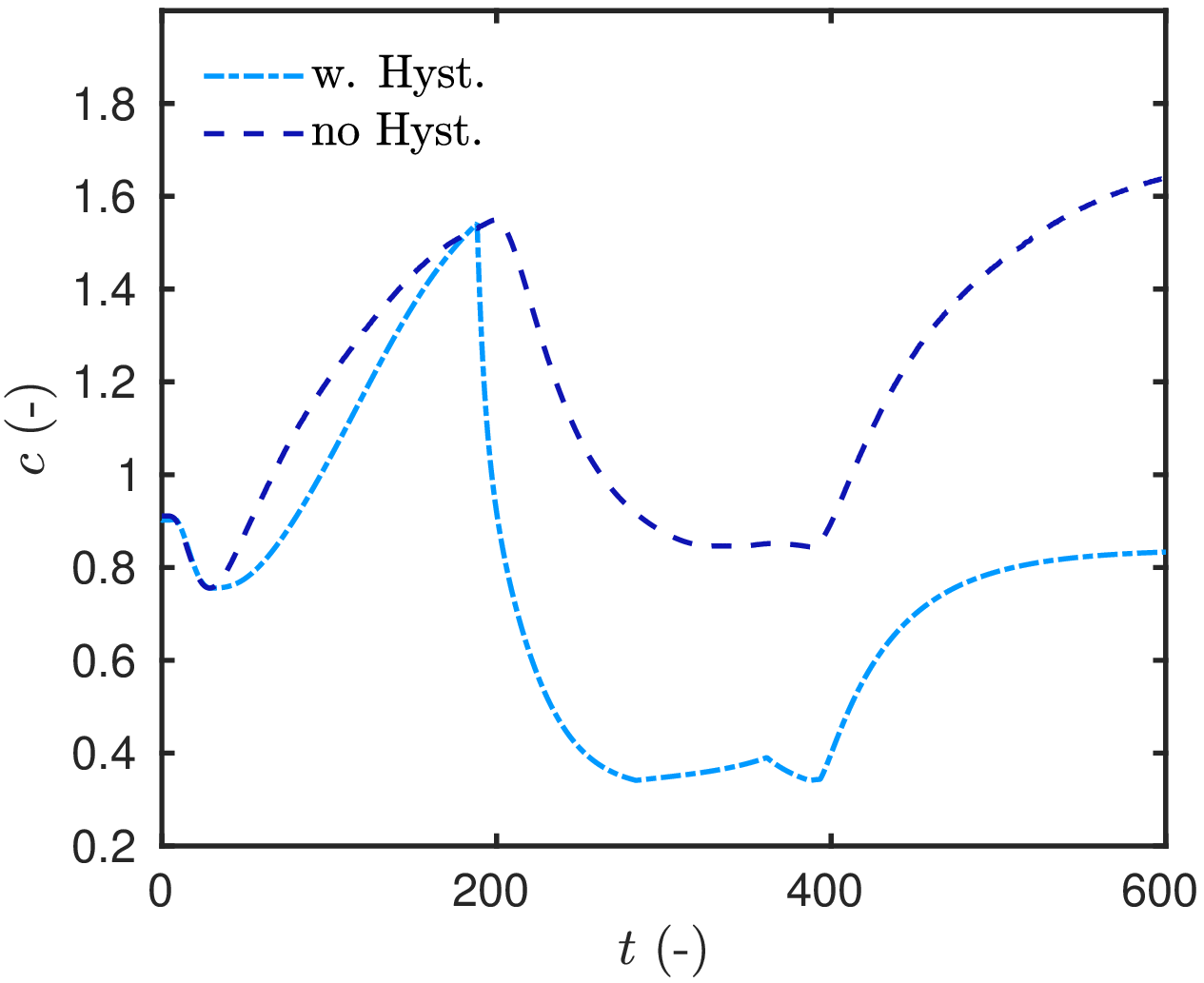}} \hspace{0.3cm}
  \subfigure[\label{fig:EST_wH_c_fu}]{\includegraphics[width=.48\textwidth]{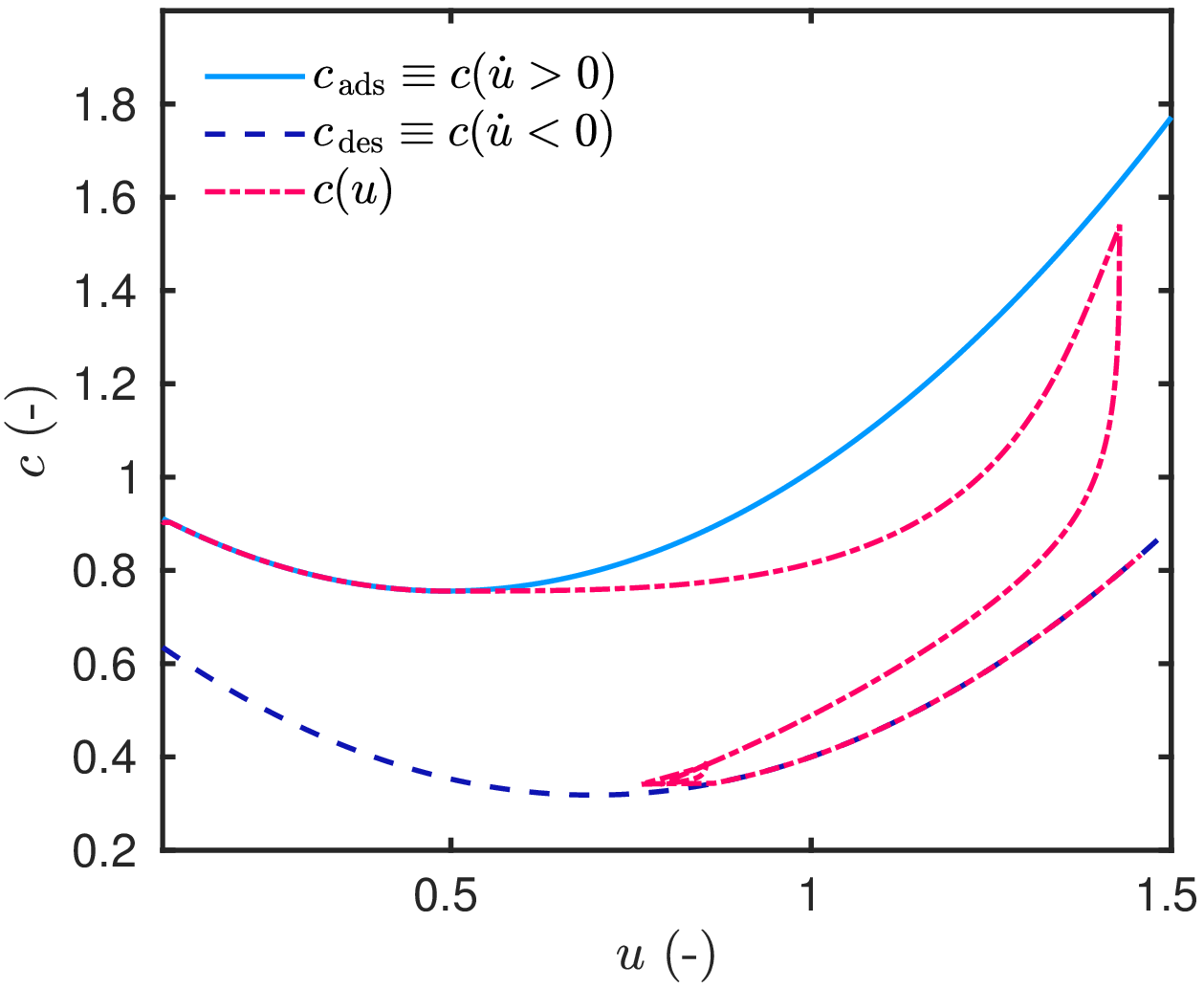}}
  \caption{\small\em Variation of the sorption coefficient $c$ according to time (a) and according to the computed field $u$ (b).}
  \label{fig:EST_wH_c}
  \end{center}
\end{figure}

The numerical predictions are compared with the experimental observations for the multiple-step experiment (design $20$). The adsorption coefficients, estimated in Section~\ref{sec:parameter_estimation_problem}, are used together with the desorption  coefficients obtained from literature \cite{Rafidiarison2015}. The \textsc{Fourier} number equals $\Fo \egal 6.1 \times 10^{\,-3}$ and the other parameters have the same numerical value as the ones mentioned in Section~\ref{sec:parameter_estimation_problem}. Figure~\ref{fig:EST_wH_v_ft} shows the comparison between the numerical predictions and the experimental data. The results from the physical model including the hysteresis effects are more satisfactory. Indeed, the residual is lower for this model than for the model without hysteresis, as shown in Figure~\ref{fig:EST_wH_res_ft}. Particularly, the importance of the hysteresis can be noted for the desorption phase $t\ \in \ \bigl[\,200 \,,\, 300 \,\bigr]\,$ and the second adsorption phase $t\ \in \ \bigl[\,400 \,,\, 600 \,\bigr]\,$.
The solution of Equation~\eqref{eq:hysteresis_model} enables to compute the time evolution of the sorption coefficient $c$ as illustrated in Figure~\ref{fig:EST_wH_c_ft}. It is compared with the function $(\, c \circ u\,)\,(\,t\,)$ where $u$ is computed using the model without hysteresis given by Eq.~\eqref{eq:moisture_dimensionlesspb_1D}. The time variation of the coefficients are similar for the first adsorption part corresponding to $t\ \in \ \bigl[\,0 \,,\, 200 \,\bigr]\,$. Indeed, for this period, the computed sorption coefficient equals the adsorption curve as noticed in Figure~\ref{fig:EST_wH_c_fu}. After this period, the coefficient computed with the hysteresis model decreases comparing to the one without hysteresis and oscillates between the adsorption and desorption curves as shown in Figure~\ref{fig:EST_wH_c_fu}.

\section{Final remarks}

\subsection{Conclusion}

In the context of building physics, inverse problems are encountered to estimate moisture dependent hygrothermal properties of porous materials, using measurements associated to heat and moisture transport. Two applications are distinguished. In the first case, concerning the diagnosis of existing building walls, there is a few \emph{a priori} estimation of material properties. Moreover, measurements must be non-intrusive and non-destructive. In the second case, measurements are performed in the laboratory to calibrate the numerical model with the experimental data. This article is encompassed in these conditions, focused on the estimation of moisture sorption isotherm coefficients of a wood fiber material.

First, the OED methodology has been described and used for searching the optimal experiment design, ensuring to provide the best accuracy of the identification method for the parameter estimation. The approach is based on the sensitivity functions of the unknown parameters, enabling to determine sensor location within the material and boundary conditions, according to an existing facility among $20$ possible designs. It has been carried out considering \emph{a priori} values of the unknown parameters. The facility allows to submit material to relative humidity steps on one surface, being all others moisture impermeable. Results have enhanced two designs: i) single step of relative humidity from $10 \unit{\%}$ to $75 \unit{\%}$ and ii) multiple steps of relative humidity  $10-75-33-75 \unit{\%}$, with a duration period of $8$ days for each step. For each design, the sensor has to be placed as close as possible to the impermeable boundary.

Then, experimental data has been provided according to the OED results for the two selected designs. The parameter estimation has been conducted by minimizing a cost function between the experimental data and the numerical results. The estimation has been accomplished using an interior point algorithm. Nine tests have been performed for the definition of the cost function $\mathrm{J} \,$. The $\mathcal{L}_{\,2}$ and $\mathcal{L}_{\,\infty}$ have been evaluated. Two series of tests aimed at estimating the three parameters using both experiments at the same time or separately with an iterative algorithm. The third series intended to estimate all five parameters of the material properties. Results have shown that the $\mathcal{L}_{\,2}$ norm provided better results of the parameter estimation problem. Moreover, it was better to consider both experiments separately to estimate only three parameters of the problem. Within this approach, the algorithm requires only two iterations to compute the solution with less than $100$ computations of the direct model. This approach has a really low computational cost compared to stochastic approaches, needing an order of $10^{\,4}$ computations for similar problems. Another advantage of this approach is to use the sensitivity functions, computed during the search of the OED, to provide an approximation of the probability distribution function of the estimated parameters at a lower computational cost.

As highlighted in Section~\ref{sec:parameter_estimation_problem}, with the estimated parameters, a better agreement between the numerical model and the experimental data is observed. However, the importance of hysteresis effects were highlighted. Particularly, when cycles of desorption-adsorption processes take place, some discrepancies remain between experimental data and numerical predictions. Therefore, a new mathematical model has been proposed to take into account the hysteresis effects on the sorption coefficients. A second differential equation has been added and increased to five the number of coefficients to be estimated. Two coefficients correspond to the adsorption curve and three to the desorption one. A regularized version of the hysteresis model was proposed to have continuous differentiable functions and, therefore, to be able to compute the sensitivity coefficients. Then, the OED was explored by computing the sensitivity coefficients of the five parameters of a family of scanning curves of adsorption and desorption processes. This clearly highlighted the possibility of including the hysteresis effects in the OED approach. The results draw attention to two designs: i) a single step of relative humidity from $10 \unit{\%}$ to $75 \unit{\%}$ and ii) multiple steps of relative humidity  $10-33-75-33 \unit{\%}$, with a $8$-day time period for each step. The sensors have to be placed near the impermeable boundary.

Finally, the numerical predictions, considering the hysteresis phenomenon, have been compared with the experimental observations of a multiple-step design. An efficient implicit--explicit numerical scheme was proposed to compute the solution of the hysteretic model. The parameters of this model correspond to the estimated ones in Section~\ref{sec:parameter_estimation_problem} for the adsorption curve and to the \emph{a priori} ones provided in the literature. The comparison has shown that the discrepancies are reduced, fitting better experimental data. During the simulation, the computed sorption coefficients oscillated between the ones from the main adsorption and desorption curves.


\subsection{Outlooks and open-problems}

An interesting perspective of improvement concerns the assumptions related to the moisture sorption isotherm coefficients $c\,(\,u\,)$. A parametrization was previously defined $c\,(\,u\,) \egal 1 \plus c_{\,1} \, u \plus c_{\,2} \, u^{\,2} $ and the parameter estimation problem aimed at identifying coefficients $c_{\,1} \,$, $c_{\,2}$ (and $\Fo$). An ambitious outlook could aim at estimating directly the function $c\,(\,u\,)$, with inspiration from the following studies \cite{Koptyug2000, Kabanikhin2008a}.


\bigskip

\subsection*{Acknowledgments}
\addcontentsline{toc}{subsection}{Acknowledgments}

This work was partly funded by the French Environment and Energy Management Agency (ADEME), the ''Assembl\'ee des Pays de Savoie`` (APS) and the French National Research Agency (ANR) through its Sustainable Cities and Buildings program (MOBAIR project ANR-12-VBDU-0009). The authors acknowledge the Junior Chair Research program ``Building performance assessment, evaluation and enhancement'' from the University of Savoie Mont Blanc in collaboration with The French Atomic and Alternative Energy Center (CEA) and Scientific and Technical Center for Buildings (CSTB) and the support  of CNRS/INSIS (Cellule \'energie) under the program ``Projets Exploratoires --- 2017''. The authors also acknowledge the Brazilian Agency CNPQ of the Ministry of Science, Technology and Innovation, for the financial support.


\section*{Nomenclature}

\begin{tabular*}{0.7\textwidth}{@{\extracolsep{\fill}} |@{} >{\scriptsize \quad} c >{\scriptsize} l >{\scriptsize} l| }
\hline
\multicolumn{3}{|c|}{\emph{Latin letters}} \\
$a$ & moisture advection coefficient & $[\mathsf{s/m}]$ \\
$a_{\,m}$ & mass transfer coefficient & $[\mathsf{m^{\,2}/s}]$ \\
$c$ & moisture storage capacity & $[\mathsf{kg/(m^3.Pa)}]$ \\
$c_{\,b}$ & specific heat & $[\unitfrac{J}{(kg.K)}]$ \\
$d$ & moisture permeability & $[\mathsf{s}]$ \\
$h$ & convective vapour transfer coefficient & $[\mathsf{s/m}]$ \\
$L$ & length & $[\mathsf{m}]$ \\
$\Ps$ & saturation pressure & $[\mathsf{Pa}]$ \\
$\Pv$ & vapor pressure & $[\mathsf{Pa}]$ \\
$r_{\,12}$ & latent heat of evaporation & $[\mathsf{J/kg}]$ \\
$\Rv$ & water gas constant & $[\mathsf{J/(kg.K)}]$\\
$T$ & temperature & $[\mathsf{K}]$ \\
$t$ & time coordinate & $[\mathsf{s}]$ \\
$U$ & relative moisture concentration &  $[\mathsf{\%}]$ \\
$x$ & space coordinate & $[\mathsf{m}]$ \\
$ \mathsf{v}$ & mass average velocity & $[\mathsf{m/s}]$ \\
$w$ & specific moisture content & $[\mathsf{kg/m^3}]$\\
\hline
\end{tabular*}

\bigskip

\begin{tabular*}{0.7\textwidth}{@{\extracolsep{\fill}} |@{} >{\scriptsize \quad} c >{\scriptsize} l >{\scriptsize} l| }
\hline
\multicolumn{3}{|c|}{\emph{Greek letters}} \\
$\lambda$ & thermal conductivity &  $[\mathsf{W/(m.K)}]$ \\
$\delta$ & thermal gradient coefficient &  $[\mathsf{K^{\,-1}}]$ \\
$\phi$ & relative humidity & $[-]$ \\
$\rho$ & specific mass & $[\mathsf{kg/m^3}]$ \\
\hline
\end{tabular*}


\newpage
\appendix

\section{Equations for the computation of the sensitivity coefficients}
\label{sec:Eq_sensitivity_coeff}

This section provides the equations derived analytically from the mathematical model (Eq.~\eqref{eq:moisture_dimensionlesspb_1D}), to compute the sensitivity coefficients.


\subsection{Model without hysteresis}

The sensitivity coefficients are denoted as follows:
\begin{align*}
  & \Theta_{\,0} \ \eqdef \ \pd{u}{\Fo} \,,
  && \Theta_{\,1} \ \eqdef \ \pd{u}{c_{\,1}} \,,
  && \Theta_{\,2} \ \eqdef \ \pd{u}{c_{\,2}} \,.
\end{align*}
For the sake of clarity, the superscript $^{\,\star}$ is omitted. The sensitivity coefficients verified the following equations. For $\Theta_{\,0}\,$:
\begin{align*}
  c\,(\,u\,)  \ \pd{\Theta_{\,0}}{t} \egal & \Fo \ d\,(\,u\,) \ \pd{^{\,2} \Theta_{\,0}}{x^{\,2}} \plus \Fo \, \Bigl(\, 2 \,  d_{\,1} \ \pd{u}{x}  \moins \Pe \,\Bigr) \ \pd{\Theta_{\,0}}{x} \plus \Fo \, d_{\,1} \ \Theta_{\,0} \ \pd{^{\,2}u}{x^{\,2}} \\[4pt]
  & \plus d_{\,1} \, \biggl(\, \pd{u}{x} \,\biggr)^{\,2} \plus d\,(\,u\,) \ \pd{^{\,2} u}{x^{\,2}} \moins \Pe \, \pd{u}{x} \moins \Bigl(\, c_{\,1} \, \Theta_{\,0} \plus 2 \, c_{\,2} \, u \, \Theta_{\,0} \,\Bigr) \ \pd{u}{t} \,.
\end{align*}
For $\Theta_{\,1}\,$:
\begin{align*}
  c\,(\,u\,)  \ \pd{\Theta_{\,1}}{t} \egal & \Fo \ d(\,u\,) \ \pd{^{\,2} \Theta_{\,1}}{x^{\,2}} \plus 2 \, \Fo \, d_{\,1} \ \pd{\Theta_{\,1}}{x} \ \pd{u}{x} \plus \Fo \, d_{\,1} \ \Theta_{\,1} \ \pd{^{\,2}u}{x^{\,2}} \\[4pt]
  & \moins \Fo \, \Pe \, \pd{\Theta_{\,1}}{x} \moins \Bigl(\, u \plus c_{\,1} \, \Theta_{\,1} \plus 2 \, c_{\,2} \, u \, \Theta_{\,1} \,\Bigr) \ \pd{u}{t} \,.
\end{align*}
and for $\Theta_{\,2}\,$:
\begin{align*}
  c\,(\,u\,)  \ \pd{\Theta_{\,2}}{t} \egal & \Fo \ d\,(\,u\,) \ \pd{^{\,2} \Theta_{\,2}}{x^{\,2}} \plus 2 \, \Fo \, d_{\,1} \ \pd{\Theta_{\,2}}{x} \ \pd{u}{x} \plus \Fo \, d_{\,1} \ \Theta_{\,2} \ \pd{^{\,2}u}{x^{\,2}} \\[4pt]
  & \moins \Fo \, \Pe \, \pd{\Theta_{\,2}}{x} \moins \Bigl(\, c_{\,1} \, \Theta_{\,2} \plus 2 \, c_{\,2} \, u\, \Theta_{\,2} \plus u^{\,2} \,\Bigr) \ \pd{u}{t} \,,
\end{align*}


\subsection{Model with hysteresis}

The model with hysteresis includes two differential equations, recalled here:
\begin{align*}
  c \ \pd{u}{t} &\egal \Fo \ \pd{}{x} \Biggl(\, d(\,u\,)  \ \pd{u}{x} \moins \Pe \ u \, \Biggr) \,, \\[3pt]
  \pd{c}{t} & \egal \beta \ \mathrm{Rsign} \ \biggl(\, \pd{u}{t} \,\biggr) \cdot \biggl(\, c \moins c_{\,\mathrm{ads}}\,(\,u\,) \,\biggr) \cdot \biggl(\, c \moins c_{\,\mathrm{des}}\,(\,u\,) \,\biggr) \,. 
\end{align*}
Therefore, two sensitivity coefficients have to be computed. The differential equations for $\Theta_{\,u,\,a,\,1} \egal \displaystyle \pd{u}{c_{\,a,\,1\,}}$ and $\Theta_{\,c,\,a,\,1} \egal \displaystyle \pd{c}{c_{\,a,\,1\,}}$ are:
\begin{align*}
  \pd{\Theta_{\,u,\,a,\,1}}{t} \egal & \Fo \, d(\,u\,) \, \pd{^{\,2} \Theta_{\,u,\,a,\,1}}{x^{\,2}} \plus \Fo \,  \biggl(\, 2 \, d_{\,1} \, \pd{u}{x} \moins \Pe \,\biggr) \, \pd{\Theta_{\,u,\,a,\,1}}{x} \plus \Fo \, d_{\,1} \, \pd{^{\,2} u}{x^{\,2}}  \, \Theta_{\,u,\,\,a,\,1}  \\[3pt]
  & \moins \Theta_{\,c,\,a,\,1} \, \pd{u}{t} \,, \\[3pt]
  \pd{\Theta_{\,c,\,a,\,1}}{t} \egal & \beta \ \mathrm{Rsign}^{\,\prime} \ \biggl(\, \pd{u}{t} \,\biggr) \pd{\Theta_{\,u,\,a,\,1}}{t} \,\, \biggl(\, c \moins c_{\,\mathrm{ads}} \,\biggr) \cdot \biggl(\, c \moins c_{\,\mathrm{des}} \,\biggr) \\[3pt]
  & \plus \beta \ \mathrm{Rsign} \ \biggl(\, \pd{u}{t} \,\biggr) \cdot \biggl(\, \Theta_{\,c,\,a,\,1} \moins \bigl(\, c_{\,1,\,a} \, \Theta_{\,u,\,a,\,1} \plus u \plus 2 \ c_{\,2,\,a} \, u \, \Theta_{\,u,\,a,\,1} \, \bigr)\, \biggr) \cdot \biggl(\, c \moins c_{\,\mathrm{des}} \,\biggr) \\[3pt]
  & \plus \beta \ \mathrm{Rsign} \ \biggl(\, \pd{u}{t} \,\biggr) \cdot \biggl(\, c \moins c_{\,\mathrm{ads}} \,\biggr) \cdot \biggl(\, \Theta_{\,c,\,a,\,1} \moins \bigl(\, c_{\,1,\,d} \, \Theta_{\,u,\,a,\,1} \plus 2 \ c_{\,2,\,d} \, u \, \Theta_{\,u,\,a,\,1} \, \bigr)\, \biggr) \,.
\end{align*}
For $\Theta_{\,u,\,a,\,2} \egal \displaystyle \pd{u}{c_{\,a,\,2\,}}$ and $\Theta_{\,c,\,a,\,2} \egal \displaystyle \pd{c}{c_{\,a,\,2\,}}\,$:
\begin{align*}
  \pd{\Theta_{\,u,\,a,\,2}}{t} \egal & \Fo \, d(\,u\,) \, \pd{^{\,2} \Theta_{\,u,\,a,\,2}}{x^{\,2}} \plus \Fo \,  \biggl(\, 2 \, d_{\,1} \, \pd{u}{x} \moins \Pe \,\biggr) \, \pd{\Theta_{\,u,\,a,\,2}}{x} \plus \Fo \, d_{\,1} \, \pd{^{\,2} u}{x^{\,2}}  \, \Theta_{\,u,\,a,\,2} \\[3pt] 
  & \moins \Theta_{\,c,\,a,\,2} \, \pd{u}{t} \,, \\[3pt]
  \pd{\Theta_{\,c,\,a,\,2}}{t} \egal & \beta \ \mathrm{Rsign}^{\,\prime} \ \biggl(\, \pd{u}{t} \,\biggr) \pd{\Theta_{\,u,\,a,\,2}}{t} \,\, \biggl(\, c \moins c_{\,\mathrm{ads}} \,\biggr) \cdot \biggl(\, c \moins c_{\,\mathrm{des}} \,\biggr) \\[3pt]
  & \plus \beta \ \mathrm{Rsign} \ \biggl(\, \pd{u}{t} \,\biggr) \cdot \biggl(\, \Theta_{\,c,\,a,\,2} \moins \bigl(\, c_{\,1,\,a} \, \Theta_{\,u,\,a,\,2} \plus u^{\,2} \plus 2 \ c_{\,2,\,a} \, u \, \Theta_{\,u,\,a,\,2} \, \bigr)\, \biggr) \cdot \biggl(\, c \moins c_{\,\mathrm{des}} \,\biggr) \\[3pt]
  & \plus \beta \ \mathrm{Rsign} \ \biggl(\, \pd{u}{t} \,\biggr) \cdot \biggl(\, c \moins c_{\,\mathrm{ads}} \,\biggr) \cdot\biggl(\, \Theta_{\,c,\,a,\,2} \moins \bigl(\, c_{\,1,\,d} \, \Theta_{\,u,\,a,\,2} \plus 2 \ c_{\,2,\,d} \, u \, \Theta_{\,u,\,a,\,2} \, \bigr)\, \biggr) \,.
\end{align*}
For $\Theta_{\,u,\,d,\,0} \egal \displaystyle \pd{u}{c_{\,d,\,0\,}}$ and $\Theta_{\,c,\,d,\,0} \egal \displaystyle \pd{c}{c_{\,d,\,0\,}}\,$:
\begin{align*}
  \pd{\Theta_{\,u,\,d,\,0}}{t} \egal & \Fo \, d(\,u\,) \, \pd{^{\,2} \Theta_{\,u,\,d,\,0}}{x^{\,2}} \plus \Fo \,  \biggl(\, 2 \, d_{\,1} \, \pd{u}{x} \moins \Pe \,\biggr) \, \pd{\Theta_{\,u,\,d,\,0}}{x} \plus \Fo \, d_{\,1} \, \pd{^{\,2} u}{x^{\,2}}  \, \Theta_{\,u,\,\,d,\,0}  \\[3pt]
  & \moins \Theta_{\,c,\,d,\,0} \, \pd{u}{t} \,, \\[3pt]
  \pd{\Theta_{\,c,\,d,\,0}}{t} \egal & \beta \ \mathrm{Rsign}^{\,\prime} \ \biggl(\, \pd{u}{t} \,\biggr) \pd{\Theta_{\,u,\,d,\,0}}{t} \,\, \biggl(\, c \moins c_{\,\mathrm{ads}} \,\biggr) \cdot \biggl(\, c \moins c_{\,\mathrm{des}} \,\biggr) \\[3pt]
  & \plus \beta \ \mathrm{Rsign} \ \biggl(\, \pd{u}{t} \,\biggr) \cdot \biggl(\, \Theta_{\,c,\,d,\,0} \moins \bigl(\, c_{\,1,\,a} \, \Theta_{\,u,\,d,\,0} \plus 2 \ c_{\,2,\,a} \, u \, \Theta_{\,u,\,d,\,0} \, \bigr)\, \biggr) \cdot \biggl(\, c \moins c_{\,\mathrm{des}} \,\biggr) \\[3pt]
  & \plus \beta \ \mathrm{Rsign} \ \biggl(\, \pd{u}{t} \,\biggr) \cdot \biggl(\, c \moins c_{\,\mathrm{ads}} \,\biggr) \cdot \biggl(\, \Theta_{\,c,\,d,\,0} \moins \bigl(\, 1 \plus c_{\,1,\,d} \, \Theta_{\,u,\,d,\,0} \plus 2 \ c_{\,2,\,d} \, u \, \Theta_{\,u,\,d,\,0} \, \bigr)\, \biggr) \,.
\end{align*}
For $\Theta_{\,u,\,d,\,1} \egal \displaystyle \pd{u}{c_{\,d,\,1\,}}$ and $\Theta_{\,c,\,d,\,1} \egal \displaystyle \pd{c}{c_{\,d,\,1\,}}\,$:
\begin{align*}
  \pd{\Theta_{\,u,\,d,\,1}}{t} \egal & \Fo \, d(\,u\,) \, \pd{^{\,2} \Theta_{\,u,\,d,\,1}}{x^{\,2}} \plus \Fo \,  \biggl(\, 2 \, d_{\,1} \, \pd{u}{x} \moins \Pe \,\biggr) \, \pd{\Theta_{\,u,\,d,\,1}}{x} \plus \Fo \, d_{\,1} \, \pd{^{\,2} u}{x^{\,2}}  \, \Theta_{\,u,\,\,d,\,1}  \\[3pt]
  & \moins \Theta_{\,c,\,d,\,1} \, \pd{u}{t} \,, \\[3pt]
  \pd{\Theta_{\,c,\,d,\,1}}{t} \egal & \beta \ \mathrm{Rsign}^{\,\prime} \ \biggl(\, \pd{u}{t} \,\biggr) \pd{\Theta_{\,u,\,d,\,1}}{t} \,\, \biggl(\, c \moins c_{\,\mathrm{ads}} \,\biggr) \cdot \biggl(\, c \moins c_{\,\mathrm{des}} \,\biggr) \\[3pt]
  & \plus \beta \ \mathrm{Rsign} \ \biggl(\, \pd{u}{t} \,\biggr) \cdot \biggl(\, \Theta_{\,c,\,d,\,1} \moins \bigl(\, c_{\,1,\,a} \, \Theta_{\,u,\,d,\,1} \plus 2 \ c_{\,2,\,a} \, u \, \Theta_{\,u,\,d,\,1} \, \bigr)\, \biggr) \cdot \biggl(\, c \moins c_{\,\mathrm{des}} \,\biggr) \\[3pt]
  & \plus \beta \ \mathrm{Rsign} \ \biggl(\, \pd{u}{t} \,\biggr) \cdot \biggl(\, c \moins c_{\,\mathrm{ads}} \,\biggr) \cdot \biggl(\, \Theta_{\,c,\,d,\,1} \moins \bigl(\, c_{\,1,\,d} \, \Theta_{\,u,\,d,\,1} \plus u \plus 2 \ c_{\,2,\,d} \, u \, \Theta_{\,u,\,d,\,1} \, \bigr)\, \biggr) \,.
\end{align*}
For $\Theta_{\,u,\,d,\,2} \egal \displaystyle \pd{u}{c_{\,d,\,2\,}}$ and $\Theta_{\,c,\,d,\,2} \egal \displaystyle \pd{c}{c_{\,d,\,2\,}}\,$:
\begin{align*}
  \pd{\Theta_{\,u,\,d,\,2}}{t} \egal & \Fo \, d(\,u\,) \, \pd{^{\,2} \Theta_{\,u,\,d,\,2}}{x^{\,2}} \plus \Fo \,  \biggl(\, 2 \, d_{\,1} \, \pd{u}{x} \moins \Pe \,\biggr) \, \pd{\Theta_{\,u,\,d,\,2}}{x} \plus \Fo \, d_{\,1} \, \pd{^{\,2} u}{x^{\,2}}  \, \Theta_{\,u,\,\,d,\,2} \\[3pt] 
  & \moins \Theta_{\,c,\,d,\,2} \, \pd{u}{t} \,, \\[3pt]
  \pd{\Theta_{\,c,\,d,\,2}}{t} \egal & \beta \ \mathrm{Rsign}^{\,\prime} \ \biggl(\, \pd{u}{t} \,\biggr) \pd{\Theta_{\,u,\,d,\,2}}{t} \,\, \biggl(\, c \moins c_{\,\mathrm{ads}} \,\biggr) \ \biggl(\, c \moins c_{\,\mathrm{des}} \,\biggr) \\[3pt]
  & \plus \beta \ \mathrm{Rsign} \ \biggl(\, \pd{u}{t} \,\biggr) \cdot \biggl(\, \Theta_{\,c,\,d,\,2} \moins \bigl(\, c_{\,1,\,a} \, \Theta_{\,u,\,d,\,2} \plus 2 \ c_{\,2,\,a} \, u \, \Theta_{\,u,\,d,\,2} \, \bigr)\, \biggr) \cdot \biggl(\, c \moins c_{\,\mathrm{des}} \,\biggr) \\[3pt]
  & \plus \beta \ \mathrm{Rsign} \ \biggl(\, \pd{u}{t} \,\biggr) \cdot \biggl(\, c \moins c_{\,\mathrm{ads}} \,\biggr) \cdot \biggl(\, \Theta_{\,c,\,d,\,2} \moins \bigl(\, c_{\,1,\,d} \, \Theta_{\,u,\,d,\,2} \plus u^{\,2} \plus 2 \ c_{\,2,\,d} \, u \, \Theta_{\,u,\,d,\,2} \, \bigr)\, \biggr) \,.
\end{align*}
The function $\mathrm{Rsign}^{\,\prime} \,: \R \ \longrightarrow \ \bigl\{\,0 \,,\, 1 \,\bigr\} $ is the regularized \textsc{Dirac} function:
\begin{align*}
\mathrm{Rsign}^{\,\prime} (\,x\,) \egal 
\begin{cases} 
\, 1 \,, & \quad  x \egal 0 \,, \\
\, 0 \,, & \quad  x \ \neq \ 0 \,.
\end{cases}
\end{align*}


\section{Proving the structural identifiability of the parameters}
\label{sec:Annex_identifiability}

This section aims at justifying the identifiability of the unknown parameters for both model with and without hysteresis. We recall that a parameter $P$ is Structurally Globally Identifiable (SGI) if the following condition is satisfied \cite{Walter1982}:
\begin{align*}
  y\, (\,P\,) \ \equiv \ y \, (\,P^{\,\prime}\,) \qquad \Longrightarrow \qquad P \ \equiv \ P^{\,\prime} \,,
\end{align*}
where $y$ is the response of the model depending on parameter $P\,$.


\subsection{Model without hysteresis}

First, the SGI is demonstrated for the estimation of parameters $\bigl(\,\Fo \,,\, c_{\,1} \,,\,c_{\,2} \,\bigr)$ in the model~\eqref{eq:moisture_dimensionlesspb_1D}, recalled here without the superscript $^{\,\star}\,$: 
\begin{align}\label{eq:pb_noH_SGI_1}
  \bigl(\,1 \plus c_{\,1} \, u \plus c_{\,2} \, u^{\,2} \,\bigr)  \ \pd{u}{t} \moins \Fo \ \pd{}{x} \Biggl(\, \bigl(\, 1 \plus d_{\,1} \, u \,\bigr)  \ \pd{u}{x} \moins \Pe \ u \, \Biggr) \egal 0 \,,
\end{align}
It is assumed that $u$ is observable. So as to prove identifiability, it is assumed that another set of parameters, denoted with a superscript $^{\,\prime}\,$, holds:
\begin{align}\label{eq:pb_noH_SGI_2}
  \bigl(\,1 \plus c^{\,\prime}_{\,1} \, u^{\,\prime} \plus c^{\,\prime}_{\,2} \, (\,u^{\,\prime}\,)^{\,2} \,\bigr)  \ \pd{u^{\,\prime}}{t} \moins \Fo^{\,\prime} \ \pd{}{x} \Biggl(\, \bigl(\, 1 \plus d_{\,1} \, u^{\,\prime} \,\bigr)  \ \pd{u}{x} \moins \Pe \ u^{\,\prime} \, \Biggr) \egal 0  \,,
\end{align}
If $u \ \equiv \ u^{\,\prime}$ then $ \displaystyle \pd{u}{t} \ \equiv \ \pd{u^{\,\prime}}{t}$ and $ \displaystyle  \pd{u}{x} \ \equiv \ \pd{u^{\,\prime}}{x}\,$. Thus, from Equations~\eqref{eq:pb_noH_SGI_1} and \eqref{eq:pb_noH_SGI_2} we have:
\begin{align*}
  \bigl(\,1 \plus c_{\,1} \, u \plus c_{\,2} \, u^{\,2} \,\bigr) \ \equiv \ \bigl(\,1 \plus c^{\,\prime}_{\,1} \, u^{\,\prime} \plus c^{\,\prime}_{\,2} \, (\,u^{\,\prime}\,)^{\,2} \,\bigr) \,,
\end{align*}
and 
\begin{align*}
  \Fo \ \pd{}{x} \Biggl(\, \bigl(\, 1 \plus d_{\,1} \, u \,\bigr)  \ \pd{u}{x} \moins \Pe \ u \, \Biggr) \ \equiv \  \Fo^{\,\prime} \ \pd{}{x} \Biggl(\, \bigl(\, 1 \plus d_{\,1} \, u^{\,\prime} \,\bigr)  \ \pd{u}{x} \moins \Pe \ u^{\,\prime} \, \Biggr) \,.
\end{align*}
Since $u \ \equiv \ u^{\,\prime}$ and $u^{\,2} \ \equiv \ (\,u^{\,\prime}\,)^{\,2}\,$, it follows that
\begin{align*}
  & c_{\,1} \ \equiv \ c_{\,1}^{\,\prime} \,, && c_{\,2} \ \equiv \ c_{\,2}^{\,\prime}  \,, && \Fo^{\,\prime} \ \equiv \ \Fo^{\,\prime} \,.
\end{align*}
Therefore, parameters $\bigl(\,\Fo \,,\, c_{\,1} \,,\,c_{\,2} \,\bigr)$ are SGI for the model without hysteresis.


\subsection{Model with hysteresis}

Now, the SGI for the five parameters $\bigl(\,c^{\,\star}_{\,a,\,1}\,,\,c^{\,\star}_{\,a,\,2}\,,\,c^{\,\star}_{\,d,\,0}\,,\,c^{\,\star}_{\,d,\,1}\,,\,c^{\,\star}_{\,d,\,2}\,\bigr)\,$ is demonstrated for the model with hysteresis. For this, Eq.~\eqref{eq:model_hysteresis} is recalled omitting the superscript $^{\,\star}\,$: 
\begin{align*}
  \pd{c}{t} & \egal \beta \ \mathrm{sign} \ \biggl(\, \pd{u}{t} \,\biggr) \, \biggl(\, c \moins \bigl(\, 1 \plus c_{\,a,\,1} \, u \plus c_{\,a,\,2} \, u^{\,2} \,\bigr) \,\biggr) \ \biggl(\, c \moins \bigl(\, c_{\,d,\,0} \plus c_{\,d,\,1} \, u \plus c_{\,d,\,2} \, u^{\,2} \,\bigr) \,\biggr) \,.
\end{align*}
Similarly, to prove the identifiability, another set of parameters is assumed:
\begin{align*}
  \pd{c^{\,\prime}}{t}  \egal \beta \ \mathrm{sign} \ \biggl(\, \pd{u^{\,\prime}}{t} \,\biggr) \, \biggl(\, c^{\,\prime} \moins \bigl(\, 1 & \plus c^{\,\prime}_{\,a,\,1} \, u^{\,\prime} \plus c^{\,\prime}_{\,a,\,2} \, (\,u^{\,\prime}\,)^{\,2} \,\bigr) \,\biggr) \ \\
  & \biggl(\, c^{\,\prime} \moins \bigl(\, c^{\,\prime}_{\,d,\,0} \plus c^{\,\prime}_{\,d,\,1} \, u^{\,\prime} \plus c^{\,\prime}_{\,d,\,2} \, (\,u^{\,\prime}\,)^{\,2} \,\bigr) \,\biggr) \,.
\end{align*}
It is assumed that $c \ \equiv \ c^{\,\prime}\,$, $u \ \equiv \ u^{\,\prime}$  and then $\displaystyle \pd{c}{t} \ \equiv \ \pd{c^{\,\prime}}{t} \,$. Using the symbolic code \texttt{Maple\texttrademark}, by expansion it can be demonstrated that: 
\begin{align*}
  & c_{\,a,\,1} \ \equiv \ c^{\,\prime}_{\,a,\,1} \,,
  && c_{\,a,\,2} \ \equiv \ c^{\,\prime}_{\,a,\,2} \,,
  && c_{\,d,\,0} \ \equiv \ c^{\,\prime}_{\,d,\,0} \,,
  && c_{\,d,\,1} \ \equiv \ c^{\,\prime}_{\,d,\,1} \,,
  && c_{\,d,\,,2} \ \equiv \ c^{\,\prime}_{\,d,\,2} \,,
\end{align*}
and that the parameters are SGI.


\section{Numerical validation of the regularized hysteresis model}
\label{sec:Annex1}

\begin{figure}
  \begin{center}
  \subfigure[\label{fig:v_ft}]{\includegraphics[width=.48\textwidth]{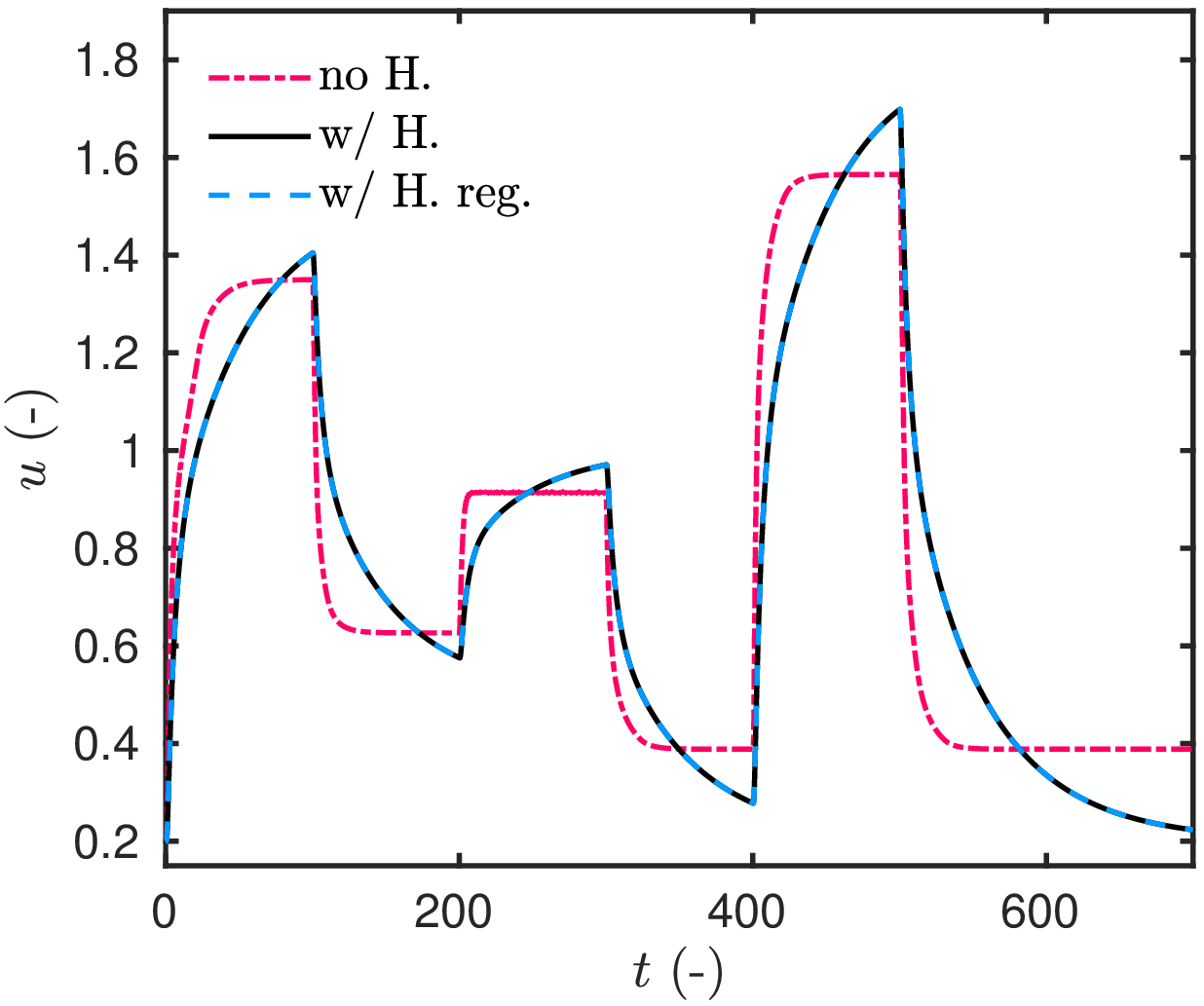}}
  \subfigure[\label{fig:c_ft}]{\includegraphics[width=.48\textwidth]{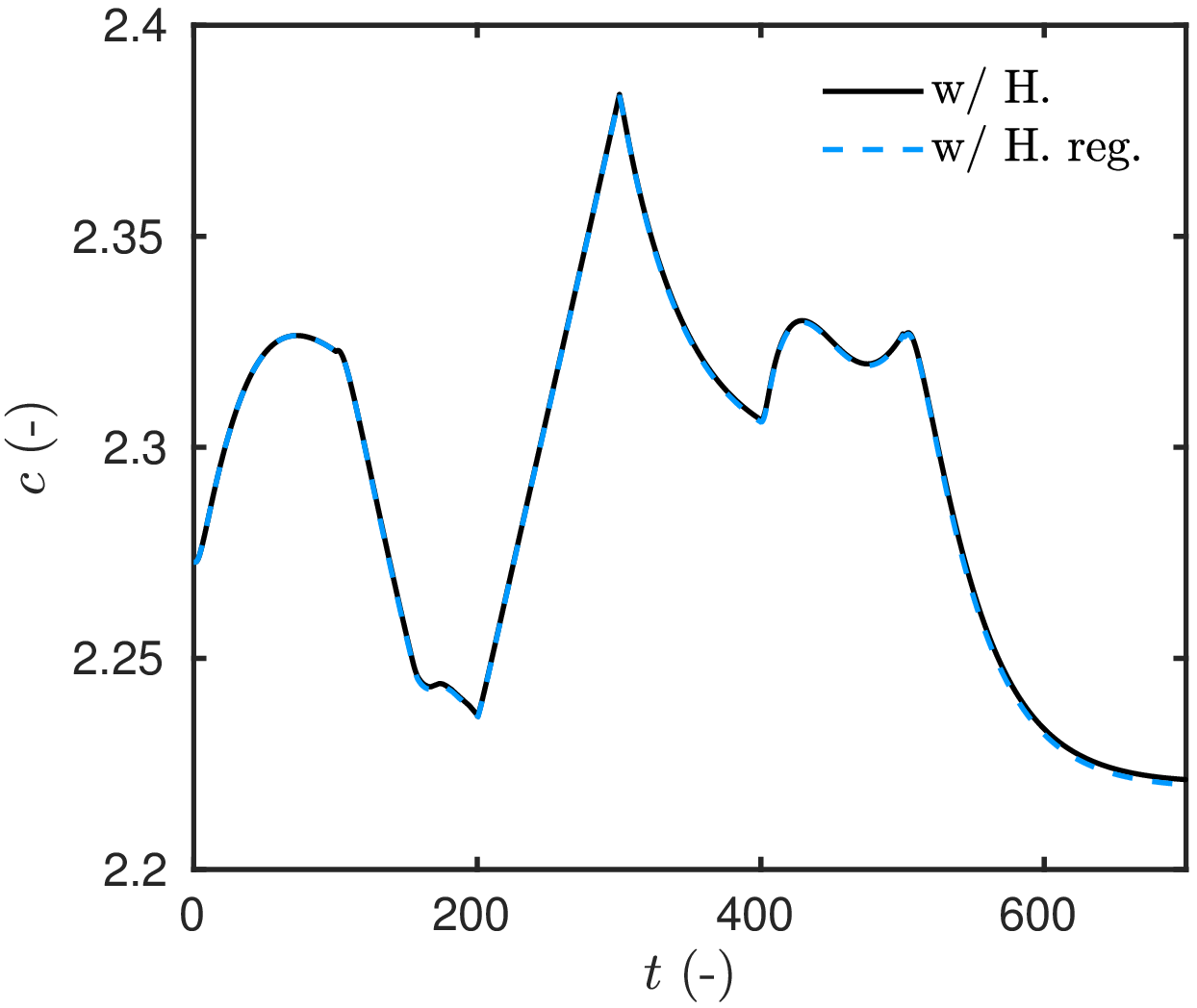}}
  \caption{\small\em Time variation of the field $u$ (a) and of the sorption capacity $c$ (b) for the model with hysteresis and the regularised one.}
  \end{center}
\end{figure}

\begin{figure}
\begin{center}
\subfigure[\label{fig:c_fv}]{\includegraphics[width=.48\textwidth]{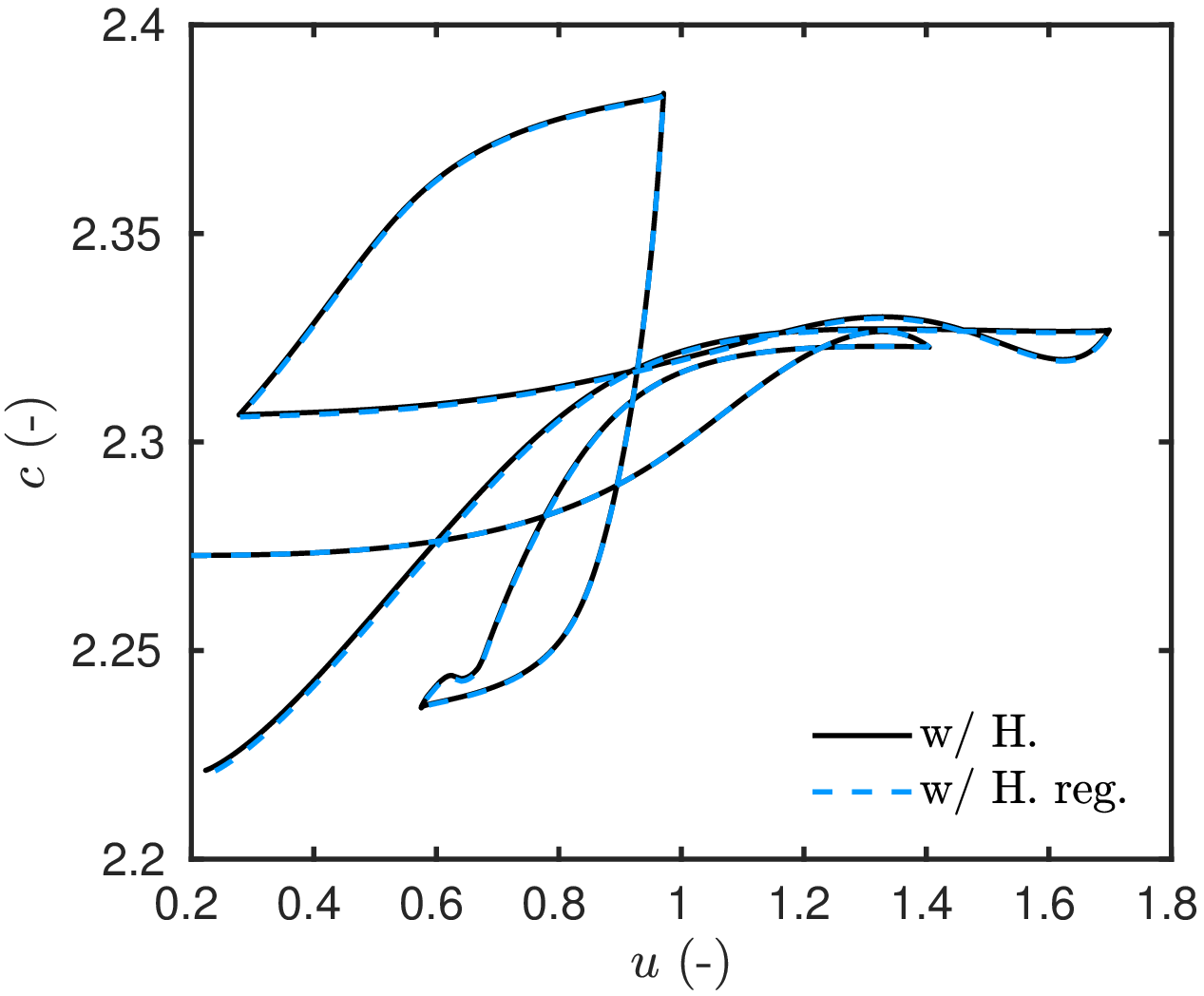}}
\subfigure[\label{fig:eps_ft}]{\includegraphics[width=.48\textwidth]{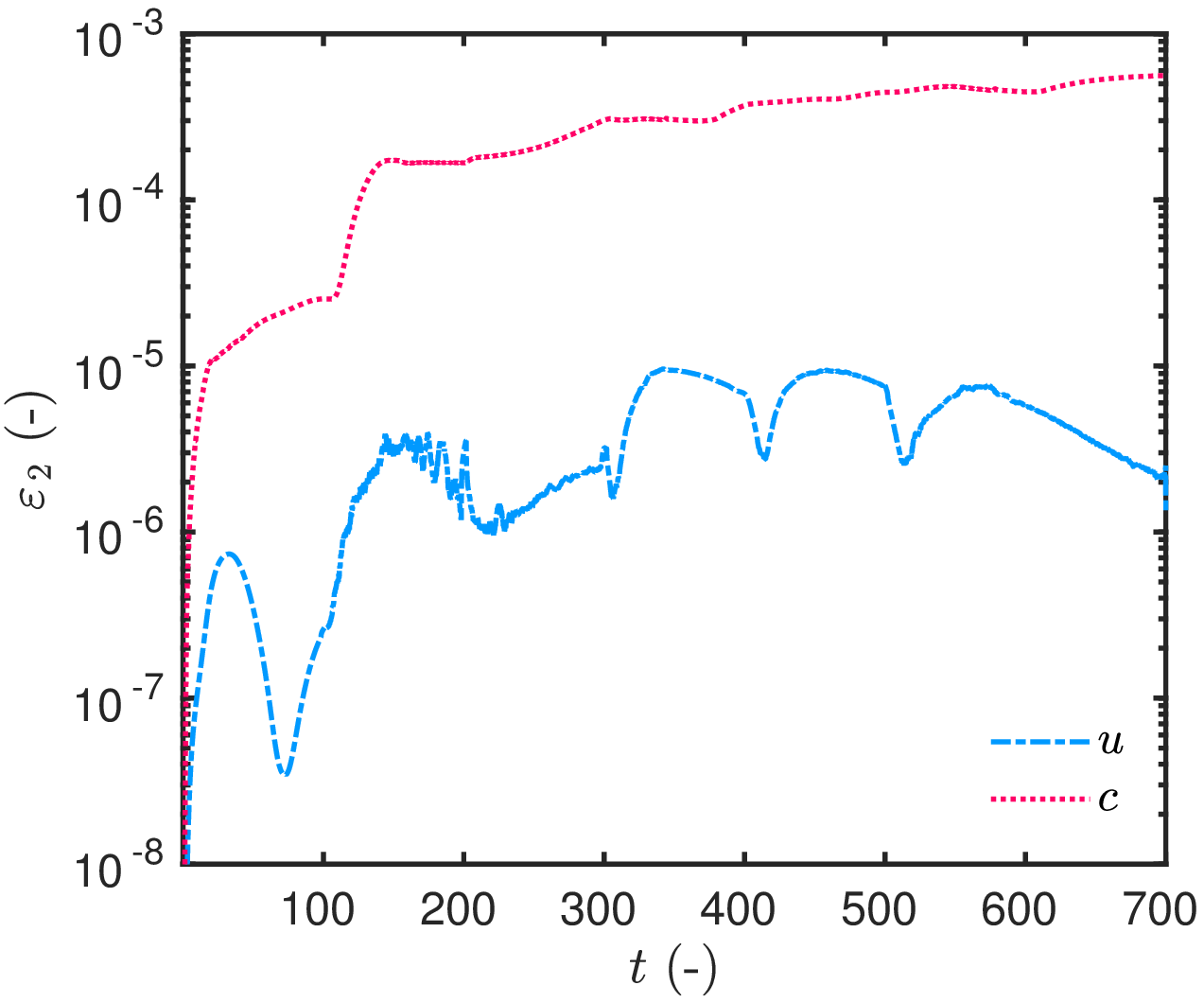}}
\caption{\small\em Variation of the sorption capacity $c$ with the field $u$ (a) according to the case study. Time variation of the $\mathcal{L}_{\,2}$ error (b) for the fields $u$ and $c$ between the model with hysteresis and the regularised one.}
\end{center}
\end{figure}

A computation using the regularized hysteresis model is carried out:
\begin{align*}
  \cs \ \pd{u}{\ts} &\egal \Fo \ \pd{}{\xs} \Biggl(\, \ds(\,u\,)  \ \pd{u}{\xs} \moins \Pe \ u \, \Biggr) \,, \\ 
  \pd{\cs}{t} & \egal \beta \ \mathrm{Rsign} \ \biggl(\, \pd{u}{t} \,\biggr) \cdot \biggl(\, \cs \moins c^{\,\star}_{\,\mathrm{ads}}\,(\,u\,) \,\biggr) \cdot \biggl(\, \cs \moins c^{\,\star}_{\,\mathrm{des}}\,(\,u\,) \,\biggr) \,,
\end{align*}
where the numerical values of the coefficients are: 
\begin{align*}
  & \Pe \egal 1 \cdot 10^{\,-2} \,,
  \quad \Fo \egal 2 \cdot 10^{\,-2} \,,
  && \ds(\,u\,) \egal 0.86 \plus 0.25 \,u \,, \quad \beta \egal 10^{\,-3} \,, \\
  & c^{\,\star}_{\,\mathrm{des}}(\,u\,) \egal 2.54 \moins 4.17 \, u \plus 3.02 \, u^{\,2} \,,
  && c^{\,\star}_{\,\mathrm{ads}}(\,u\,) \egal 3.36 \moins 6.11 \, u \plus 3.37 \, u^{\,2} \,.
\end{align*}
The initial and boundary conditions are defined in Eq.~\eqref{eq:moisture_dimensionlesspb_1D} with the following numerical values:
\begin{align*}
  \Bi \egal 13.7 \,, 
  && \ui \egal 0.2 \,,
  && \uL \egal \begin{cases} 
  \, 1.5 \,, & \quad  t \ \geq \ 0 \,, \quad t \ < \ 100  \,, \\
  \, 0.5 \,, & \quad  t \ \geq \ 100 \,, \quad t \ < \ 200 \,, \\
  \, 1.0 \,, & \quad  t \ \geq \ 200 \,, \quad t \ < \ 300 \,, \\
  \, 0.2 \,, & \quad  t \ \geq \ 300 \,, \quad t \ < \ 400 \,, \\
  \, 1.8 \,, & \quad  t \ \geq \ 400 \,, \quad t \ < \ 500 \,. \\
  \end{cases}
\end{align*}
The case study is similar to the experimental designs investigated in this work. The simulation time horizon is $t \egal 700\,$. The time variations of the field $u$ is given in Figure~\ref{fig:v_ft}. The results of the regularized and non-regularized model are compared to the model with no-hysteresis, considering only the adsorption curve. The importance of the hysteresis in the time variations of $u$ can be noted. Figure~\ref{fig:c_ft} shows the time variation of $c$ by computation of its differential equation. The variations of the sorption coefficient $c$ with the computed values of $u$ are shown in Figure~\ref{fig:c_fv}. In both Figures~\ref{fig:c_ft} and \ref{fig:c_fv}, a perfect agreement is observed between the regularized and non-regularized models. Moreover, as noticed in Figure~\ref{fig:eps_ft}, the $\mathcal{L}_{\,2}$ error of the fields $u$ and $c\,$, computed between the regularized and non-regularized models, scales with $10^{\,-5}$ and $10^{\,-3}\,$, respectively. The agreement between the results of the two models is very satisfactory, validating the implementation of the regularized hysteretic model.

\section{Implicit-Explicit numerical scheme for the hysteresis model}
\label{sec:Annex2}

To relax stability restriction, an implicit--explicit numerical scheme is used to compute the solution of the hysteresis model:
\begin{align}\label{eq:hysteresis_model_annex}
  \pd{c}{t} & \egal \beta \ \mathrm{sign} \ \biggl(\, \pd{u}{t} \,\biggr) \cdot \biggl(\, c \moins c_{\,\mathrm{ads}}\,(\,u\,) \,\biggr) \cdot \biggl(\, c \moins c_{\,\mathrm{des}}\,(\,u\,) \,\biggr) \,.
\end{align}
For the sake of clarity, the $\star$ superscript have been omitted. A uniform discretisation is considered for time intervals. The discretisation parameter is denoted using $\Delta t$ for the time.  The discrete values of function $c\,(\,t\,)$ are denoted by $c^{\,n} \ \eqdef \ c\,(\,t^{\,n}\,)$ with $n \ \in \ \bigl\{\, 1 \,, \ldots \,, N_{\,t} \,\bigr\}\,$.

When $\displaystyle \pd{u}{t} \ < \ 0\,$, Equation~\eqref{eq:hysteresis_model_annex} is discretised according to:
\begin{align*}
  \frac{1}{\Delta t} \, \bigl(\,c^{\,n+1} \moins c^{\,n} \,\bigr) \egal \beta \ \mathrm{sign} \ \biggl(\, \pd{u}{t} \,\biggr) \cdot \bigl(\,c^{\,n+1} \moins c_{\,\mathrm{ads}} \,\bigr)\cdot \bigl(\,c^{\,n} \moins c_{\,\mathrm{des}} \,\bigr) \,,
\end{align*}
which gives the explicit expression of $c^{\,n+1}\,$:
\begin{align*}
  c^{\,n+1} \egal \frac{c^{\,n} \moins \Delta t \, \beta \,  \mathrm{sign} \ \biggl(\, \displaystyle \pd{u}{t} \,\biggr) \ \bigl(\,c^{\,n} \moins c_{\,\mathrm{des}} \, \bigr) \, c_{\,\mathrm{ads}}}{1 \moins \Delta t \, \beta \, \mathrm{sign} \ \biggl(\, \displaystyle \pd{u}{t} \,\biggr)  \ \bigl(\,c^{\,n} \moins c_{\,\mathrm{des}} \, \bigr)}\  \,.
\end{align*}
Similarly, when $\displaystyle \pd{u}{t} \ > \ 0\,$, Equation~\eqref{eq:hysteresis_model_annex} is discretised according to:
\begin{align*}
  \frac{1}{\Delta t} \, \bigl(\,c^{\,n+1} \moins c^{\,n} \,\bigr) \egal \beta \ \mathrm{sign} \ \biggl(\, \pd{u}{t} \,\biggr) \, \bigl(\,c^{\,n} \moins c_{\,\mathrm{ads}} \,\bigr)\, \bigl(\,c^{\,n+1} \moins c_{\,\mathrm{des}} \,\bigr) \,,
\end{align*}
to obtain the explicit expression of $c^{\,n+1}\,$:
\begin{align*}
  c^{\,n+1} \egal \frac{c^{\,n} \moins \Delta t \, \beta \,  \mathrm{sign} \ \biggl(\, \displaystyle \pd{u}{t} \,\biggr) \ \bigl(\,c^{\,n} \moins c_{\,\mathrm{ads}} \, \bigr) \, c_{\,\mathrm{des}}}{1 \moins \Delta t \, \beta \, \mathrm{sign} \ \biggl(\, \displaystyle \pd{u}{t} \,\biggr)  \ \bigl(\,c^{\,n} \moins c_{\,\mathrm{ads}} \, \bigr)} \,.
\end{align*}
This numerical scheme provides robust and stable results as already shown in Figures~\ref{fig:EST_wH_v} and \ref{fig:EST_wH_c}.


\bigskip\bigskip
\addcontentsline{toc}{section}{References}
\bibliographystyle{abbrv}
\bibliography{biblio}

\begin{thebibliography}{10}

\bibitem{Administration2015}
U.~E.~I. Administration.
\newblock {\em {Annual Energy Outlook 2015, with projections to 2040}}.
\newblock EIA, Washington, DC, 2015.

\bibitem{Alifanov1995}
O.~M. Alifanov, E.~A. Artioukhine, and S.~V. Rumyantsev.
\newblock {\em {Extreme Methods for Solving Ill-Posed Problems with
  Applications to Inverse Heat Transfer Problems}}.
\newblock Begellhouse, New York, 1995.

\bibitem{Anderson2005}
M.~L. Anderson, W.~Bangerth, and G.~F. Carey.
\newblock {Analysis of parameter sensitivity and experimental design for a
  class of nonlinear partial differential equations}.
\newblock {\em Int. J. Num. Meth. Fluids}, 48(6):583--605, jun 2005.

\bibitem{Artyukhin1985}
E.~A. Artyukhin and S.~A. Budnik.
\newblock {Optimal planning of measurements in numerical experiment
  determination of the characteristics of a heat flux}.
\newblock {\em Journal of Engineering Physics}, 49(6):1453--1458, dec 1985.

\bibitem{BauklimatikDresden2011}
B.~{Bauklimatik Dresden}.
\newblock {Simulation program for the calculation of coupled heat, moisture,
  air, pollutant, and salt transport}.
\newblock {\em http://www.bauklimatik-dresden.de/delphin/index.php?aLa=en},
  2011.

\bibitem{Beck1977}
J.~V. Beck and K.~J. Arnold.
\newblock {\em {Parameter Estimation in Engineering and Science}}.
\newblock John Wiley {\&} Sons, Inc., New York, 1977.

\bibitem{Belleudy2016}
C.~Belleudy, M.~Woloszyn, M.~Chhay, and M.~Cosnier.
\newblock {A 2D model for coupled heat, air, and moisture transfer through
  porous media in contact with air channels}.
\newblock {\em Int. J. Heat Mass Transfer}, 95:453--465, apr 2016.

\bibitem{Berger2017b}
J.~Berger, T.~Busser, D.~Dutykh, and N.~Mendes.
\newblock {On the estimation of moisture permeability and advection
  coefficients of a wood fibre material using the optimal experiment design
  approach}.
\newblock {\em Experimental Thermal and Fluid Science}, 90:246--259, jan 2018.

\bibitem{Berger2017}
J.~Berger, D.~Dutykh, and N.~Mendes.
\newblock {On the optimal experiment design for heat and moisture parameter
  estimation}.
\newblock {\em Exp. Therm. Fluid Sci.}, 81:109--122, feb 2017.

\bibitem{Berger2017a}
J.~Berger, S.~Gasparin, D.~Dutykh, and N.~Mendes.
\newblock {Accurate numerical simulation of moisture front in porous material}.
\newblock {\em Building and Environment}, 118:211--224, jun 2017.

\bibitem{Berger2018a}
J.~Berger, S.~Gasparin, D.~Dutykh, and N.~Mendes.
\newblock {On the Solution of Coupled Heat and Moisture Transport in Porous
  Material}.
\newblock {\em Transport in Porous Media}, 121(3):665--702, feb 2018.

\bibitem{Berger2016b}
J.~Berger, H.~R.~B. Orlande, N.~Mendes, and S.~Guernouti.
\newblock {Bayesian inference for estimating thermal properties of a historic
  building wall}.
\newblock {\em Building and Environment}, 106:327--339, sep 2016.

\bibitem{Biddulph2014}
P.~Biddulph, V.~Gori, C.~A. Elwell, C.~Scott, C.~Rye, R.~Lowe, and
  T.~Oreszczyn.
\newblock {Inferring the thermal resistance and effective thermal mass of a
  wall using frequent temperature and heat flux measurements}.
\newblock {\em Energy and Buildings}, 78:10--16, aug 2014.

\bibitem{Busser2016}
T.~Busser, A.~Piot, M.~Pailha, T.~Bejat, and M.~Woloszyn.
\newblock {From materials properties to modelling hygrothermal transfers of
  highly hygroscopic walls}.
\newblock In {\em CESBP}, Dresden, Germany, 2016.

\bibitem{Byrd2000}
R.~H. Byrd, J.~C. Gilbert, and J.~Nocedal.
\newblock {A Trust Region Method Based on Interior Point Techniques for
  Nonlinear Programming}.
\newblock {\em Math. Progr.}, 89(1):149--185, 2000.

\bibitem{Czel2012}
B.~Cz{\'{e}}l and G.~Gr{\'{o}}f.
\newblock {Inverse identification of temperature-dependent thermal conductivity
  via genetic algorithm with cost function-based rearrangement of genes}.
\newblock {\em Int. J. Heat Mass Transfer}, 55(15-16):4254--4263, jul 2012.

\bibitem{DelBarrio2003}
E.~P. {Del Barrio}.
\newblock {Multidimensional inverse heat conduction problems solution via
  lagrange theory and model size reduction techniques}.
\newblock {\em Inverse Problems in Engineering}, 11(6):515--539, dec 2003.

\bibitem{Delleur2006}
J.~W. Delleur.
\newblock {\em {The handbook of groundwater engineering}}.
\newblock CRC Press, Boca Raton, FL, 2 edition, 2006.

\bibitem{Derluyn2012}
H.~Derluyn, D.~Derome, J.~Carmeliet, E.~Stora, and R.~Barbarulo.
\newblock {Hysteretic moisture behavior of concrete: Modeling and analysis}.
\newblock {\em Cement and Concrete Research}, 42(10):1379--1388, oct 2012.

\bibitem{Dubois2014}
S.~Dubois, F.~McGregor, A.~Evrard, A.~Heath, and F.~Lebeau.
\newblock {An inverse modelling approach to estimate the hygric parameters of
  clay-based masonry during a Moisture Buffer Value test}.
\newblock {\em Building and Environment}, 81:192--203, nov 2014.

\bibitem{Emery1998}
A.~F. Emery and A.~V. Nenarokomov.
\newblock {Optimal experiment design}.
\newblock {\em Measurement Science and Technology}, 9(6):864--876, jun 1998.

\bibitem{Fadale1995}
T.~D. Fadale, A.~V. Nenarokomov, and A.~F. Emery.
\newblock {Two Approaches to Optimal Sensor Locations}.
\newblock {\em Journal of Heat Transfer}, 117(2):373, 1995.

\bibitem{Finsterle2015}
S.~Finsterle.
\newblock {Practical notes on local data-worth analysis}.
\newblock {\em Water Resources Research}, 51(12):9904--9924, dec 2015.

\bibitem{IBP2005}
I.~Fraunhofer.
\newblock {Wufi}.
\newblock {\em http://www.hoki.ibp.fhg.de/wufi/wufi{\_}frame{\_}e.html}, 2005.

\bibitem{Gasparin2017b}
S.~Gasparin, J.~Berger, D.~Dutykh, and N.~Mendes.
\newblock {An improved explicit scheme for whole-building hygrothermal
  simulation}.
\newblock {\em Building Simulation}, pages 1--40, nov 2017.

\bibitem{Gasparin2017}
S.~Gasparin, J.~Berger, D.~Dutykh, and N.~Mendes.
\newblock {Stable explicit schemes for simulation of nonlinear moisture
  transfer in porous materials}.
\newblock {\em J. Building Perf. Simul.}, 11(2):129--144, 2018.

\bibitem{James2010}
C.~James, C.~J. Simonson, P.~Talukdar, and S.~Roels.
\newblock {Numerical and experimental data set for benchmarking hygroscopic
  buffering models}.
\newblock {\em Int. J. Heat Mass Transfer}, 53(19-20):3638--3654, sep 2010.

\bibitem{Janssen2007}
H.~Janssen, B.~Blocken, and J.~Carmeliet.
\newblock {Conservative modelling of the moisture and heat transfer in building
  components under atmospheric excitation}.
\newblock {\em Int. J. Heat Mass Transfer}, 50(5-6):1128--1140, mar 2007.

\bibitem{Kabanikhin2008}
S.~I. Kabanikhin.
\newblock {Definitions and examples of inverse and ill-posed problems}.
\newblock {\em J. Inv. Ill-posed Problems}, 16(4):317--357, jan 2008.

\bibitem{Kabanikhin2011}
S.~I. Kabanikhin.
\newblock {\em {Inverse and ill-posed problems: theory and applications}}.
\newblock Walter De Gruyter, Berlin, 2011.

\bibitem{Kabanikhin2008a}
S.~I. Kabanikhin, A.~Hasanov, and A.~V. Penenko.
\newblock {A gradient descent method for solving an inverse coefficient heat
  conduction problem}.
\newblock {\em Numerical Analysis and Applications}, 1(1):34--45, jan 2008.

\bibitem{Kalagasidis2007}
A.~S. Kalagasidis, P.~Weitzmann, T.~R. Nielsen, R.~Peuhkuri, C.-E. Hagentoft,
  and C.~Rode.
\newblock {The International Building Physics Toolbox in Simulink}.
\newblock {\em Energy and Buildings}, 39(6):665--674, jun 2007.

\bibitem{Kanevce2005}
G.~H. Kanevce, L.~P. Kanevce, G.~S. Dulikravich, and H.~R.~B. Orlande.
\newblock {Estimation of thermophysical properties of moist materials under
  different drying conditions}.
\newblock {\em Inverse Problems in Science and Engineering}, 13(4):341--353,
  aug 2005.

\bibitem{Karalashvili2015}
M.~Karalashvili, W.~Marquardt, and A.~Mhamdi.
\newblock {Optimal experimental design for identification of transport
  coefficient models in convection-diffusion equations}.
\newblock {\em Computers {\&} Chemical Engineering}, 80:101--113, sep 2015.

\bibitem{Koptyug2000}
I.~V. Koptyug, S.~I. Kabanikhin, K.~T. Iskakov, V.~B. Fenelonov, L.~Y.
  Khitrina, R.~Z. Sagdeev, and V.~N. Parmon.
\newblock {A quantitative NMR imaging study of mass transport in porous solids
  during drying}.
\newblock {\em Chemical Engineering Science}, 55(9):1559--1571, may 2000.

\bibitem{Luikov1966}
A.~V. Luikov.
\newblock {\em {Heat and mass transfer in capillary-porous bodies}}.
\newblock Pergamon Press, New York, 1966.

\bibitem{Mendes2017}
N.~Mendes, M.~Chhay, J.~Berger, and D.~Dutykh.
\newblock {\em {Numerical methods for diffusion phenomena in building
  physics}}.
\newblock PUCPRess, Curitiba, Parana, 2017.

\bibitem{Mendes2004}
N.~Mendes and P.~C. Philippi.
\newblock {Multitridiagonal-Matrix Algorithm for Coupled Heat Transfer in
  Porous Media: Stability Analysis and Computational Performance}.
\newblock {\em Journal of Porous Media}, 7(3):193--212, 2004.

\bibitem{Mendes2005}
N.~Mendes and P.~C. Philippi.
\newblock {A method for predicting heat and moisture transfer through
  multilayered walls based on temperature and moisture content gradients}.
\newblock {\em Int. J. Heat Mass Transfer}, 48(1):37--51, 2005.

\bibitem{Mendes1999}
N.~Mendes, I.~Ridley, R.~Lamberts, P.~C. Philippi, and K.~Budag.
\newblock {Umidus: A PC program for the Prediction of Heat and Mass Transfer in
  Porous Building Elements}.
\newblock In {\em IBPSA 99}, pages 277--283, Japan, 1999. International
  Conference on Building Performance Simulation.

\bibitem{Mualem2009}
Y.~Mualem and A.~Beriozkin.
\newblock {General scaling rules of the hysteretic water retention function
  based on Mualem's domain theory}.
\newblock {\em European Journal of Soil Science}, 60(4):652--661, aug 2009.

\bibitem{Nassiopoulos2013}
A.~Nassiopoulos and F.~Bourquin.
\newblock {On-Site Building Walls Characterization}.
\newblock {\em Numerical Heat Transfer, Part A: Applications}, 63(3):179--200,
  jan 2013.

\bibitem{Nenarokomov2005}
A.~V. Nenarokomov and D.~V. Titov.
\newblock {Optimal experiment design to estimate the radiative properties of
  materials}.
\newblock {\em Journal of Quantitative Spectroscopy and Radiative Transfer},
  93(1-3):313--323, jun 2005.

\bibitem{Ozisik2000}
M.~N. Ozisik and H.~R.~B. Orlande.
\newblock {\em {Inverse Heat Transfer: Fundamentals and Applications}}.
\newblock CRC Press, New York, 2000.

\bibitem{Rafidiarison2015}
H.~Rafidiarison, R.~R{\'{e}}mond, and E.~Mougel.
\newblock {Dataset for validating 1-D heat and mass transfer models within
  building walls with hygroscopic materials}.
\newblock {\em Building and Environment}, 89:356--368, jul 2015.

\bibitem{Rouchier2017}
S.~Rouchier, T.~Busser, M.~Pailha, A.~Piot, and M.~Woloszyn.
\newblock {Hygric characterization of wood fiber insulation under uncertainty
  with dynamic measurements and Markov Chain Monte-Carlo algorithm}.
\newblock {\em Building and Environment}, 114:129--139, mar 2017.

\bibitem{Rouchier2013}
S.~Rouchier, M.~Woloszyn, G.~Foray, and J.-J. Roux.
\newblock {Influence of concrete fracture on the rain infiltration and thermal
  performance of building facades}.
\newblock {\em Int. J. Heat Mass Transfer}, 61:340--352, jun 2013.

\bibitem{Rouchier2016}
S.~Rouchier, M.~Woloszyn, Y.~Kedowide, and T.~B{\'{e}}jat.
\newblock {Identification of the hygrothermal properties of a building envelope
  material by the covariance matrix adaptation evolution strategy}.
\newblock {\em J. Building Perf. Simul.}, 9(1):101--114, jan 2016.

\bibitem{Shampine1997}
L.~F. Shampine and M.~W. Reichelt.
\newblock {The MATLAB ODE Suite}.
\newblock {\em SIAM J. Sci. Comput.}, 18:1--22, 1997.

\bibitem{Ucinski2004}
D.~Ucinski.
\newblock {\em {Optimal Measurement Methods for Distributed Parameter System
  Identification}}.
\newblock 2004.

\bibitem{VandeWouwer2000}
A.~{Vande Wouwer}, N.~Point, S.~Porteman, and M.~Remy.
\newblock {An approach to the selection of optimal sensor locations in
  distributed parameter systems}.
\newblock {\em Journal of Process Control}, 10(4):291--300, aug 2000.

\bibitem{Vololonirina2014}
O.~Vololonirina, M.~Coutand, and B.~Perrin.
\newblock {Characterization of hygrothermal properties of wood-based products -
  Impact of moisture content and temperature}.
\newblock {\em Construction and Building Materials}, 63:223--233, jul 2014.

\bibitem{Walter1982}
E.~Walter and Y.~Lecourtier.
\newblock {Global approaches to identifiability testing for linear and
  nonlinear state space models}.
\newblock {\em Math. Comp. Simul.}, 24(6):472--482, dec 1982.

\bibitem{Walter1990}
E.~Walter and L.~Pronzato.
\newblock {Qualitative and quantitative experiment design for phenomenological
  models - A survey}.
\newblock {\em Automatica}, 26(2):195--213, mar 1990.

\bibitem{Woloszyn2008}
M.~Woloszyn and C.~Rode.
\newblock {Tools for performance simulation of heat, air and moisture
  conditions of whole buildings}.
\newblock {\em Building Simulation}, 1(1):5--24, mar 2008.

\bibitem{Xu2007}
X.~Xu and S.~Wang.
\newblock {Optimal simplified thermal models of building envelope based on
  frequency domain regression using genetic algorithm}.
\newblock {\em Energy and Buildings}, 39(5):525--536, may 2007.

\end{thebibliography}
\bigskip\bigskip

\end{document}